%% file: main.tex
\documentclass[12pt]{article}

\usepackage[utf8]{inputenc}
\usepackage[T1]{fontenc}
\usepackage[english]{babel}

\usepackage{charter} 
\usepackage{fullpage}
\usepackage{microtype} 
\usepackage{verbatim}
\usepackage{setspace}
\usepackage{fancyhdr}
\usepackage{longtable}
\usepackage{amsmath, amsthm, amstext, amsfonts, amssymb}
\usepackage{titlesec}
\usepackage{graphicx}
\usepackage[format=plain,labelsep=period,justification=justified,singlelinecheck=false,font=small]{caption}
\usepackage{ragged2e}
\usepackage{float}
\usepackage{natbib}
\usepackage{multibib}
\usepackage{bigstrut}
\usepackage{rotating}
\usepackage{multirow}
\usepackage{comment}
\usepackage{array}
\usepackage{booktabs}
\usepackage{wasysym}
\usepackage{dsfont}
\usepackage{bbm}
\usepackage[hyperfootnotes=false,citebordercolor={0 0 1}, citecolor={customdblue}, linkcolor={customdred}, urlcolor={customdred}, colorlinks]{hyperref}
\usepackage[margin=1in]{geometry}
\usepackage{afterpage}
\usepackage{pdflscape}
\usepackage{footnote}
\makesavenoteenv{tabular}
\usepackage{url}
\usepackage{mathtools}
\usepackage{overpic}
\usepackage{calc}
\usepackage{ifthen}
\usepackage{subcaption}
\usepackage{centernot}
\usepackage{physics}
\usepackage{etoolbox}
\usepackage{graphics}
\usepackage{nicefrac}
\usepackage{dcolumn}
\usepackage{adjustbox}
\usepackage{color, colortbl}
\usepackage{MnSymbol}
\usepackage{cleveref}
\usepackage{xcolor}
\usepackage{soul}
\usepackage{indentfirst} 
\usepackage{bm}
\usepackage[toc,page]{appendix} 
\usepackage{tocloft} 
\usepackage{pythonhighlight} 
\usepackage{fontawesome} 
\usepackage{xr} 
\usepackage{enumitem}

\usepackage[most]{tcolorbox}
\tcbuselibrary{skins, breakable}

\usepackage{tikz}
\usetikzlibrary{decorations.pathreplacing, arrows.meta, positioning, fit, calc}

\hypersetup{
    bookmarksdepth=subsection,
    bookmarksopen=false,
    bookmarksnumbered=true,
    pdftitle={AI Financial Advice: Supply, Demand, and Life Cycle Implications},
    pdfauthor={Taha Choukhmane, Tim de Silva, Weidong Lin, and Matthew Akuzawa},
    pdfsubject={Personal financial advice from large language models},
    pdfkeywords={large language models, financial advice, life cycle, household finance},
}

\setcitestyle{aysep={}} 

\newcolumntype{L}{>{$}l<{$}}
\newcolumntype{C}{>{$}c<{$}} 

\definecolor{darkmidnightblue}{rgb}{0.0, 0.2, 0.4}
\definecolor{customorange}{RGB}{170,86,2}
\definecolor{customdorange}{RGB}{92,44,1}
\definecolor{customdred}{RGB}{138,1,51}
\definecolor{customdblue}{RGB}{1,50,110}

\renewcommand{\thetable}{\arabic{table}}

\titlespacing*{\section}{0pt}{0.05in}{0.03in}
\titlespacing*{\subsection}{0pt}{0.05in}{0.03in}
\titlespacing*{\subsection}{0pt}{0.03in}{0.03in}
\titlespacing*{\paragraph}{0pt}{2pt}{10pt}

\allowdisplaybreaks

\newtcolorbox{LLMPrompt}[1][]{%
  breakable,
  colframe=customdred,
  colback=customdred!10!white,
  title=\textsc{Prompt Example},
  box align=center,
  #1
}

\newcites{app}{References }

\usepackage{xspace}
\newcommand{\GPT}{\texttt{GPT-5.2}\xspace}
\newcommand{\GPTMini}{\texttt{GPT-5 Mini}\xspace}
\newcommand{\Gemini}{\texttt{Gemini 3 Flash}\xspace}
\newcommand{\GPTNew}{\texttt{GPT-5.6 Terra}\xspace}

\newcommand{\papertitle}{AI Financial Advice:\\ Supply, Demand, and Life Cycle Implications}

\newtcolorbox{surveycard}[1]{%
  enhanced,
  breakable,
  colback=white,
  colframe=black!25,
  boxrule=0.6pt,
  arc=2mm,
  left=6pt,right=6pt,top=6pt,bottom=6pt,
  title=\textcolor{blue!70!black}{\bfseries #1},
  coltitle=black,
  fonttitle=\normalsize,
}

\newcommand{\surveytextbox}[1][4.0cm]{%
  \vspace{4pt}
  \begin{tcolorbox}[
    colback=black!3,
    colframe=black!25,
    boxrule=0.6pt,
    arc=1.5mm,
    height=#1,
    left=0pt,right=0pt,top=0pt,bottom=0pt
  ]
  \end{tcolorbox}
}

\begin{document}



\renewcommand{\thefootnote}{\fnsymbol{footnote}}

\title{\textsc{\papertitle\footnote{
	First draft: November 2025. We thank John Beshears, James Choi, Anastassia Fedyk, Alberto Rossi, Jon Reuter, Darrell Duffie, Fiona Greig, Jonathan Parker, Chris Tonetti, Tony Cookson, Aizhan Anarkulova, Steffen Andersen, Antonio Coppola, Eric So, and conference and seminar participants at MIT Sloan, Vanguard, Stanford IFDM, Erasmus, CEPR Household Finance, SFS Cavalcade, FIRS, WFA, and the NBER Household Finance and Economics of Aging meetings, for helpful comments and feedback. An earlier version of this paper was circulated with the title ``How Good is Generative AI Personal Financial Advice?''. 
}}}
\author{
	\href{https://www.tahachoukhmane.com/}{Taha Choukhmane}\footnote{MIT Sloan School of Management and NBER, \href{mailto:tahac@mit.edu}{tahac@mit.edu}.} \hspace{-0.15in}
		\and \href{https://www.timdesilva.me}{Tim de Silva}\footnote{Stanford University, Graduate School of Business and HAI, \href{mailto:tdesilva@stanford.edu}{tdesilva@stanford.edu}.} \hspace{-0.15in}
    \and \href{https://mitsloan.mit.edu/programs/phd/weidong-weldon-lin}{Weidong Lin}\footnote{MIT Sloan School of Management, \href{mailto:wel@mit.edu}{wel@mit.edu}.} \hspace{-0.15in}
    \and \href{https://www.linkedin.com/in/matthew-akuzawa-4182a21ba}{Matthew Akuzawa}\footnote{MIT Sloan School of Management, \href{mailto:maku49@mit.edu}{maku49@mit.edu}.}
}
\date{July 29, 2026}
\onehalfspacing
\maketitle
\singlespacing

\begin{abstract}
	\noindent
	\input{abstract.tex}
\end{abstract}



\thispagestyle{empty}

\clearpage

\renewcommand{\thefootnote}{\arabic{footnote}}
\setcounter{footnote}{0}

\pagenumbering{arabic}

\onehalfspacing


\input{introduction}

\section{Methodology for Simulating LLM Advice}\label{sec:methodology}

\begin{figure}[!t]
	\caption{Overview of Methodology for Simulating LLM Advice}
	\vspace{-0.2in}
	\label{fig:methodology}
    \vspace{0.2cm}
    \begin{center}
		\makebox[\linewidth][c]{\resizebox{\linewidth}{!}{\includegraphics[]{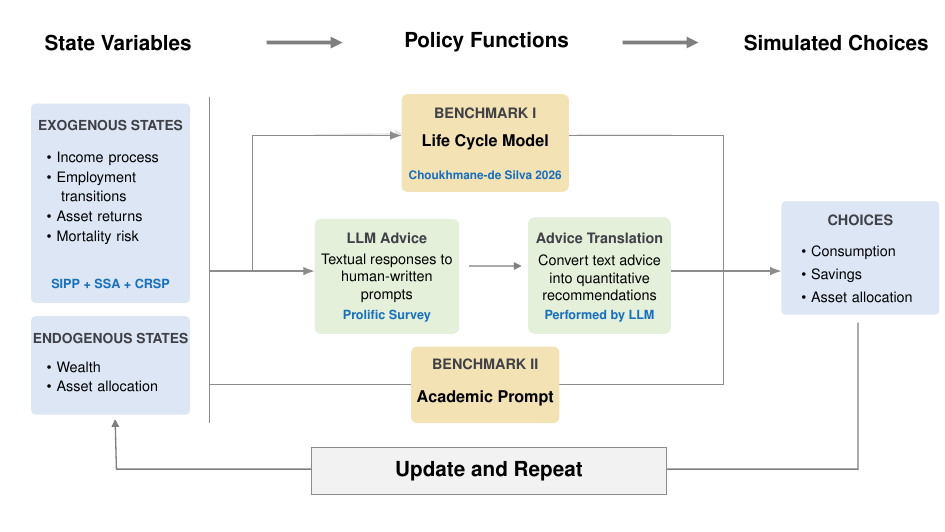}}}
    \end{center}
    \vspace{-0.2in}
    \scriptsize{\emph{Notes:} This figure summarizes the methodology for simulating LLM advice. State variables (left) feed into policy functions (center), which produce simulated choices (right). The life cycle model (Benchmark I) provides optimal policy functions derived from dynamic optimization. LLM advice uses survey prompts to generate textual recommendations, which are then translated into quantitative choices. Academic prompts (Benchmark II) replace survey prompts with researcher-designed prompts. Endogenous state variables are updated each period and the process repeats over the life cycle. \Cref{sec:fact1} also defines a one-shot advice benchmark that is not shown in the schematic.}
\end{figure}

We begin by describing our methodology, which consists of three steps that are summarized in \autoref{fig:methodology}. First, we conduct a survey and ask a demographically balanced sample to write prompts seeking saving and investing advice from an LLM (\Cref{sec:survey}). Second, we calibrate a quantitative life cycle model to U.S.\ income dynamics, employment transitions, asset returns, and tax rules. This model provides both the economic environment for our simulations and a normative benchmark grounded in economic theory (\Cref{sec:life_cycle_model}). Finally, we combine the survey with the model to simulate the life cycle paths of individuals who would follow an LLM's advice each year (\Cref{sec:simulation}). After describing the methodology and benchmarks, which apply in general to any LLM, we describe our model selection.

\subsection{Survey}\label{sec:survey}

The first step in our methodology is to conduct a survey that elicits prompts in which individuals seek financial guidance from an LLM. We design the survey in Qualtrics and administer it to a demographically balanced U.S. sample recruited through the Prolific online platform. Our initial sample consists of 1,000 complete responses, of which 952 pass both Prolific's authenticity check to screen out AI-generated responses and a minimum specificity requirement. \Cref{app:prompt_selection} provides more details on this sample selection.

\hypertarget{back:survey_prompts}{}
The survey asks each respondent to write three free-text prompts to an LLM financial advisor (see \autoref{fig:survey_questions} for the exact wording). The first prompt asks respondents to describe their financial situation, including whatever details they think would help an LLM provide useful advice. Respondents are told to assume the LLM has no prior information about them. The second prompt asks respondents to seek advice on how much to spend over the coming year, while the third asks for advice on how to invest their savings between stocks and safer options. For the second and third prompts, respondents are told that the LLM already has the information from their first response. We focus on spending and investing advice because saving, budgeting, and investing are the domains in which people most commonly seek AI financial guidance, both in our survey and others \citep{JDPower2025_AI_FinancialAdvice}.\footnote{Among the 48\% of respondents to our survey who report having used an AI tool for financial advice or information in the past three months, the most common topics are saving (49\%) and investing (49\%), followed by budgeting (42\%) and financial education (35\%) (\autoref{fig:ai_topics}).}

In addition to the prompts, we survey respondents about the state variables that will be used in our life cycle simulation (including age, income, employment status, and financial wealth), as well as demographic characteristics such as gender, race, financial literacy, and prior experience with AI. \hypertarget{back:survey_demographics}{}\autoref{tab:survey_demographics} compares the demographics of our Prolific sample with those of the Current Population Survey (CPS). The sample broadly matches the CPS across age, gender, and income, though it over-represents the unemployed, under-represents the lowest-income earners, and under-represents respondents older than 70.

\hypertarget{back:summary_prompt}{}Summary statistics for the prompts written by our survey respondents---including word counts, character counts, and the prevalence of numerical and dollar-amount mentions---are reported in \autoref{tab:summary_prompt}. The financial situation prompt is the longest of the three (mean of 43 words), with 65\% of respondents mentioning specific numbers and 34\% mentioning dollar amounts. The spending and investment advice prompts are shorter (mean of 27 words each), with lower rates of quantitative detail.

\subsection{Life Cycle Model}\label{sec:life_cycle_model}

In the second step, we build a quantitative life cycle model of consumption, saving, and portfolio choices based on \cite{Choukhmane2026}. Additional model details and the calibration are presented in \Cref{app:life_cycle_model}.

Each model period corresponds to one year. Individuals enter working life at age 22, retire deterministically at age 65, and face age-dependent mortality risk based on the Social Security Actuarial Life Tables until a maximum age of 90. They have time-separable CRRA utility over consumption with annual discount factor $\beta$ and relative risk aversion $\gamma$.

Individuals face employment risk and move across four labor-market states: employed, transitioning from job-to-job, unemployed, and retired. While employed, they earn stochastic labor income that combines a deterministic age profile, a persistent AR(1) component, and a transitory shock. Job-to-job transitions are associated, in expectation, with persistent wage gains, whereas unemployment spells lead to persistent wage losses upon reemployment. The parameters of the income process and the employment transition probabilities are estimated using Survey of Income and Program Participation (SIPP) data. The government provides unemployment benefits to the unemployed and Social Security retirement benefits tied to average lifetime earnings to the retired. It also taxes income according to the 2025 U.S.\ federal schedule and taxes capital returns at a flat rate.

Investors can allocate their savings across four assets: a risk-free bond, a diversified stock market index, an individual stock, and another risky asset representing holdings such as crypto, gold, commodities, and collectibles. The returns of the diversified index are calibrated using historical CRSP total market returns. We refer to the individual stock and the other risky asset as non-diversified assets, since they capture risky positions outside the diversified market index that an LLM can recommend. Both non-diversified assets have the same return process, but independent shock realizations, so they represent distinct risky positions. We calibrate their return process using evidence on individual stock returns from \cite{Bessembinder2018}. We construct these returns such that neither can improve the Sharpe ratio of a portfolio that combines the bond and the diversified stock market index. 

In each period, investors choose consumption, saving, and portfolio shares across the four assets subject to no-borrowing and no-leverage constraints. We define the following portfolio shares, which we refer to throughout.
\begin{align*}
    \text{Equity Share} &= \frac{\text{Diversified Stock Holdings} + \text{Individual Stock Holdings}}{\text{Total Wealth}} \\[0.2in]
    \text{Diversified Equity Share} &= \frac{\text{Diversified Stock Holdings}}{\text{Total Wealth}} \\[0.2in]
    \text{Non-Diversified Asset Share} &= \frac{\text{Individual Stock Holdings} + \text{Other Risky Asset Holdings}}{\text{Total Wealth}}
\end{align*}
We keep track of seven state variables when solving the life cycle model: age, labor productivity, transitory income shock, employment status, tenure, average lifetime income, and savings. 

\subsection{Simulating Advice over the Life Cycle}\label{sec:simulation}

The final step in our methodology combines the survey responses with the life cycle model to simulate the life cycle paths of individuals who follow the advice provided by an LLM, as illustrated in \autoref{fig:methodology}. Each simulated life path proceeds by iterating, every year from age 22 to 89, through five steps: states, prompt, advice, choices, and shocks. \autoref{tab:simulation_example} illustrates these steps with a concrete example.

\paragraph*{Step 1: States.} At the beginning of each period, each simulated individual is characterized by the state variables of our life cycle model (\autoref{tab:simulation_example}, Panel A). In our example, the individual is 26 years old, employed, earning \$68,810, and holding \$24,380 in financial assets.

\paragraph*{Step 2: Prompt selection and preparation.} \hypertarget{back:prompt_bucketing}{}We assign each simulated individual to one of 12 buckets defined by employment status, age, and income. We then randomly draw a prompt written by a survey respondent in the same bucket, so that the LLM advice assigned to the simulated individual is based on a prompt from someone with similar observable characteristics (\autoref{fig:prompt_bucketing}).\footnote{In choosing which characteristics to match on, we face a trade-off between matching simulated individuals more closely to survey respondents and maintaining sufficient sample size within each bucket.} Within each prompt bucket, the characteristics of the simulated individual may still differ from those of the sampled survey respondent. We therefore identify any references to state variables in the prompt and replace them with the corresponding values from the simulation. For example, \autoref{tab:simulation_example}, Panel B shows the raw prompt as originally written by a 25-year-old survey respondent earning approximately \$75,000 per year. Panel C shows the same prompt after variable insertion: every mention of a state variable---including age, income, wealth levels, and portfolio allocations---is replaced with the simulated individual's current values. Our insertion rules, described in \Cref{app:survey_description}, preserve the author's writing style and concerns while ensuring that the quantitative details match the simulated individual's situation. The three survey prompts (financial situation, spending advice, and investment advice) are then concatenated into a single message and sent to an LLM with an instruction to respond in 200 words or fewer.

\paragraph*{Step 3: LLM advice.} The LLM responds to the prepared prompt (\autoref{tab:simulation_example}, Panel D) with textual advice. The qualitative properties of this advice, including which topics the LLM emphasizes and how they compare with what individuals discuss in their prompts, are described in \Cref{sec:textual_analysis}.

\paragraph*{Step 4: Translating advice into quantitative choices.} Next, the LLM's textual advice is converted into the quantitative recommendations for the choice variables of our model: consumption and savings held in the four asset classes. We do this by passing the LLM's advice back to another LLM, along with deterministic instructions to extract numerical recommendations and output them in a structured plain-text format. This second LLM call maps qualitative advice (e.g., ``Keep an emergency fund of 3-6 months of essential expenses in a HYSA/money market'' and ``Automate investing each payday (e.g., 15--25\% of gross)'') into dollar amounts for consumption and contributions to each asset class (\autoref{tab:simulation_example}, Panel E). The extraction follows a pre-defined ordering; the full translation prompt, including the asset class definitions, extraction ordering, budget constraints, and deterministic rules that ensure all outputs satisfy accounting identities, is in \Cref{app:json_prompt}.

\paragraph*{Step 5: Shocks and state transitions.} After extracting the quantitative recommendations, we draw realizations of the exogenous shocks in our life cycle model: labor market transitions, persistent and transitory income shocks, and asset returns for each of the four asset classes. The realized shocks, combined with the LLM's choices, determine the state variables at the beginning of the next period (\autoref{tab:simulation_example}, Panel F). These updated states are then used as the input for the next iteration.

\paragraph*{Simulation design.} Each query to the LLM is independent, with no memory of past responses; the only link across periods is the evolution of state variables that depend on prior choices. In each run of the model, we simulate the same 1,000 individuals, each starting life at age 22 with no wealth, retiring at age 65, and dying by age 90. A new prompt is drawn and prepared in every period for every individual, and the distribution from which prompts are drawn evolves as an individual's state variables change.

\afterpage{
\begin{landscape}
\begin{table}[!t]
	\caption{From States to Choices: Simulation Example for One Period}
	\label{tab:simulation_example}
	\scriptsize
	\makebox[\linewidth][c]{
		\begin{tabular}{p{3.5cm} p{\dimexpr\linewidth-3.5cm-4\tabcolsep\relax}}
			\toprule
			\multicolumn{2}{l}{\textbf{Panel A: State variables}} \\
			\midrule
			\multicolumn{2}{l}{
				\begin{tabular}{lccccccc}
					Age & Employment status & Pre-tax annual income & Post-tax annual income & Bonds & Diversified stocks & Individual stocks & Other risky assets \\
					26 & Employed & \$68,810 & \$62,223 & \$17,938 & \$6,442 & \$0 & \$0
				\end{tabular}
			} \\
			\midrule
			\multicolumn{2}{l}{\textbf{Panel B: Raw survey prompt (as written by respondent)}} \\
			\midrule
			Financial situation & ``I make around 75000 a year before taxes and pay 1700 a month in rent. That is my primary income and I make some money completing paid research surveys. I have around 80000 saved. I am 25.'' \\[2pt]
			Spending advice & ``I am looking for ways to go out more and enjoy my life without sacrificing my savings. I have a good grip on money and actually struggle to justify spending.'' \\[2pt]
			Investment advice & ``How should I put away money now that interest rates are going down. Safe bets like my money market and CDs are no longer yield much money.'' \\
			\midrule
			\multicolumn{2}{l}{\textbf{Panel C: Prepared prompt (after variable insertion)}} \\
			\midrule
			Financial situation & ``I make around \textbf{68810} a year before taxes and pay 1700 a month in rent. That is my primary income and I make some money completing paid research surveys. I have around \textbf{24380} saved. I am \textbf{26}.'' \\[2pt]
			Spending advice & ``I am looking for ways to go out more and enjoy my life without sacrificing my savings. I have a good grip on money and actually struggle to justify spending.'' \\[2pt]
			Investment advice & ``How should I put away money now that interest rates are going down. Safe bets like my money market and CDs are no longer yield much money.'' \\
			\midrule
			\multicolumn{2}{l}{\textbf{Panel D: LLM advice}} \\
			\midrule
			\multicolumn{2}{p{20.5cm}}{``Given your income/savings, you can safely `permission' yourself to spend more without derailing goals by separating \textbf{spending from saving}. \begin{enumerate}[label=\arabic*), itemsep=0pt, parsep=0pt]
                \item \textbf{Set a floor, then automate:} Keep an emergency fund of 3-6 months of essential expenses in a HYSA/money market (don't chase yield; it's for safety). Automate investing each payday (e.g.,15-25\% of gross, adjust to comfort). What's left is guilt-free.
                \item \textbf{Create a ``fun'' system:} Open a separate checking/savings called ``FUN''. Auto-transfer a fixed amount (start \$300-\$600/month). If it's there, you must use it or roll it to a bigger experience. 
                \item \textbf{Where to put money as rates fall:} For long-term (5+ years): low-cost broad index funds/ETFs (e.g., total US + total international) in 401(k)/Roth IRA first. For medium-term goals (1-5 years): HYSA + Treasuries/T-bill ladder (often tax-advantaged vs CDs), or a short-term bond fund if you can tolerate modest fluctuation.
                \item \textbf{Psychology:} Treat spending from the FUN account as a goal, not a failure-your system already protected savings.
            \end{enumerate}''} \\
			\midrule
			\multicolumn{2}{l}{\textbf{Panel E: Extracted choices}} \\
			\midrule
			\multicolumn{2}{l}{
				\begin{tabular}{lccccc}
					Consumption & Total saving & Bonds contribution & Diversified stocks contribution & Individual stocks contribution & Other risky assets contribution \\
					\$44,378 & \$17,845 & +\$5,400 & +\$12,445 & \$0 & \$0
				\end{tabular}
			} \\
			\midrule
			\multicolumn{2}{l}{\textbf{Panel F: Updated states (beginning of next period)}} \\
			\midrule
			\multicolumn{2}{l}{
				\begin{tabular}{lccccccc}
					Age & Employment status & Pre-tax annual income & Post-tax annual income & Bonds & Diversified stocks & Individual stocks & Other risky assets \\
					27 & Employed (no transition) & \$73,016 (income shocks) & \$65,504 & \$23,805 & \$19,505 & \$0 & \$0
				\end{tabular}
			} \\
			\bottomrule
		\end{tabular}
	}
	\vskip 0.1in
	\scriptsize{\emph{Notes:} This table illustrates the five steps used to simulate LLM advice for a single individual in a single period. Panel A shows the state variables at the start of the period. Panel B shows the raw survey prompt as originally written by a respondent. Respondent text is reproduced verbatim; spelling and grammar errors are preserved. Panel C shows the same prompt after variable insertion, with boldface values indicating state variables substituted into the respondent's text. Panel D shows the full LLM advice. Panel E shows the quantitative choices extracted from the translation LLM call that are required to satisfy the budget constraint: consumption + total net contributions = post-tax annual income. Panel F shows the state variables at the start of the next period after labor market, income, and asset return shocks are realized.}
\end{table}
\end{landscape}
}

\subsection{Benchmarks to Compare with LLM Advice}\label{sec:benchmarks}
We compare our simulated LLM advice to three benchmarks. The first is respondents' status-quo self-reported behavior. The second is optimal behavior in the life cycle model, which provides a normative benchmark. The third is LLM advice generated from a structured researcher-designed prompt, which tests whether providing more complete and standardized information moves advice closer to the model benchmark. The latter two benchmarks are illustrated in \autoref{fig:methodology}.

\paragraph*{Empirical benchmark: observed behavior.} The observed-behavior benchmark uses respondents' self-reported savings and portfolio choices from our survey. The behavior of our survey respondents is broadly consistent with evidence on saving and investing over the life cycle prior to the widespread adoption of AI financial advice tools, such as the Survey of Consumer Finances \citep{Gomes2021b}.

\paragraph*{Advice benchmark I: life cycle model.} The second benchmark is constructed by simulating the life cycle model described above and solving for optimal behavior under the same realizations of exogenous shocks used in the LLM simulation. Because the aggregate stock index is calibrated to be mean-variance efficient, we assume that individuals in this benchmark hold neither non-diversified asset class. Portfolio choice, therefore, reduces to an allocation decision between the risk-free bond and the diversified stock index.

\paragraph*{Advice benchmark II: academic prompts.} \hypertarget{back:academic_prompt_table}{}The third benchmark replaces the respondent-written survey prompts with a more structured prompt that provides the LLM with explicit information about the individual's current state variables and makes explicit assumptions about the economic environment. The prompt includes individual-specific information on all the state variables, along with baseline assumptions about expected asset returns, taxes, and Social Security rules. It also instructs the LLM to act as a professional financial advisor trained in portfolio theory and life cycle planning. The full prompt and assumptions are presented in \Cref{app:prompt_baseline} and \autoref{tab:prompt_table}.\footnote{For simplicity, the academic prompt assumes the individual is single with no dependents. In contrast, household composition does vary across respondents and we incorporate changes in household composition by age in the life cycle model through an equivalence scale.}

\subsection{LLM Selection}\label{sec:model_selection}

\hypertarget{back:model_selection}{}
We choose an OpenAI model as our primary LLM because survey evidence---both from our survey (\autoref{fig:ai_model_usage}) and others \citep{Lloyds2025_AI_Money,JDPower2025_AI_FinancialAdvice}---indicates that ChatGPT is the most commonly used LLM for financial advice.\footnote{Using a closed-source model has clear disadvantages from the perspective of reproducibility. However, we chose to prioritize using a model that households are more likely to use in practice.} Among OpenAI's model family, we use \GPT \citep{OpenAI2025a}, which was the latest available model at the time we ran our analysis. We set its reasoning effort to ``Low'', which approximates the default ChatGPT experience: most consumer queries are served by fast, low-compute models. For the task in Step 4 of \Cref{sec:simulation} that translates the LLM's textual advice into quantitative recommendations, we use \GPTMini, which is a smaller and faster model in the same family. We choose this smaller model because this task involves deterministic extraction rules rather than open-ended reasoning. As a robustness check, we repeat our main analysis using \Gemini \citep{gemini2025b}, a model of similar vintage from the second-most-commonly-used model provider in our survey, and \GPTNew \citep{OpenAI2026}, a more recent model from the same provider. We report these results in \Cref{app:results} and summarize them in \Cref{sec:robustness}. 

\section{Describing Prompts and Baseline LLM Advice}\label{sec:descriptives}

This section first describes the qualitative content of the prompts written by our survey respondents and the LLM's corresponding textual responses. We then describe the LLM's baseline advice over the life cycle, and examine its robustness to alternative methodological choices.

\subsection{What Respondents Write and What the LLM Recommends}\label{sec:textual_analysis}

\paragraph*{Measuring topics in prompts and advice.}
We perform a simple textual analysis by matching keywords in each prompt and LLM response to a dictionary of 27 pre-defined topic categories. For each prompt (or LLM response) and category, we record whether any keyword from that category appears. When comparing prompts with LLM responses, we report the share of respondents who mention each category in their prompts and the corresponding mean mention rate in the advice generated from their prompts, weighting each respondent equally. \hypertarget{back:dictionaries}{}Full methodological details, including the complete dictionary and topic categories (\autoref{tab:dictionaries}), and prompt-level summary statistics (\autoref{tab:summary_prompt}), are in \Cref{app:textual_analysis} and \Cref{app:results}.

\begin{figure}[!t]
	\caption{Word Clouds: Respondent Prompts and LLM Advice}
	\vspace{-0.2in}
	\label{fig:wordclouds}
	\begin{center}
	\makebox[\linewidth][c]{%
	\begin{tabular}{@{}c@{\hskip 0.02\linewidth}c@{}}
		\small{\textbf{Combined Prompts}} & \small{\textbf{LLM Responses}} \\[2pt]
		\includegraphics[width=0.48\linewidth]{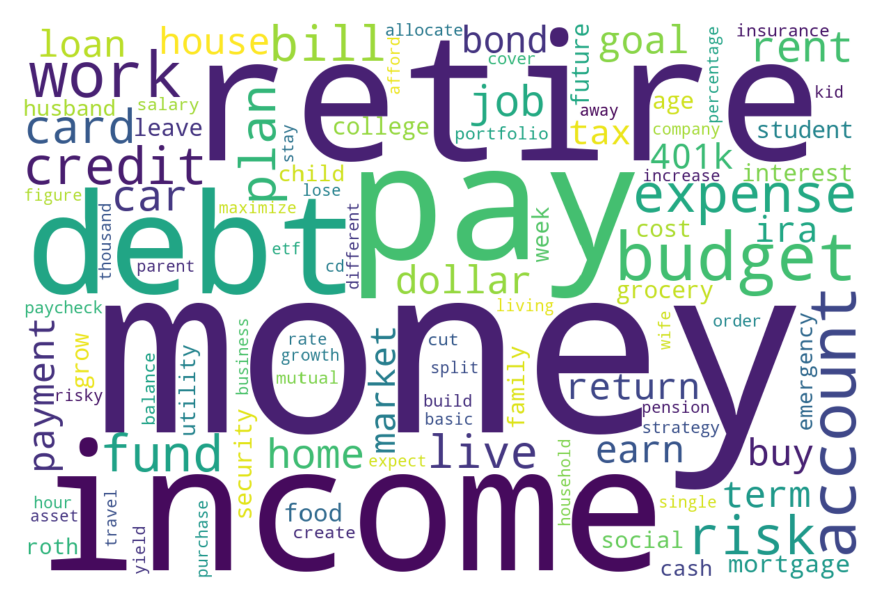} &
		\includegraphics[width=0.48\linewidth]{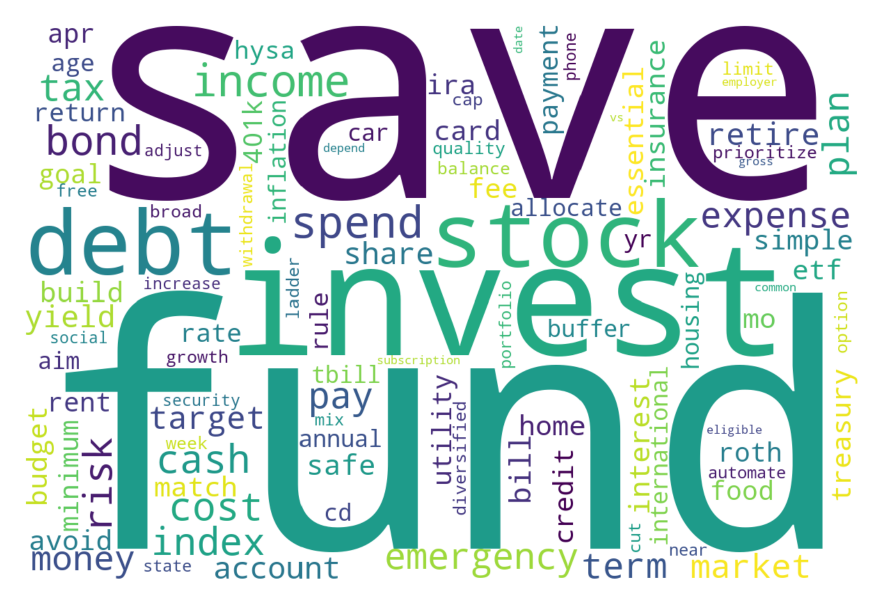}
	\end{tabular}%
	}
	\end{center}
	\vspace{-0.2in}
		\scriptsize{\emph{Notes:} This figure shows word clouds for respondent-written prompts, pooling across the three prompt questions (left) and for the corresponding LLM advice (right). Word size is proportional to word frequency. Common stopwords and words that mechanically reflect the survey questions are excluded. \autoref{fig:wordclouds_by_prompt} shows analogous word clouds separately for each prompt question. Details are provided in \Cref{app:textual_analysis}.}
\end{figure}

\paragraph*{What respondents write about.}
\hypertarget{back:textual_topics}{}
\autoref{fig:wordclouds} shows word clouds for the survey prompts and LLM's responses. The most prominent words in the prompts are ``money,'' ``debt,'' ``income,'' ``pay,'' and ``retire.'' The topics respondents discuss vary across the three prompts they write (\autoref{fig:dict_all}). When respondents describe their financial situation, the most common categories are Income (35\%) and Retirement (30\%). When asking for spending advice, Budgeting is the most frequent (45\%). When asking for investment advice, Investment Assets (53\%) and Risk Preference (40\%) are the most common categories.

\paragraph*{How LLM advice responds to these prompts.}
\hypertarget{back:prompt_advice_topics}{}
The LLM advice closely reflects the topics survey respondents write about: the five most common topics in the prompts (i.e., Budgeting, Investment Assets, Income, Risk Preference, and Retirement) all appear at substantially higher rates in the LLM's advice (\autoref{fig:dict_combined_vs_advice}). At the same time, the LLM also introduces topics that are less commonly mentioned by respondents. For example, Liquidity appears in 84\% of advice responses despite only 6\% of respondents mentioning it explicitly, reflecting the model's high propensity to recommend building an emergency fund. Consistent with this pattern, the most prominent words in the LLM advice word cloud are ``fund,'' ``invest,'' ``save,'' ``stock,'' and ``debt,'' along with terms such as ``essential,'' ``insurance,'' and ``emergency'' (\autoref{fig:wordclouds}). Despite these differences, topic rankings remain broadly aligned between prompts and responses: the rank correlation across all 27 categories is 0.72.

\begin{figure}[!t]
	\caption{Asset Classes and Products: Prompts vs.\ Advice}
	\label{fig:keywords_vs_advice}
	\parbox{\textwidth}{
		\small{\textbf{Panel A}: Investment Asset Classes} \vskip 0.01in
		\makebox[\linewidth][c]{
			\resizebox{1\linewidth}{!}{
				\includegraphics[]{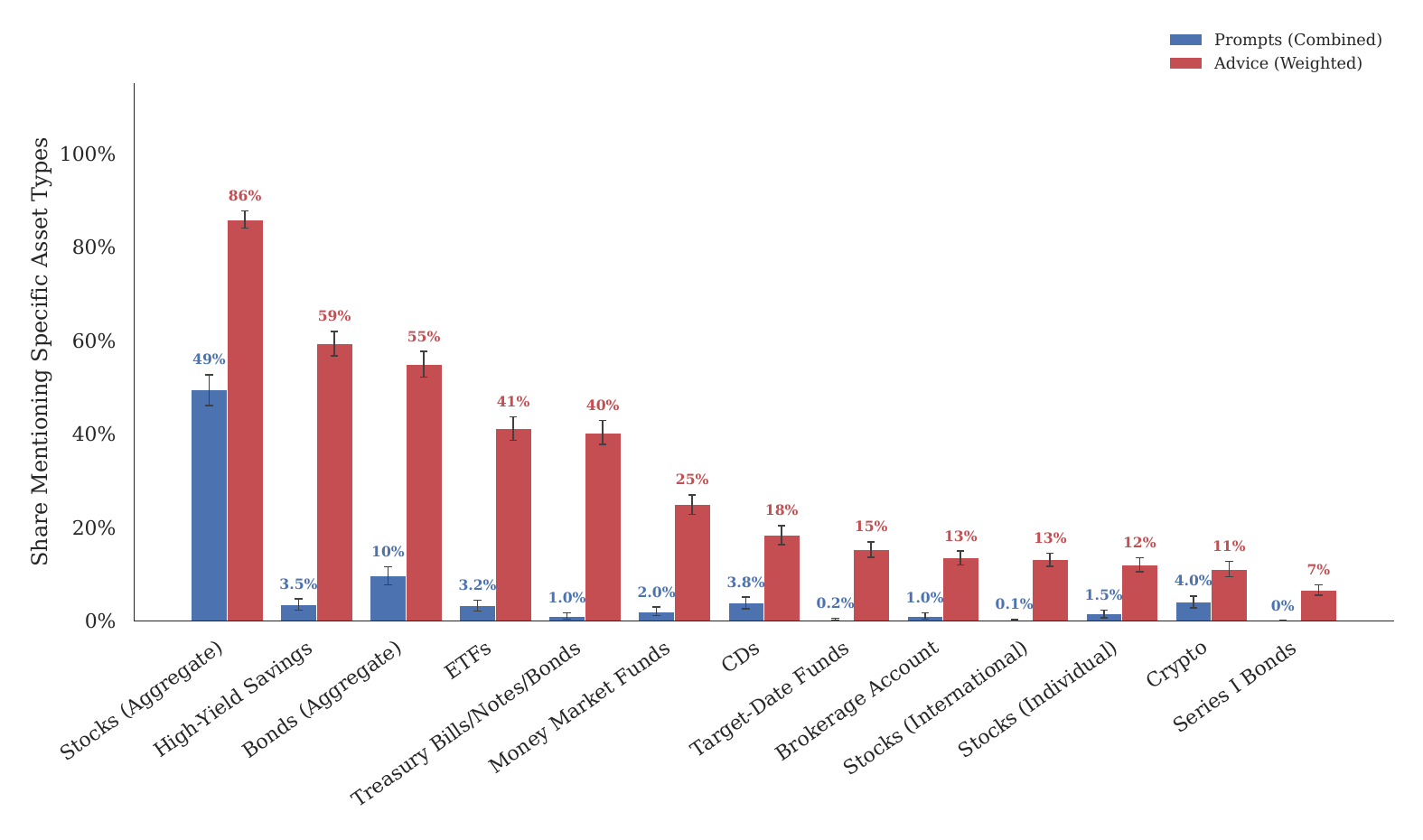}
		}}
	}
	\vskip 0.1in
	\parbox{\textwidth}{
		\small{\textbf{Panel B}: Specific Products by Provider} \vskip 0.01in
		\makebox[\linewidth][c]{
			\resizebox{1\linewidth}{!}{
				\includegraphics[]{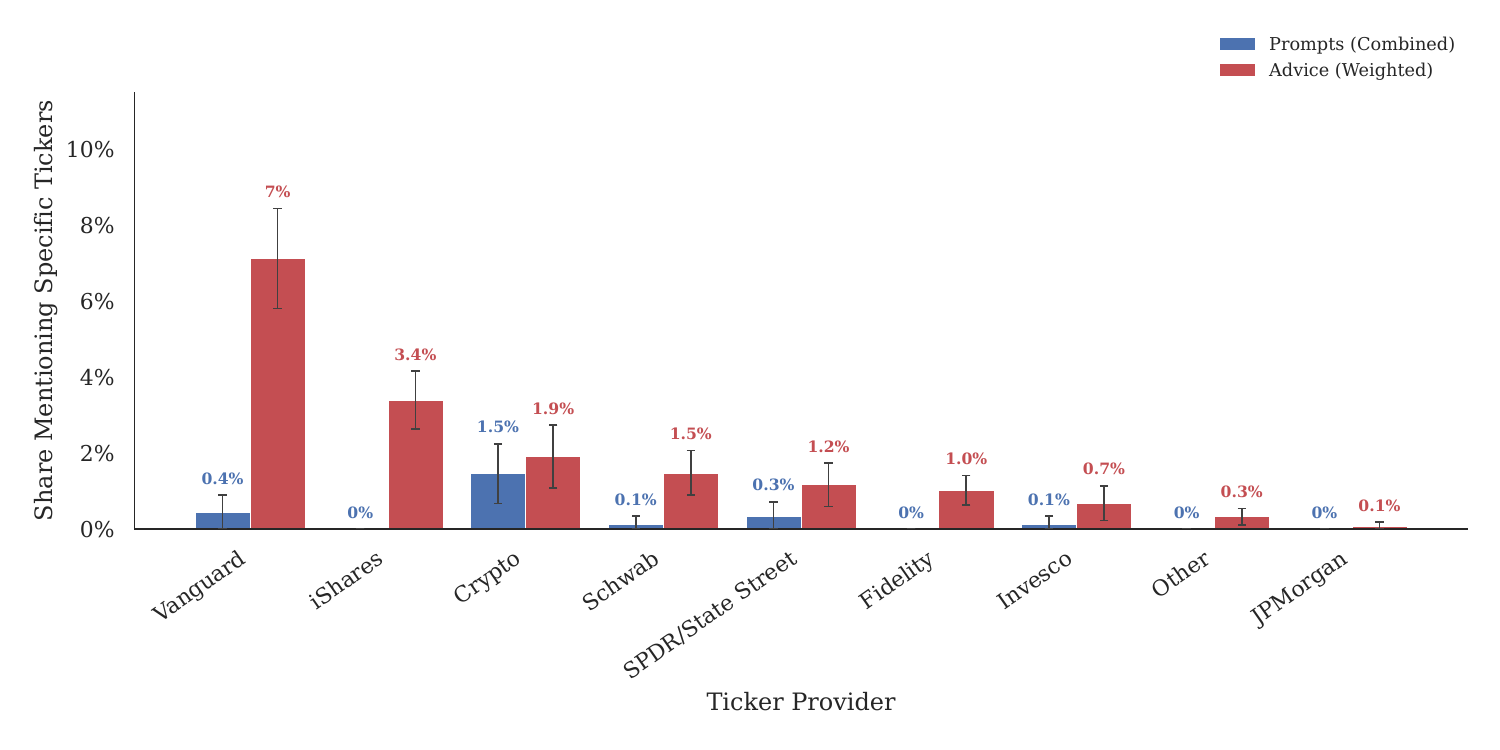}
		}}
	}
	\scriptsize{\emph{Notes:} This figure shows grouped bar charts comparing the mention rates of asset classes and products in respondents' prompts and in baseline LLM advice, as described in \Cref{sec:simulation}. Panel A shows specific investment asset classes with at least 5\% mention rate in either prompts or advice, ordered by descending advice mention rate. Panel B shows specific financial products, including tickers and fund names, aggregated by issuing provider and ordered by descending advice mention rate. Blue bars show the share of respondents mentioning the category in any of the three prompts. Red bars show the share of advice responses mentioning the category, with each respondent's prompt weighted equally. Error bars denote 95\% confidence intervals.}
\end{figure}

\paragraph*{LLM advice provides specific investment recommendations.}
One important way in which the LLM advice goes beyond respondents' prompts is by naming specific asset classes, financial products, and providers (\autoref{fig:keywords_vs_advice}). For example, high-yield savings accounts appear in 59\% of advice responses, compared with 3.5\% of prompts; bonds appear in 55\%, compared with 10\%; and Treasury bills, notes, or bonds appear in 40\%, compared with 1\%. At the product level, fewer than 3\% of respondents name any specific ticker, yet LLM advice sometimes recommends branded products: Vanguard investment products appear in 7\% of responses; iShares appears in 3.4\%; and cryptocurrency tokens such as Bitcoin or Ethereum appear in 1.9\%. The fact that this advice often contains specific recommendations about asset classes, account types, commercial products, and providers raises broader questions about how AI financial advice might affect product competition and the regulation of new sources of financial advice.

\subsection{Baseline Quantitative Recommendations}

\begin{figure}[!t]
    \caption{Life Cycle Profiles of LLM-Recommended Consumption and Equity Shares}
    \vspace{-0.2in}
    \label{fig:lifecycle_llm}
    \begin{center}
		\makebox[\linewidth][c]{\resizebox{\linewidth}{!}{\includegraphics[]{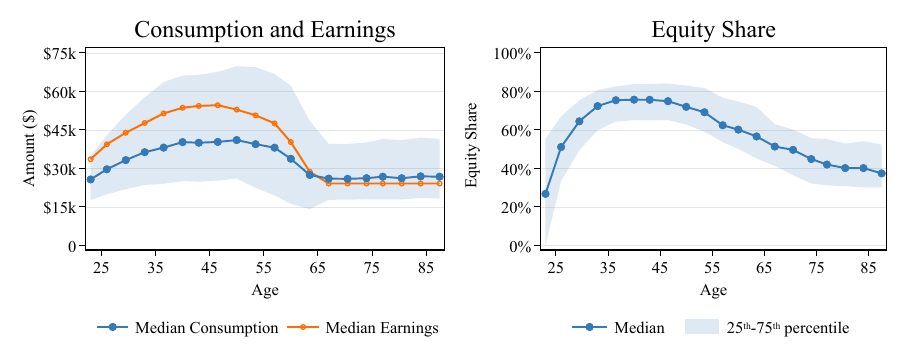}}}
    \end{center}
    \vspace{-0.2in}
    \singlespacing
    \scriptsize{\emph{Notes:} This figure plots the life cycle profiles for individuals following the LLM's recommended choices using the survey prompts, as described in \Cref{sec:simulation}. Ages are grouped into 20 equally sized bins. The left panel shows consumption and post-tax earnings, and the right panel shows the equity share. Dots denote median values within each bin, and shaded areas represent the interquartile range (25$^{\text{th}}$--75$^{\text{th}}$ percentile). All values are in 2025 dollars.}
\end{figure}

\hypertarget{back:lifecycle_profiles}{}
\autoref{fig:lifecycle_llm} plots the median consumption and equity share by age for individuals following the LLM's advice, along with 25th--75th percentile bands.\footnote{We find quantitatively similar results when using \Gemini and \GPTNew, which are shown in \autoref{fig:lifecycle_alternative_models}.} \hypertarget{back:summary_stats}{}\autoref{tab:summary_stats} presents additional summary statistics. 

\paragraph*{Consumption and saving advice.}
The left panel of \autoref{fig:lifecycle_llm} shows that the LLM recommends that individuals smooth their income over working life: consumption is flatter than income, though less flat than our life cycle model prescribes, as we show in \Cref{sec:LLCDeparture}. Individuals following the LLM's advice achieve this by saving more during working life and consuming less than their income, while withdrawing and consuming (slightly) more than their income during retirement. This is reflected in their wealth accumulation, with individuals building up a large stock of over \$1 million in assets by the time they retire at age 65 (\autoref{tab:summary_stats}).

\paragraph*{Investment advice.}
\hypertarget{back:nondiversified_assets}{}
The right panel of \autoref{fig:lifecycle_llm} shows that the recommended equity shares are relatively high, around 65\% on average, and decline with age after approximately age 45. Qualitatively, these portfolio choices are consistent with textbook theories of portfolio choice with human capital \citep[e.g.,][]{Merton1969,Gomes2020}. While less steep than a typical target-date fund glide path, the LLM's recommended equity allocation in working life is similar to the average equity shares that retirement savers choose when making active portfolio choices \citep{Choukhmane2026}. The LLM also recommends small allocations to non-diversified assets (e.g., individual stocks and other risky assets such as crypto, gold, commodities, and collectibles). As shown in \autoref{fig:nondiv_lifecycle}, participation in these asset classes increases with age, reaching approximately 35\% for individual stocks and 50\% for other risky assets by retirement. However, conditional portfolio shares remain small, averaging 2--3\% of financial wealth, so that the vast majority of the LLM's recommended risky asset allocation is in diversified equities.

\section{LLM Advice and Life Cycle Theory}

This section evaluates the financial behavior implied by following LLM advice against the prescriptions of life cycle theory. We first show that following the advice would move respondents toward broad theoretical prescriptions: higher participation in diversified equities, equity shares that decline with age later in life, and larger savings buffers. We then document systematic departures from the normative theory's more subtle prescriptions: the advice implies unusually high patience, relies on simple saving and withdrawal heuristics, smooths consumption imperfectly, and allows portfolios to drift passively with realized returns. Finally, we show that a more structured prompt reduces some (but not all) of these departures.

\subsection{Fact 1: Following LLM Advice Would Move Individuals Toward Life Cycle Theory} \label{sec:fact1}

We begin by asking whether following LLM advice would move respondents closer to the broad prescriptions of life cycle theory relative to observed behavior. This comparison is intentionally qualitative: precise quantitative recommendations from a life cycle model depend on the model specification, preference parameters, the calibration of asset and income risk, and other features of the environment. Therefore, we focus on general principles that are robust across calibrations of standard life cycle portfolio choice models: broad participation in diversified equities, equity shares that decline later in life, and the accumulation of precautionary savings buffers \citep[e.g.,][]{Gomes2020}.

\begin{figure}[!t]
    \caption{Observed Behavior vs.\ LLM-Recommended Behavior}
    \vspace{-0.2in}
    \label{fig:status_quo}
    \begin{center}
		\makebox[\linewidth][c]{\resizebox{\linewidth}{!}{\includegraphics[]{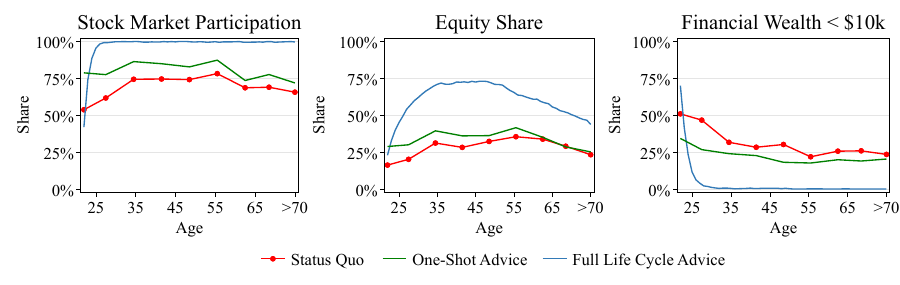}}}
    \end{center}
    \vspace{-0.2in}
    \scriptsize{\emph{Notes:} This figure compares respondents' observed financial behavior with the behavior implied by following our simulated LLM advice. The three panels show average stock market participation rate (left), average equity share (middle), and the share of individuals with financial wealth below \$10,000 (right). Ages are grouped into eight 6-year bins and one last bin that captures every observation age 70 and above. Red lines show respondents' observed behavior. Green lines show the one-shot advice benchmark, which applies the LLM advice generated from each respondent's own prompt to that respondent's current financial situation. Blue lines show the full life cycle simulation, in which simulated individuals follow LLM advice in each period, and prompts are updated to reflect the evolving state variables in the simulation, as described in \Cref{sec:simulation}.}
\end{figure}

\hypertarget{back:observed_behavior_alternative_models}{}
We compare respondents' observed behavior with two versions of LLM-recommended behavior. The first is a one-shot benchmark, which uses the advice generated from each respondent's original survey prompt and current self-reported financial situation, without replacing state variables or simulating subsequent choices.\footnote{We convert the qualitative advice into numerical choices using each respondent's self-reported age, income, employment status, and asset holdings, together with the same translation procedure described in Step 4 of \Cref{sec:simulation}.} The second is the full life cycle simulation from \Cref{sec:simulation}, in which individuals follow LLM advice each period using prompts drawn and updated to reflect evolving state variables in the life cycle simulations. \autoref{fig:status_quo} plots respondents' observed behavior in red, the one-shot advice benchmark in green, and the full life cycle simulation in blue.\footnote{We find quantitatively similar results when using \Gemini and \GPTNew, as shown in \autoref{fig:status_quo_alternative_models}.}

\paragraph*{Advice would raise participation in diversified equities.} As shown in the left panel of \autoref{fig:status_quo}, one-shot advice already points respondents toward broader stock market participation. Consistent with the empirical evidence on limited stock market participation, one-third of respondents in our survey report holding no equities, and their average equity share is around 30\%. Relative to observed behavior, the green line shows that one-shot advice would raise participation by approximately 5--10 percentage points and raise average equity shares by approximately 5 percentage points. The full life cycle simulation amplifies this movement: following the advice each period leads to near-universal participation and raises equity shares by as much as 40 percentage points relative to observed behavior.

\paragraph*{Advice would generate higher equity shares that decline later in life.}
As shown in the middle panel of \autoref{fig:status_quo}, recommended portfolio allocations also move closer to the age profile implied by standard life cycle theory. Respondents' observed equity shares are roughly flat by age and, if anything, increase through age 55. By contrast, in the full life cycle simulation, equity shares are higher, rising early in working life as individuals build emergency buffers, then declining after approximately age 45. As we discuss in the next section, however, this decline is quantitatively more modest than what a standard life cycle model predicts.

\paragraph*{Advice would build larger savings buffers.}
The right panel of \autoref{fig:status_quo} shows that following LLM advice would lead respondents to build financial buffers more quickly. More than 20\% of respondents in every age group report holding less than \$10,000 in financial wealth. One-shot recommendations reduce this share in all age groups, but cannot fully offset low accumulated wealth within a single year. In contrast, when we simulate individuals following LLM advice every period, virtually all build a savings buffer above \$10,000 by age 30.

\begin{figure}[!t]
	\caption{Association between Prompt Topics and LLM Recommendations}
	\vspace{-0.2in}
	\label{fig:phrase_effect}
	\begin{center}
		\makebox[\linewidth][c]{
			\resizebox{\linewidth}{!}{
				\includegraphics[]{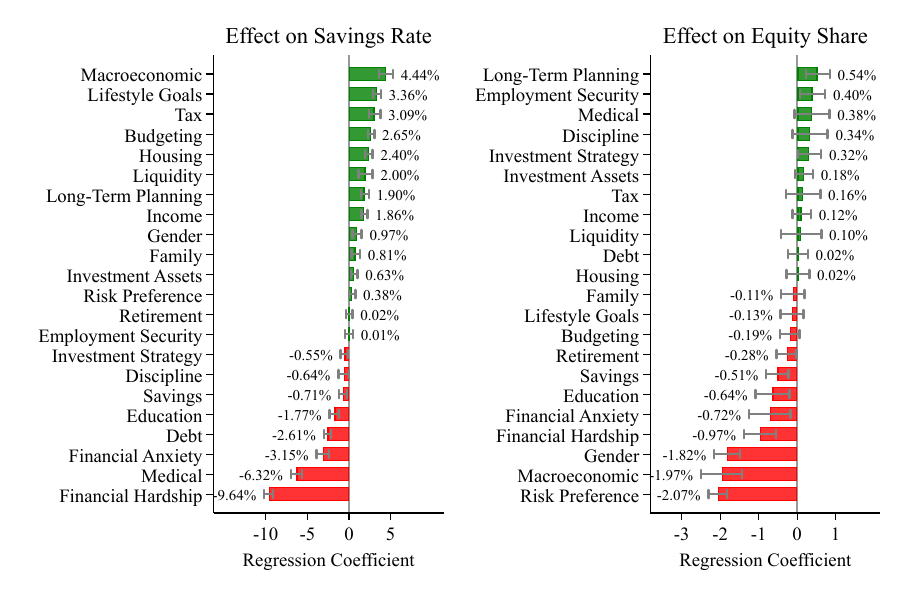}
		}}
	\end{center}
	\scriptsize{\emph{Notes:} This figure plots the estimated association between the dictionary categories mentioned in each prompt and the LLM's recommended net savings rate (left) and change in equity share (right), using the baseline results described in \Cref{sec:simulation}. Each bar is the coefficient on an indicator for whether the prompt mentions the category, from a regression (one per outcome) that includes all category indicators and controls for income, wealth, age fixed effects, and prompt-bucket fixed effects. Dictionary categories and corresponding keywords are reported in \autoref{tab:dictionaries}. Only categories that appear in at least 5\% of prompts are included in the regression and in the figure. Error bars denote 95\% confidence intervals.}
\end{figure}

\paragraph*{Advice responds sensibly to the topics respondents raise.} 
Beyond these average patterns, LLM recommendations also vary systematically with the topics respondents mention in their prompts, often in ways that align with life cycle theory. To document this, we exploit the fact that, within each employment-by-income-by-age prompt bucket, the specific prompt assigned to a simulated individual is randomly drawn. This randomness allows us to compare recommendations generated by different prompts for otherwise similar simulated individuals. Specifically, we regress two outcomes, the net saving rate and the change in equity share, on indicators for each of the dictionary categories defined in \Cref{sec:textual_analysis}, controlling for income, wealth, age fixed effects, and prompt-bucket fixed effects.\footnote{We use the change in equity share rather than its level because of the substantial inertia documented in \autoref{fig:shocks}.} \autoref{fig:phrase_effect} reports coefficients for categories mentioned in at least 5\% of prompts.

The results suggest that the LLM responds to prompt content in ways that are broadly consistent with life cycle theory. When prompts mention macroeconomic conditions, typically uncertainty about inflation or recessions, the LLM raises the recommended saving rate by around 4pp and lowers the recommended equity share by 2pp, consistent with a greater precautionary motive. Mentions of financial hardship lower the recommended saving rate by around 10pp and the equity share by 1pp, consistent with immediate liquidity needs. Mentions of risk preferences modestly raise the saving rate by 0.4pp but lower the recommended equity share by 2.1pp. This last pattern echoes \cite{rumpf2026humans}, who find that LLM portfolio recommendations are more sensitive to stated risk preferences than human advisors.

\paragraph*{Summary.}
Taken together, these results show that following LLM advice would move respondents toward broad prescriptions of life cycle theory with higher participation in diversified equities, high equity shares that decline later in life, and larger savings buffers. This is not only an average pattern: the LLM also adjusts its recommendations for saving and risk-taking in response to economically relevant information in respondents' prompts. These findings should be interpreted as the effect of following advice, not the effect of receiving advice. Whether LLM recommendations change actual financial decisions, and how those effects compare to other forms of financial advice, are important questions for future work.

\subsection{Fact 2: LLM Advice Departs from Life Cycle Theory in Systematic Ways}\label{sec:LLCDeparture}

We next compare LLM advice to the more quantitative predictions from our calibrated life cycle model. While Fact 1 shows that following the advice would move respondents toward broad principles of life cycle theory, this comparison highlights some systematic departures from the model's more specific prescriptions. Specifically, the advice implies unusually high patience, relies on simple saving and withdrawal heuristics, smooths consumption imperfectly in response to income shocks, and allows portfolios to drift passively rather than be actively rebalanced.

\paragraph*{Advice implies unusually high patience.}
We summarize the distance between LLM advice and the life cycle model by estimating the preference parameters that best fit the LLM's recommendations. Specifically, we use the simulated method of moments (SMM) to estimate the intertemporal discount factor ($\beta$) and coefficient of relative risk aversion ($\gamma$) that minimize the distance between model-generated and LLM-generated life cycle profiles.\footnote{This approach is similar to \cite{Cook2026}, who estimate structural preference parameters from LLM behavior in dictator games and job search scenarios. \cite{haq2026revealing} pursue an alternative approach, eliciting LLM preferences directly, and show they align closely with those of human subjects.} We target two sets of moments: the wealth-to-income ratio during working years and the equity share over the full life cycle. Additional details are provided in \Cref{app:smm}.

The first row of \autoref{tab:smms} reports estimates targeting moments from the baseline advice simulation with survey prompts. The estimated risk aversion, $\hat{\gamma}=5.3$, is moderate and within the range of standard estimates. By contrast, the estimated discount factor, $\hat{\beta}=1.034$, is above one. This high value reflects the LLM's recommendation to consume less than income throughout working life, which generates substantial wealth accumulation by retirement. One possible explanation is unmodeled bequest motives, although bequests are rarely mentioned in respondents' prompts. Another possible interpretation is that the LLM advice reflects a form of paternalism: it may recommend a higher saving target than the standard model would prescribe because individuals may systematically undershoot the targets they receive. In that sense, conservative advice can partly offset imperfect implementation. However, if recommendations are followed exactly, a standard life cycle model can rationalize this advice only with an unrealistically high discount factor.

\hypertarget{back:model_fit}{}
Panel A of \autoref{fig:model_fit} shows the fit of the estimated model on the targeted moments. The model matches the LLM's wealth accumulation during working life and its equity shares after age 45, but cannot match the relatively low and upward-sloping equity shares that the LLM recommends between ages 25 and 45.\footnote{This mismatch reflects a standard implication of life cycle portfolio models: absent participation costs \citep{GomesMichaelides2005} or correlated labor income risk \citep{Benzoni2007,Catherine2020}, optimal equity shares are high early in life.} 

\begin{table}[!t]
    \caption{Simulated Method of Moments Estimation Results}
    \label{tab:smms}
    \footnotesize
    \makebox[\linewidth][c]{
		\input{table_smm_results_data}
    }
    \vskip 0.1in
    \scriptsize{\emph{Notes:} This table reports simulated method of moments (SMM) estimates of the annual time discount factor, $\beta$, and coefficient of relative risk aversion, $\gamma$, for each prompt-model combination shown. Survey prompts are respondent-written prompts from the survey; academic prompts are researcher-designed prompts described in \Cref{sec:benchmarks}. The estimates minimize the weighted sum of squared errors between a set of moments from simulations with LLM advice and another from simulations using the model's policy functions. For a full description of the targeted moments and estimation procedure, see \Cref{app:smm}.}
\end{table}

\begin{figure}[!t]
    \caption{Saving and Withdrawal Heuristics in LLM Advice: Survey Prompts}
    \vspace{-0.2in}
    \label{fig:heuristics_survey}
    \begin{center}
		\makebox[\linewidth][c]{\resizebox{\linewidth}{!}{\includegraphics[]{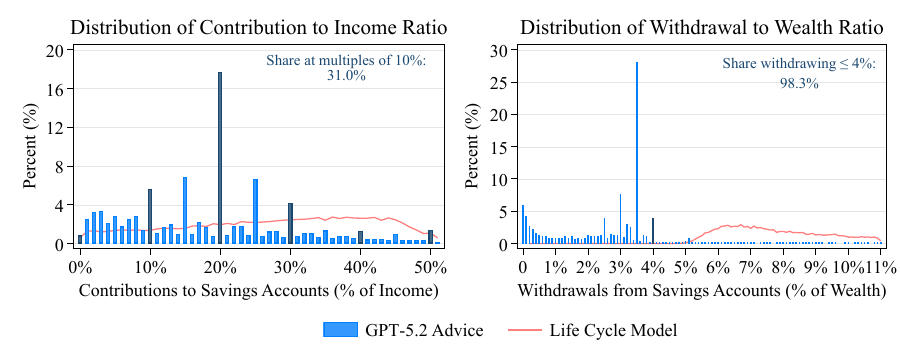}}}
    \end{center}
    \vspace{-0.2in}
    \singlespacing
    \scriptsize{\emph{Notes:} This figure plots the distributions of saving rates and withdrawal-to-wealth ratios implied by the baseline LLM advice described in \Cref{sec:simulation}. The left panel restricts the sample to ages below 65 and plots saving contributions as a share of income. The right panel restricts the sample to ages 65 and above and plots withdrawals as a share of wealth. Blue histograms show the choices implied by the LLM advice. Dark blue bars highlight mass at common heuristic values, including saving rates at multiples of 10\% and withdrawal-to-wealth ratios at 4\%. Red lines show the corresponding distributions generated by the estimated life cycle model using the preference parameters reported in \autoref{tab:smms}.}
\end{figure}

\paragraph*{Advice relies on simple saving and withdrawal heuristics.}
We next examine the LLM's individual-level recommendations that underlie the aggregate life cycle profiles. \autoref{fig:heuristics_survey} plots the distributions of LLM-recommended saving rates, saving amounts, and withdrawal rates in the baseline advice simulation (blue histograms), along with the corresponding distributions implied by the life cycle model estimated in \autoref{tab:smms} (red lines). The LLM recommendations exhibit bunching at round numbers: 31\% of saving rates are multiples of 10\%, and 34\% of saving amounts are multiples of \$5,000. These spikes are absent by construction in the life cycle model, where choices vary continuously with state variables. Withdrawal advice displays an even stronger heuristic pattern. More than 98\% of LLM-recommended withdrawals are at or below 4\% of assets, echoing the 4\% retirement withdrawal rule popularized by \cite{Cooley1998} and commonly used by financial advisors. By contrast, almost no individuals in the life cycle model choose withdrawal rates below 4\%, consistent with the model prescribing faster decumulation than the LLM. 

\begin{figure}[!t]
    \caption{Consumption Smoothing: Survey vs.\ Academic Prompts}
    \vspace{-0.2in}
    \label{fig:smoothing}
    \begin{center}
		\makebox[\linewidth][c]{\resizebox{\linewidth}{!}{\includegraphics[]{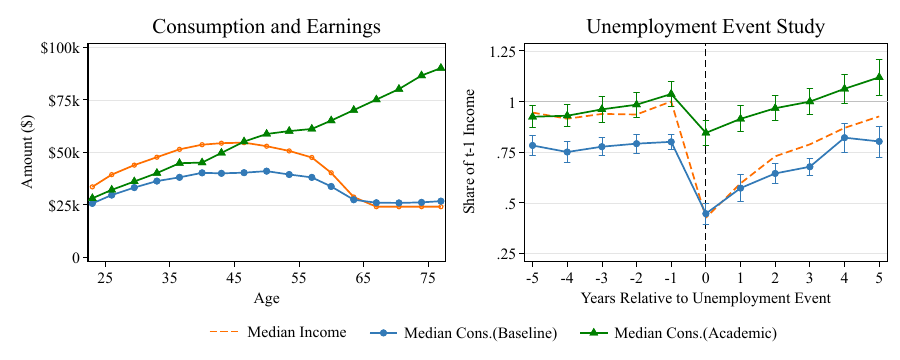}}}
    \end{center}
    \vspace{-0.2in}
    \scriptsize{\emph{Notes:} This figure compares consumption profiles from the baseline using the survey prompts with profiles from the academic-prompt benchmark described in \Cref{sec:benchmarks}. The left panel plots median consumption and median income over the life cycle. The right panel plots the median ratio of consumption relative to income one year prior to unemployment within five years before and after the unemployment event, restricting to individuals between ages 30 and 55. The blue and green lines show consumption implied by the baseline LLM advice and the academic benchmark, respectively; the orange line shows the corresponding median income. Error bars denote 95\% confidence intervals. All values are in 2025 dollars.}
\end{figure}

\paragraph*{Consumption is insufficiently smoothed.}
\hypertarget{back:consumption_smoothing}{}
As shown in the left panel of \autoref{fig:smoothing}, the LLM's recommended consumption (blue line) closely tracks income (orange line) over the life cycle. The right panel shows that this limited consumption smoothing also appears around simulated transitions into unemployment. After job loss, simulated income falls by roughly 50\%, and the LLM's recommended consumption falls by a similar percentage, consistent with empirical evidence on consumption responses to unemployment in \cite{Gruber1997} and \cite{Ganong2019}. In the life cycle model, by contrast, consumption barely falls after job loss because individuals draw down buffer stocks to absorb the shock (\autoref{fig:shocks_professor}). The fact that LLM advice smooths consumption insufficiently both over predictable life cycle changes in income and around employment and income shocks is particularly notable, given that simulated individuals hold sizable liquid balances.\footnote{Results are similar in our baseline specification, where asset holdings are mentioned in the prompt only if the original survey respondent included such information, and in a robustness exercise that appends all relevant state variables to the prompt; see discussion in \Cref{sec:robustness} and \autoref{fig:pipeline_comp}.}

\begin{figure}[!t]
    \caption{Passive Portfolio Drift in LLM Advice}
    \vspace{-0.2in}
    \label{fig:shocks}
    \begin{center}
		\makebox[\linewidth][c]{\resizebox{0.8\linewidth}{!}{\includegraphics[]{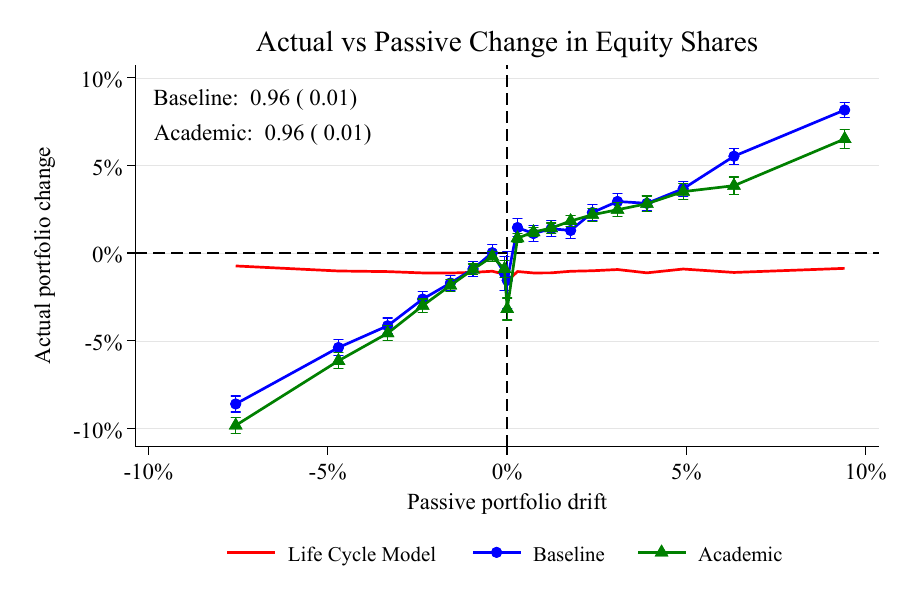}}}
    \end{center}
    \vspace{-0.2in}
    \scriptsize{\emph{Notes:} This figure compares actual portfolio changes with passive portfolio drift for the baseline LLM advice, the academic-prompt benchmark, and the estimated life cycle model. Blue denotes baseline advice using survey prompts, green denotes advice using the academic prompt, and red denotes the life cycle model. Actual portfolio change is the adjustment in the equity share before returns are realized from one period to the next. Passive portfolio drift is the mechanical change in the equity share induced by asset returns over the period, assuming no rebalancing. The sample is restricted to observations with equity shares strictly between 0\% and 100\%, and observations are grouped into 20 equal-sized bins of passive portfolio drift. Error bars denote 95\% confidence intervals. Regression slopes estimated on the raw data and standard errors for the survey and academic prompts are reported in the top left.}
\end{figure}

\paragraph*{Portfolio recommendations exhibit passive drift.}
Following \cite{Calvet2009}, we measure passive portfolio drift as the change in an individual's equity share from period $t$ to $t+1$ that would occur if the individual made no active portfolio adjustment and simply let returns change the portfolio shares. \autoref{fig:shocks} plots actual changes in equity shares against this passive-drift measure. In the life cycle model with CRRA preferences, these two measures are essentially uncorrelated: target allocations are not sensitive to wealth, and individuals actively choose their portfolio each period. In contrast, LLM-recommended portfolio changes move almost one-for-one with passive drift, exhibiting substantial inertia consistent with evidence from household portfolios \citep{Calvet2009,Choukhmane2026}.\footnote{Quantitatively, \cite{Calvet2009} estimate a regression coefficient of 0.5 between actual and passive portfolio changes using observational data, suggesting that LLM recommendations exhibit roughly twice as much inertia.} This inertia persists even when we skip the translation step and provide the LLM with all relevant state variables, including the existing portfolio allocation (as we discuss later in \Cref{sec:robustness}). This suggests that the inertia in the LLM's advice is not an artifact of respondents omitting information about their current portfolios. Instead, the LLM tends to advise individuals on how to allocate new savings rather than explicitly recommending that they rebalance existing holdings.

\subsection{Structured Prompts Reduce Some Departures from Life Cycle Theory}
\hypertarget{back:structured_prompt_outputs}{}
We next examine whether replacing the prompts written by survey respondents with a more structured researcher-designed prompt, which we call the \emph{academic prompt}, moves LLM advice closer to life cycle theory. As described in \Cref{sec:benchmarks}, the academic prompt provides explicit values for all current state variables, clarifies assumptions about the economic environment, and frames the task as providing professional financial advice.\footnote{The academic prompt also skips the translation step by instructing the LLM to provide quantitative recommendations directly. As we discuss in \Cref{sec:robustness}, skipping the translation step and providing all state variables while using survey prompts generates advice similar to our baseline. Therefore, the differences in this section can primarily be attributed to the content and framing of the academic prompt, rather than to differences in the translation pipeline or information about state variables. The corresponding life cycle profiles and comparison with observed behavior are shown in \autoref{fig:lifecycle_professor} and \autoref{fig:status_quo_professor}.}

\begin{figure}[!t]
    \caption{Saving and Withdrawal Heuristics in LLM Advice: Academic Prompts}
    \vspace{-0.2in}
    \label{fig:heuristics_professor}
    \begin{center}
		\makebox[\linewidth][c]{\resizebox{\linewidth}{!}{\includegraphics[]{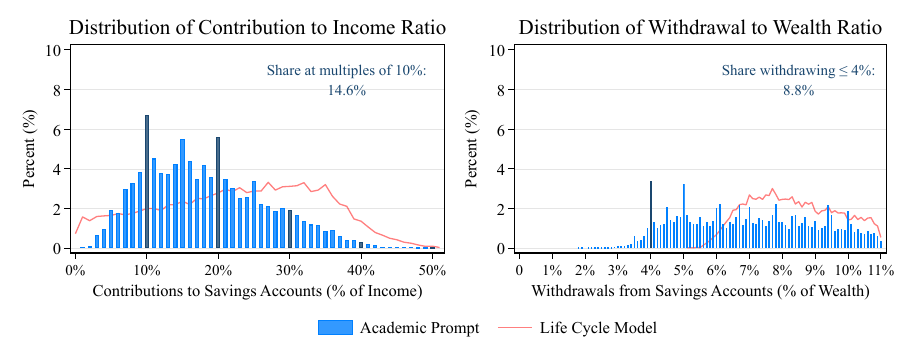}}}
    \end{center}
    \vspace{-0.2in}
    \singlespacing
    \scriptsize{\emph{Notes:} This figure reproduces \autoref{fig:heuristics_survey} using the LLM advice generated in response to the academic prompt. See the notes to \autoref{fig:heuristics_survey} for additional details.}
\end{figure}

\paragraph*{Structured prompts reduce heuristics.}
The LLM's advice produced in response to the academic prompt exhibits less reliance on simple saving and withdrawal heuristics. \autoref{fig:heuristics_professor} shows that the distribution of recommended saving rates exhibits less bunching at round numbers than under respondent-written survey prompts, and withdrawal recommendations are less concentrated at values at or below 4\% of assets. These changes translate into a more reasonable estimated discount factor: the academic prompt estimate is $\hat{\beta}=0.99$, relative to $\hat{\beta}=1.034$ under survey prompts (second row of \autoref{tab:smms}). Portfolio choices, by contrast, remain broadly similar and imply a similar estimate of risk aversion, $\hat{\gamma}=4.7$.

\paragraph*{Structured prompts improve consumption smoothing.} The advice produced in response to the academic prompt exhibits greater consumption smoothing over the life cycle and in response to income shocks. The left panel of \autoref{fig:smoothing} shows that consumption no longer tracks income over the life cycle (green line), and individuals decumulate more substantially in retirement. The right panel shows a parallel improvement around transitions into unemployment. With the academic prompt (green), the drop in consumption is smaller and reverses more quickly, consistent with greater use of accumulated savings buffers to absorb income shocks.

\paragraph*{Structured prompts do not reduce portfolio inertia.}
In contrast, structured prompts do not meaningfully reduce portfolio inertia. As shown in \autoref{fig:shocks} (green line), the relationship between actual changes in equity shares and passive portfolio drift remains close to one under both survey and academic prompts. This suggests that the lack of active rebalancing is not primarily due to missing state variables or the informal wording of respondent prompts. Instead, it appears to reflect a more prevalent pattern in the LLM's financial advice.

\subsection{Robustness of the Patterns in Baseline LLM Advice}\label{sec:robustness}

Next, we show that the broad patterns of advice documented above are robust to alternative methodological choices, including using a different model, eliciting quantitative recommendations directly, or repeating identical queries multiple times. 

\paragraph*{Alternative models.}
\hypertarget{back:alternative_models}{}
To assess whether our findings are specific to \GPT, we rerun the advice-generation step using \Gemini and \GPTNew while retaining \GPTMini for the translation task. The advice provided by these alternative models yields very similar patterns. Relative to respondents' observed behavior, both models recommend higher stock market participation, higher equity shares, and larger savings buffers (\autoref{fig:status_quo_alternative_models}). Quantitatively, the advice provided by both models implies similar structural time and risk preference parameters (\autoref{tab:smms}) and a similar model fit (\autoref{fig:model_fit_alternative_models}, Panels C and D). Like \GPT, \Gemini and \GPTNew rely on simple saving and withdrawal heuristics (\autoref{fig:heuristics_alternative_models}), and exhibit limited consumption smoothing around job loss and substantial passive portfolio drift (\autoref{fig:shocks_alternative_models}). Collectively, these results suggest that our findings are not specific to a single model.

\paragraph*{Eliciting quantitative advice directly.}
Our baseline simulation first elicits qualitative advice using survey-based prompts and then uses a second LLM call to translate that advice into quantitative choices, with the full set of state variables entering only at the translation stage. As a robustness check of this two-step pipeline, we consider an alternative pipeline that elicits quantitative recommendations directly by appending the state variables to the survey prompt and instructing the LLM to return recommendations, without a separate translation step. \autoref{fig:pipeline_comp} shows that the direct quantitative pipeline reproduces the patterns we document in this section for our baseline advice. Consumption and stock-share life cycle profiles are very similar to the baseline two-step pipeline (Panel A), passive portfolio drift is essentially unchanged, and the consumption response to job loss is nearly identical. Reliance on heuristics persists, though it is somewhat attenuated (Panel B): the share of saving rates at multiples of 10\% falls from 31.0\% to 18.7\%, and the share of withdrawals at or below 4\% of assets falls from 98.3\% to 92.4\%. Thus, our main results are not driven by the separate translation step or by withholding state variables during advice generation.

\paragraph*{Repeated queries.}
\hypertarget{back:llm_query_variation}{}
LLM advice is inherently stochastic: the same prompt may produce different textual advice across queries, and the same textual advice may be translated into different quantitative recommendations.\footnote{\GPT is a reasoning model and does not allow users to set the temperature to zero. \cite{rumpf2026humans} document similar stochasticity in the financial advice provided by LLMs.} To quantify the importance of randomness in different steps in our methodology, we repeat the advice-generation and translation steps for the one-shot advice (described in \Cref{sec:fact1}).\footnote{We use one-shot advice due to its significant cost advantage.} For each individual, we conduct two exercises. First, we pass the same prompt through the full pipeline five times. Second, we hold the textual advice fixed and translate it into quantitative recommendations five times. In each exercise, we calculate the standard deviation of the five resulting quantitative recommendations. \autoref{fig:LLM_variation} reports the distribution of these standard deviations across individuals. When we repeat the full pipeline, the median standard deviation is \$3,220 for recommended annual consumption and 6.5 percentage points for the recommended equity share. When we repeat only the translation of a fixed piece of advice, the corresponding medians fall to \$335 and 1.6 percentage points. At the median, translation-only variation is therefore approximately one-tenth as large as full-pipeline variation for consumption and one-quarter as large for equity shares. Thus, although the translation step introduces some randomness, most of the variation in quantitative recommendations arises when the model generates new textual advice in response to the same prompt.

\section{Heterogeneity in LLM Advice: Supply and Demand}
Our analysis thus far has focused on characterizing the average properties of LLM advice. In this section, we examine whether this advice varies systematically across groups of individuals, and how much of this variation arises from demand or supply. Demand-side differences reflect variation in the prompts respondents write: what topics they raise, what information they provide, and how they frame their financial decisions. Supply-side differences reflect how the LLM responds to the same or similar prompts when observable characteristics change. We first show that simulated individuals accumulate more wealth when advice is generated from prompts written by respondents with higher financial literacy or prior AI experience, or by men. We then use randomized variation in prompt labels to determine the roles of demand and supply in determining gender and racial differences in advice.

\subsection{Fact 3: LLM Advice Varies Systematically Across Groups of Individuals}
To quantify heterogeneity in LLM advice, we repeat the life cycle simulation separately for groups of survey respondents based on their financial literacy, prior AI experience, and gender. For each characteristic, we split respondents into two groups and simulate life cycle paths using only prompts written by respondents in one group at a time. Importantly, all simulations use the same realizations of exogenous shocks (i.e., income, employment, mortality, and asset returns), so any differences in simulated outcomes are driven entirely by differences in the prompts written by each group and the LLM's responses. 

\begin{figure}[!t]
    \caption{Heterogeneity in LLM Financial Advice}
    \label{fig:heterogeneity_gaps}
    \makebox[\linewidth][c]{%
    \parbox{0.48\linewidth}{
        \centering\small{\textbf{Financial Literacy: Low vs.\ High}} \vskip 0.01in
        \makebox[\linewidth][c]{
            \resizebox{\linewidth}{!}{
				\includegraphics[]{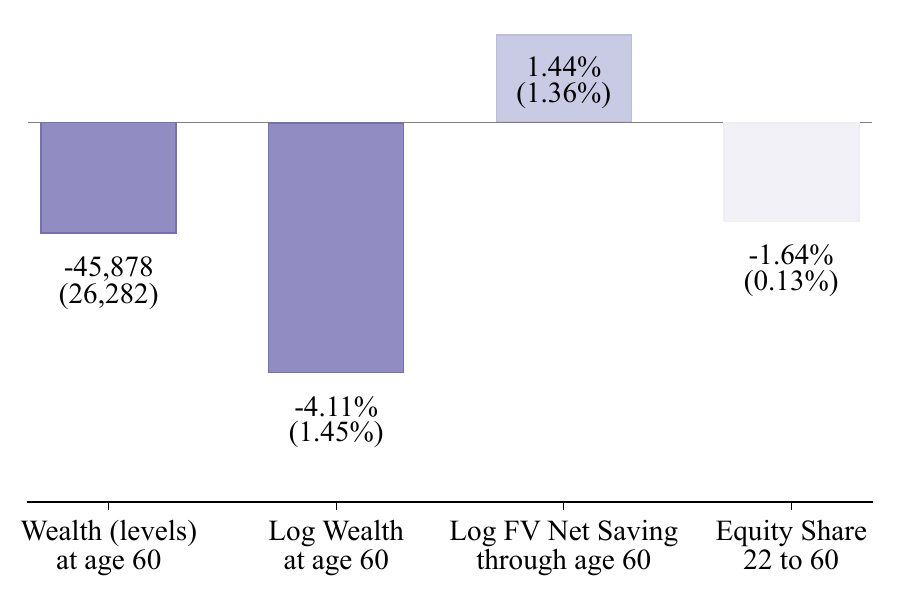}
        }}
    }%
    \hfill
    \parbox{0.48\linewidth}{
        \centering\small{\textbf{Prior AI Experience: No vs.\ Yes}} \vskip 0.01in
        \makebox[\linewidth][c]{
            \resizebox{\linewidth}{!}{
				\includegraphics[]{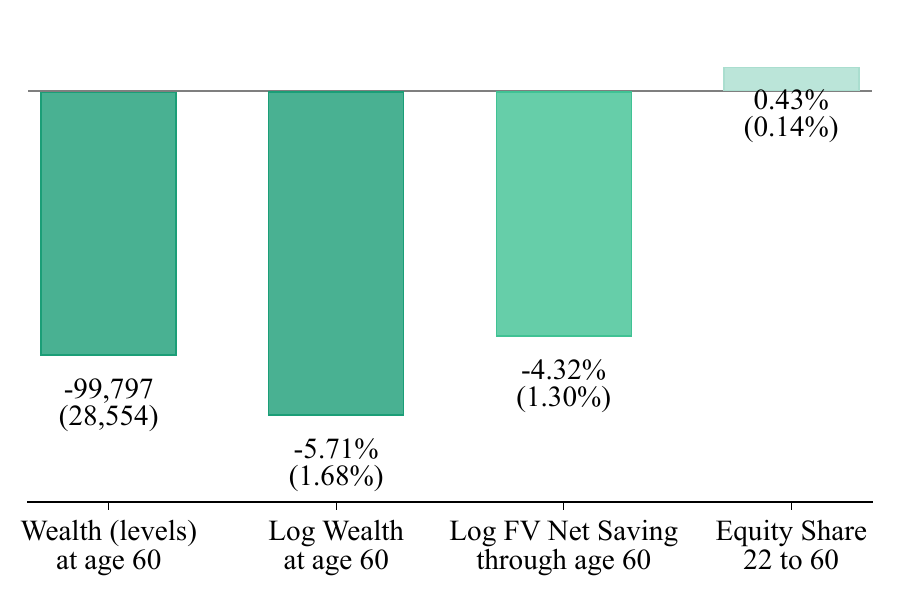}
        }}
    }%
    }
    \vskip 0.05in
    \makebox[\linewidth][c]{%
    \parbox{0.48\linewidth}{
        \centering\small{\textbf{Gender: Female vs.\ Male}} \vskip 0.01in
        \makebox[\linewidth][c]{
            \resizebox{\linewidth}{!}{
				\includegraphics[]{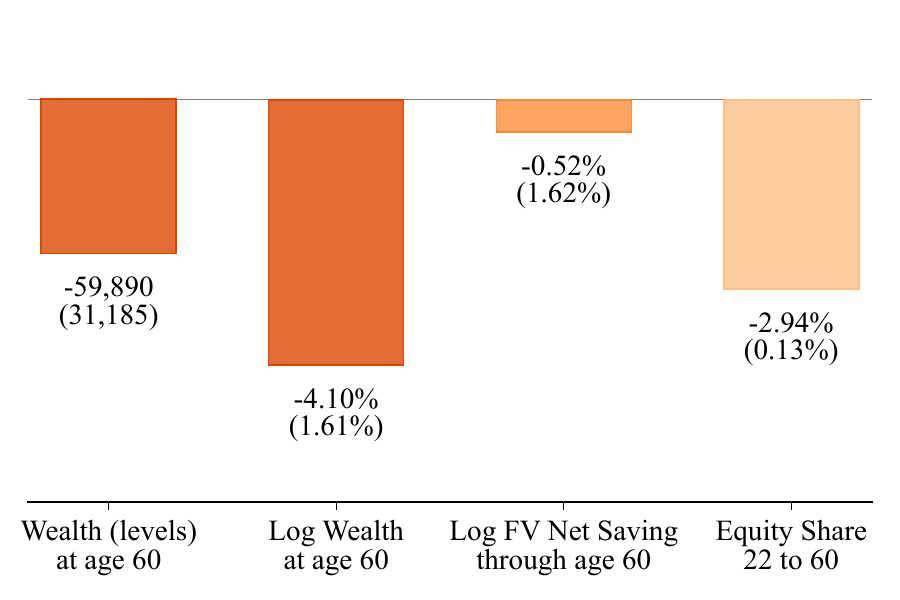}
        }}
    }%
    }
    \scriptsize{\emph{Notes:} This figure compares life cycle outcomes implied by LLM advice generated from prompts written by different groups of survey respondents. All exogenous shocks are held fixed across prompt groups. The top-left panel compares simulations with prompts written by respondents with high versus low financial literacy, measured using the Big Five financial literacy questions. The top-right panel compares simulations with prompts written by respondents who have versus have not previously used AI for financial advice. The bottom panel compares simulations with prompts written by male versus female respondents. In each panel, the four bars report the difference in simulated wealth at age 60 (in levels), log wealth at age 60, the log future value of net saving through age 60 cumulated at the risk-free rate, and the average equity asset share between ages 22 and 60. All values are in 2025 dollars. Standard errors are reported in parentheses.}
\end{figure}

\autoref{fig:heterogeneity_gaps} summarizes the resulting differences in life cycle outcomes. In each panel, the first two bars report the implied difference in average wealth at age 60 across the two groups, in levels and in logs. The next two bars relate these wealth differences to differences in cumulative savings, measured as the future value of net saving compounded at the risk-free rate, and differences in portfolio allocation, measured by the average equity share between ages 22 and 60.\footnote{We compound net saving at the risk-free rate rather than at ex-post portfolio returns, so that this measure is not directly affected by portfolio choices. Additionally, we stop at age 60 to ensure that these differences entirely reflect working-life advice.}

\paragraph*{Low-financial-literacy prompts accumulate less wealth.} \hypertarget{back:financial_literacy_heterogeneity}{}The top-left panel of \autoref{fig:heterogeneity_gaps} splits respondents into high- and low-literacy groups based on their performance on the Big Five financial literacy test \citep{lusardi2023importance}. Prompts written by respondents who answer at least one of these questions incorrectly generate advice that results in wealth at age 60 that is \$46,000, or 4.1\%, lower on average. This difference is driven primarily by portfolio choice: prompts from the less financially literate group lead to a 1.64pp lower average allocation to equity between ages 22 and 60. Differences in cumulative savings go in the opposite direction and would, if anything, partially offset the wealth difference, since the less financially literate group is advised to save more through age 60. The word clouds in \autoref{fig:finlit_wordclouds} suggest that this difference in investment advice reflects both what the two groups ask about in their prompts and the textual advice they receive; \autoref{fig:finlit_examples} provides illustrative examples.

\paragraph*{Prompts without AI experience accumulate less wealth.} \hypertarget{back:prior_ai_heterogeneity}{}In our survey, 48\% of respondents report having used an AI tool for financial advice or information in the past three months (\autoref{fig:ai_incidence}), a share consistent with other recent industry surveys \citep{Lloyds2025_AI_Money,JDPower2025_AI_FinancialAdvice}. The top-right panel of \autoref{fig:heterogeneity_gaps} shows that prompts written by respondents who have not previously used AI for financial advice generate recommendations that result in wealth at age 60 that is \$100,000, or 5.7\%, lower on average. Unlike the financial literacy result, this difference is driven almost entirely by saving behavior: the future value of cumulative net saving is 4.32\% lower in the non-AI-user simulation, while the difference in average equity asset shares is economically small (0.43pp).\footnote{Although AI use and financial literacy are correlated, the distinct mechanisms behind the wealth gaps (i.e., portfolio allocation versus cumulative saving) suggest that these two comparisons capture different sources of heterogeneity in the advice.} \autoref{fig:aiuse_examples} provides illustrative prompts from each group, and word clouds in \autoref{fig:aiuse_wordclouds} show the corresponding differences in the LLM's textual advice.

\paragraph*{Prompts from women accumulate less wealth.} \hypertarget{back:gender_heterogeneity}{}The bottom panel of \autoref{fig:heterogeneity_gaps} compares advice given by the LLM to men and women. Prompts written by women generate recommendations that result in wealth at age 60 that is \$60,000, or 4.1\%, lower on average than prompts written by men. This difference is mainly driven by portfolio choice: the average equity share recommended to women is 2.94pp lower than the share recommended to men, while the future value of cumulative net saving is nearly identical across the two groups. \autoref{fig:gender_men} shows that this difference in equity share recommendations appears across all age and income groups, and that men are also advised to rebalance their portfolios more actively. This gender difference in investment advice is consistent with \cite{foltyn2026worth}, who find an average gender gap in equity allocations of 1.8pp across 33 LLMs in researcher-designed scenarios. It is also consistent with patterns observed in observational data \citep{Agnew2003} and in professional financial advice \citep{bucherkoenen2025gender}.

\subsection{Demographic Differences Reflect Both Demand and Supply}\label{sec:gender}
The heterogeneity results above raise a natural question: do differences in LLM advice reflect differences in what respondents ask (demand), or differences in how the LLM responds to otherwise similar prompts (supply)? We answer this question by first focusing on gender specifically because it has been a focus of study in the literature on human financial advice \citep{bucherkoenen2025gender}, and because our prompt design allows us to experimentally separate demand- from supply-side drivers of gender differences in a way that is difficult in human-advisor settings \citep{reuter2024demand}. Two features of this exercise are worth stating upfront. First, it separates the two channels rather than decomposing the gap in \autoref{fig:heterogeneity_gaps}: identifying the supply channel requires inserting a gender label that respondents did not write, so the experiment measures the LLM's response to an explicit gender statement rather than to the subtler ways gender may be conveyed in naturally written prompts (or be reflected in an LLM memory of past interactions with the user). Second, the estimates capture the effect of gender on the advice generated in a single period, which are not directly comparable to the cumulative life cycle differences reported above.

\begin{figure}[!t]
    \caption{Gender Differences in Prompts and Advice}
    \label{fig:gender_wordclouds}
    \parbox{\textwidth}{
        \small{\textbf{Panel A}: Words Overrepresented in Prompts by Gender} \vskip 0.05in
        \makebox[\linewidth][c]{%
        \begin{tabular}{@{}p{0.48\linewidth}@{\hskip 0.04\linewidth}p{0.48\linewidth}@{}}
            \centering\small{\textbf{Men}} & \centering\small{\textbf{Women}} \tabularnewline
        \end{tabular}%
        }
        \vskip 0.05in
        \makebox[\linewidth][c]{
            \resizebox{\linewidth}{!}{
				\includegraphics[]{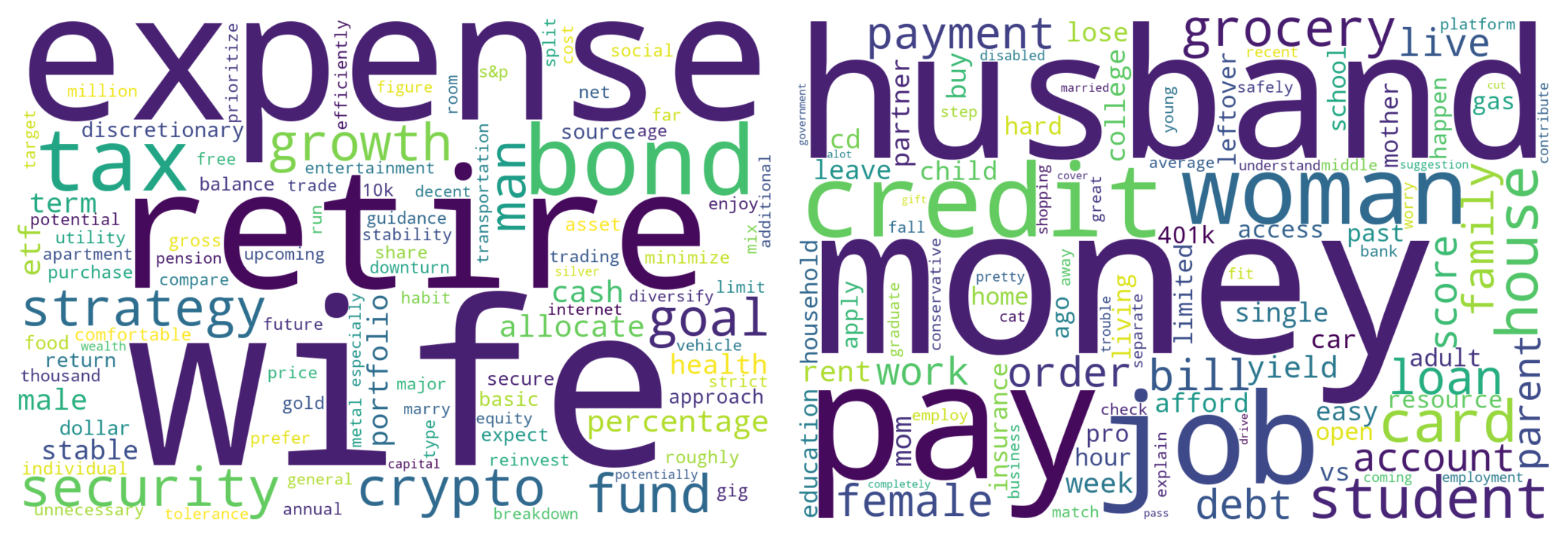}
        }}
    }
    \vskip 0.15in
    \parbox{\textwidth}{
        \small{\textbf{Panel B}: Words Overrepresented in Advice by Gender} \vskip 0.05in
        \makebox[\linewidth][c]{%
        \begin{tabular}{@{}p{0.48\linewidth}@{\hskip 0.04\linewidth}p{0.48\linewidth}@{}}
            \centering\small{\textbf{Men}} & \centering\small{\textbf{Women}} \tabularnewline
        \end{tabular}%
        }
        \vskip 0.05in
        \makebox[\linewidth][c]{
            \resizebox{\linewidth}{!}{
				\includegraphics[]{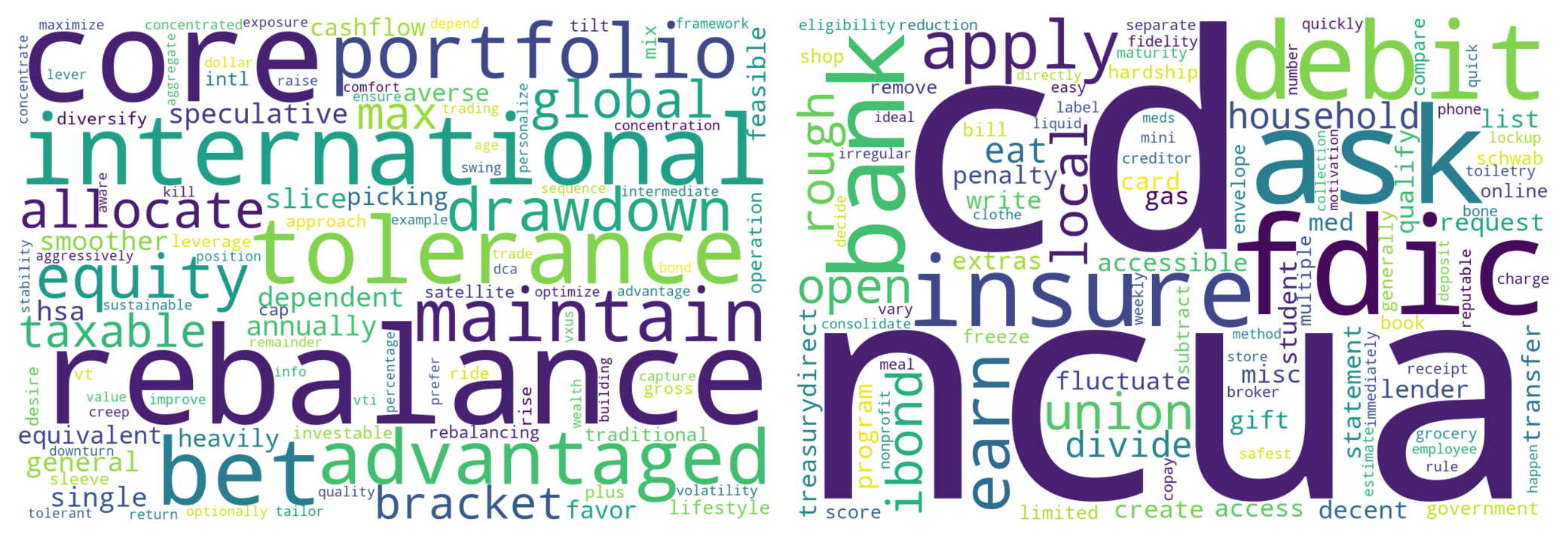}
        }}
    }
    \scriptsize{\emph{Notes:} This figure compares gender differences in respondent-written prompts and the corresponding LLM advice. Panel A pools the three prompt questions and shows words overrepresented in prompts written by men (left) and women (right). Panel B shows words overrepresented in the corresponding LLM advice for prompts written by men (left) and women (right). Word size is proportional to the difference in word frequency between groups. Common stopwords and words that mechanically reflect the survey questions are excluded.}
\end{figure}

\paragraph*{Men and women write prompts and receive advice with different language.} \autoref{fig:gender_wordclouds} shows word clouds of terms that are overrepresented in prompts (Panel A) written by men versus women and in the corresponding advice (Panel B). Men are much more likely to mention stocks and discuss investment strategies, while women are more likely to explicitly mention their gender, discuss debt, and mention safe, liquid investment options such as certificates of deposit (CDs), deposits, and credit unions. The LLM's responses also differ systematically, with more aggressive investment language in advice to men. One of the most common words overrepresented among men is ``rebalance,'' consistent with the gender differences in recommended portfolio rebalancing shown in \autoref{fig:gender_men}.

\paragraph*{Randomized labels to separate demand from supply.}
To separate demand from supply, we use the 83\% of prompts that do not explicitly reveal the respondent's gender, whether directly or through gendered self-descriptions (e.g., prompts written by a woman referencing being a ``mother'' or a ``wife''), and randomly insert either ``I am a man'' or ``I am a woman'' at the beginning of the prompt each time it is used in the simulation. Because the inserted label is randomized, its effect identifies the supply channel: how the LLM responds to gender holding the content of the prompt fixed. The effect of the prompt author's gender, holding the inserted label fixed, identifies the demand channel: how prompts written by men and women differ in ways that affect advice.

Using these labeled prompts, we simulate life cycles as in the baseline and estimate the following regression during working life
\begin{equation}
    Y_{i,t} = \beta_D \, \text{FemaleAuthor}_{i,t} + \beta_S \, \text{FemaleLabel}_{i,t} + \delta^{\text{bucket}}_{i,t} + \varepsilon_{i,t}\label{eq:gender_decomposition}
\end{equation}
where $Y_{i,t}$ is the outcome of interest for individual $i$ and age $t$, $\text{FemaleAuthor}_{i,t}$ indicates that the prompt used to generate advice in period $t$ was written by a woman, and $\text{FemaleLabel}_{i,t}$ indicates that the prompt used to generate advice was randomly assigned a female label. The fixed effects $\delta^{\text{bucket}}_{i,t}$ correspond to the age $\times$ income $\times$ employment cells within which prompts are randomly drawn, ensuring that comparisons are made among prompts assigned to similar simulated individuals. The coefficient $\beta_D$ captures demand (the effect of prompts written by women rather than men, holding labeled gender fixed), while $\beta_S$ captures supply (the effect of a gender label, holding prompt content fixed). Both coefficients measure period-by-period effects on recommended choices from a single prompt draw; they do not cumulate advice differences over the life cycle as in \autoref{fig:heterogeneity_gaps}.

\paragraph*{Gender gap in diversified holdings is mostly demand-driven.} The first three columns in the top row of \autoref{tab:demographic_demand_supply} show the results from estimating (\ref{eq:gender_decomposition}) with the change in diversified equity share as the dependent variable.\footnote{We use the change in diversified equity share rather than its level because of the substantial inertia documented in \autoref{fig:shocks}. Results are quantitatively similar when using the level of diversified equity shares.} Moving from a prompt written by a man and assigned a male label to a prompt written by a woman and assigned a female label reduces the recommended diversified equity share by 1.50pp in that period. Two-thirds of this combined effect is demand-driven: drawing a prompt written by a woman reduces the recommended diversified equity share by $\hat{\beta}_D=-$0.96pp relative to drawing a prompt written by a man, holding the randomized gender label fixed. The remaining one-third is supply-driven: holding prompt content fixed, a female label reduces the recommended diversified equity share by $\hat{\beta}_S=-$0.54pp. Therefore, both what respondents write and how the LLM responds to stated gender contribute to the gender gap in portfolio advice, with demand accounting for more of the difference.

\input{table_demographic_demand_supply}

\paragraph*{Gender gap in non-diversified holdings is entirely demand-driven.} We next estimate (\ref{eq:gender_decomposition}) with the change in non-diversified assets as the dependent variable, which includes individual stocks, crypto, gold, commodities, and collectibles, and accounts for approximately 1\% of baseline allocations. In contrast to the gender gap in diversified holdings, the middle three columns in the top row of \autoref{tab:demographic_demand_supply} show that the gender difference in non-diversified holdings is entirely demand-driven ($\hat{\beta}_D = -0.19$pp, $\hat{\beta}_S \approx 0$pp). Finally, in the last three columns, we estimate the same specification for the net saving rate. This gap is not statistically distinguishable from zero and is not the result of offsetting forces: both the demand coefficient ($\hat{\beta}_D = -0.88$pp) and the supply coefficient ($\hat{\beta}_S = -1.76$pp) are negative, though neither is statistically distinguishable from zero.

\clearpage

\paragraph*{Racial differences are smaller and are demand-driven.} Finally, we apply the same analysis to two comparisons based on race: Black versus White respondents, and Asian versus White respondents.\footnote{We do not have data on Hispanic ethnicity, so White, Black, and Asian are the only racial or ethnic groups in our survey with a sufficient number of prompts to perform the separate simulations by group.} As with gender, we drop any prompts that explicitly mention the respondent's race, which respondents disclose far less often than gender, and we randomly insert ``I am Black'', ``I am Asian'', or ``I am White'' at the beginning of each prompt. Using these labeled prompts, we simulate life cycles as in the baseline and estimate two versions of (\ref{eq:gender_decomposition}): one comparing prompts written by Black and White respondents, and another comparing prompts written by Asian and White respondents.

The second and third rows of \autoref{tab:demographic_demand_supply} show that the LLM's portfolio recommendations also differ by race: moving from a White-written, White-labeled prompt to a Black-written, Black-labeled prompt raises the recommended diversified equity share by 0.69pp, while the corresponding Asian--White difference is $-$0.81pp. Both differences are smaller than the gender difference of $-$1.50pp and are driven by demand. For the Black--White comparison, the demand effect is 0.66pp while the supply effect is approximately zero. For the Asian--White comparison, the supply effect is also small and statistically insignificant, though it is estimated less precisely than the gender benchmark. Differences in recommended non-diversified asset shares are similar in magnitude to the gender differences and are entirely demand-driven, whereas differences in saving rates are not statistically distinguishable from zero.

\paragraph*{Interpretation.}
Distinguishing demand- versus supply-driven differences in LLM advice matters because the two channels have distinct interpretations. The demand-driven differences across gender and race reflect real differences in what these demographic groups choose to write about, which could be amplified or dampened by model design. The supply-driven difference for gender, in which the LLM gives different advice to identical prompts labeled as coming from a man versus a woman, may reflect implicit inference: the model may use gender as a signal for characteristics correlated with gender, such as risk preferences or labor-market risk \citep{croson2009gender}. But even when such correlations exist, gender is a coarse signal, and this type of inference would ideally be made explicit to users \citep{filippin2016reconsideration}. Alternatively, supply-driven differences may reflect algorithmic bias, with the model reproducing gender stereotypes present in its training data.

The near-absence of a supply-side effect for race, in contrast to gender, could reflect how these models are trained: providers oftentimes evaluate and align their models to avoid differential treatment on the basis of race, and these safeguards may be enforced more stringently for race than for gender. However, this interpretation comes with a caveat. An explicit statement, such as ``I am Black'', is precisely the kind of salient signal such safeguards are designed to detect, so the null supply effect does not rule out differential responses to subtler markers of race, such as names, dialect, or contextual cues, that are more common in actual usage.

\vspace{0.3cm}

\section{Conclusion}

Large language models have, within a few years, become a major source of personal financial advice that around half of respondents report using in both our survey and industry surveys \citep{Lloyds2025_AI_Money,JDPower2025_AI_FinancialAdvice}. This paper develops a method to characterize the advice that these models provide and documents three facts. First, following LLM advice would move most respondents closer to the broad prescriptions of life cycle theory: they would hold larger savings buffers, be more likely to own diversified equity funds, and reduce their equity shares as they get older. Second, the advice departs from the theory's more demanding prescriptions: it implies unusually high patience, relies on simple saving and withdrawal heuristics, smooths consumption imperfectly after income shocks, and allows portfolios to drift passively with realized returns. Third, the advice varies systematically across groups, including by gender, financial literacy, and prior AI experience. These differences reflect both demand (what respondents write) and supply (how the model responds to otherwise identical prompts).

These findings suggest that generative AI could offer an affordable, widely accessible source of financial guidance and help overcome the costs, biases, and conflicts of interest associated with traditional human financial advisors \citep{reuter2024demand}. The fact that LLM advice broadly aligns with life cycle theory is promising, especially because these general-purpose models are not explicitly optimized to improve household financial decisions. At the same time, our results show that the content of AI financial advice depends not only on model capabilities but also on what users ask. The departures from life cycle theory that we document provide concrete diagnostics to guide model development, while the demand-side constraints we identify are likely to limit the quality of the advice users receive even as models improve. More broadly, our methodology, which combines survey responses with simulations, provides a systematic way to track how AI financial advice changes as models evolve, as users learn to write different prompts, and as firms and regulators decide whether this advice should mirror, amplify, or mitigate heterogeneity in what individuals ask. Our results characterize the effect of following LLM advice, not of receiving it: whether people act on these recommendations, and how any effects compare with those of human advisors, robo-advisors, and other sources of guidance, remain important open questions for future work.\footnote{\cite{Moss2026} provide early evidence: using data from an AI-powered investment adviser, they find that investors act on its recommendations, but that greater user intervention is associated with worse risk-adjusted performance.}


\clearpage
\begin{spacing}{0.20}
\small{
    \bibliographystyle{te}
    \bibliography{library,library_added}
}
\end{spacing}


\clearpage
\thispagestyle{empty}
\appendix
\setcounter{page}{1}
\setcounter{footnote}{0}
\titleformat{\section}[hang]{\Large\bfseries}{Appendix \Alph{section}.}{12pt}{}{}
\setcounter{figure}{0}
\counterwithin{figure}{section}
\renewcommand{\thefigure}{\Alph{section}\arabic{figure}}
\setcounter{table}{0}
\counterwithin{table}{section}
\renewcommand{\thetable}{\Alph{section}\arabic{table}}
\setcounter{assumption}{0}
\renewcommand{\theassumption}{\Alph{section}\arabic{assumption}}
\setcounter{proposition}{0}
\renewcommand{\theproposition}{\Alph{section}\arabic{proposition}}
\renewcommand{\thepage}{A.\arabic{page}}

\makeatletter
\immediate\write\@auxout{\string\setcounter{tocdepth}{-5}}
\makeatother

\begin{appendices}

\tableofcontents
\addtocontents{toc}{\protect\setcounter{tocdepth}{3}} 

\clearpage

\small

\input{appendix}


\end{appendices}

\end{document}

%% file: abstract.tex
We ask a representative sample to write prompts seeking spending and investing advice from LLMs, then simulate the lifetime effects of following the advice under realistic asset and labor market conditions. Applying this method to \GPT, we find following the advice would move respondents toward life cycle theory: broader participation in diversified equity funds, age-declining equity shares, and larger savings buffers. Recommendations vary systematically by gender, prior AI experience, and financial literacy. For gender, two-thirds of recommended equity-share differences arise from men and women writing different prompts (demand), while one-third arise from gender labels attached to otherwise identical prompts (supply).

%% file: introduction.tex
Millions of people now turn to large language models (LLMs) for personal financial advice. Within a few years of these tools becoming widely available, around half of adults in the U.S.\ and the U.K.\ report using them for financial guidance \citep{Lloyds2025_AI_Money,JDPower2025_AI_FinancialAdvice}, a share that might already exceed the proportion who consult a human financial advisor \citep{Saad2025GallupAdvice}. This rapid adoption raises the prospect that affordable, widely accessible, high-quality financial advice may be closer than ever. However, whether that promise is realized depends on two distinct forces: the content of the advice these models give, and how people change their behavior in response to it. A natural first step, and the focus of this paper, is to characterize the advice itself: what LLMs recommend, and how those recommendations vary across individuals.

We develop and implement a method to describe quantitatively the personal financial advice that LLMs give. Studying this advice systematically is challenging because it depends on what individuals choose to ask about, on features of the models that supply the advice, and on how uncertain financial outcomes accumulate over the life cycle. We address these three challenges by (i) surveying a demographically balanced sample to elicit realistic prompts seeking financial advice from an LLM, (ii) simulating the quantitative implications of the advice received in response to these prompts over the life cycle under realistic asset and labor market conditions, and (iii) evaluating why the advice differs across heterogeneous groups by randomizing features of prompts, which allows us to separate supply-side differences in how models respond to otherwise identical prompts from demand-side differences in what individuals ask.

We apply our method to the spending, saving, and investing advice provided by \GPT. We focus on these domains both because they involve some of the most consequential financial decisions households make and because our survey shows they are the topics people most commonly seek AI financial guidance on. We have three main results. First, following LLM advice would move most survey respondents closer to the broad recommendations of standard life cycle theory relative to their observed behavior, including greater participation in diversified equity funds, equity shares that decline with age after 45, and sizable savings buffers. Second, LLM advice departs from life cycle theory in systematic ways: it implies unusually high patience, relies on simple heuristics, smooths consumption imperfectly after income shocks, and allows portfolios to drift passively with realized returns. Third, LLM advice varies systematically with individual characteristics, such as gender, financial literacy, and prior AI experience. These differences accumulate over the life cycle into wealth differences at age 60 of approximately 4--6\% between groups. As in the market for human financial advice, such differences can reflect both demand and supply \citep{reuter2024demand}. For example, we find that men receive higher recommended equity shares than women, and by randomizing gender labels, we show that two-thirds of this difference in investment advice comes from differences in what men and women write in their prompts, while one-third comes from the model supplying different advice to otherwise identical prompts when labeled as coming from men rather than women.

Taken together, our results highlight the potential of generative AI to improve financial decision-making, but suggest that its impact is likely to vary across individuals. The finding that LLM advice broadly aligns with life cycle theory should not be taken for granted. Unlike other forms of automated financial advice, such as robo-advisors, these models are not optimized to improve household financial decisions \citep{DAcunto2019,Rossi2020}. One might therefore worry that, in pursuit of engagement, LLMs would instead reinforce existing behaviors, validate biases, or simply tell people what they want to hear \citep{Fedyk2026}. Our results suggest a more nuanced conclusion: current general-purpose LLMs often provide advice that aligns with standard financial guidance, but this advice is sensitive to demand and, therefore, differs systematically across users.

While we primarily use \GPT, we show that our main results are quantitatively similar using \Gemini---a similar generation model from a different provider---and \GPTNew---a more recent model from the same provider. Nevertheless, the advice provided by these models may differ from that of future models. Our contribution is therefore not only to characterize the advice of a particular generation of models, but also to provide a framework for studying LLM financial advice. As these models continue to evolve rapidly, three features of our analysis are likely to remain relevant. First, differences in advice arise not only from model capabilities, but also from demand: what users ask, what information they provide, and how they frame their financial problems. These demand-side constraints are likely to remain important even as models continue to improve rapidly. Second, our survey-and-simulation approach can be reapplied to track how both model and user behavior change over time. Third, progress in applying generative AI to household finance will depend in part on the availability of benchmarks and diagnostic tests against which models can be evaluated and improved. While many such benchmarks exist for other tasks, we provide a set of diagnostics for personal financial advice grounded in normative life cycle theory (e.g., consumption smoothing, diversification, portfolio rebalancing).

\paragraph*{Methodology.} Our method for studying LLM financial advice proceeds in three steps. First, we survey a demographically balanced sample of 1,000 U.S. adults recruited through Prolific. We ask respondents to write three prompts to an LLM. The first prompt is a description of their financial situation, the second is for advice on spending versus saving, and the last is for investing advice. This is the main innovation of our approach: using prompts written by regular users introduces meaningful heterogeneity in both the topics people raise and how they express themselves. To our knowledge, our survey provides the first systematic evidence on how prompts vary across individuals seeking financial advice from LLMs. Alongside the prompts, we collect data on respondents' demographics, financial situation, financial literacy, and prior experience seeking financial guidance from LLMs. 

In the second step, we build a quantitative life cycle model in the tradition of \cite{Gourinchas2002} and \cite{Cocco2005} (and most closely related to \citealt{Choukhmane2026}). The model is calibrated to match historical U.S. data and incorporates stochastic income, stochastic asset returns, job-to-job and unemployment transitions, mortality risk, and a tax and social insurance system. We use the model to simulate households' consumption-saving decisions and their allocation of savings across four asset classes: fixed income, diversified equity funds, individual stock holdings, and other risky assets, including crypto, gold, commodities, and collectibles. 

The final step in our method combines the prompts written by our survey respondents with our life cycle model. Rather than simulating choices based on policy functions derived from dynamic optimization, we simulate choices based on the LLM recommendations. Specifically, at each age, we randomly match each agent in our simulation with a prompt written by a survey respondent with a similar age, income, and employment status. We then replace the state variables in that prompt with those of the simulated individual and submit two successive LLM queries: first, we submit the prompt and receive textual financial advice from an LLM; then, we use a second LLM to translate this textual advice into quantitative recommendations for saving, consumption, and asset allocations. We repeat this process at each age over the individual's life cycle and across many simulated individuals. Importantly, each query to the LLM is independent, with no memory of past responses; the only link across periods is the evolution of state variables that depend on prior choices.

Our method can be reapplied to other populations, advice topics, or models. A key design choice is the trade-off between specificity and external validity. Asking people to write prompts about defined topics, in our case spending, saving, and investing, allows us to compare responses across individuals and integrate them into a parsimonious life cycle framework. We view this as a feature of the approach that can be adjusted depending on the research question. Greater specificity would allow studying decisions that we abstract from (e.g., insurance, housing, credit management), while more open-ended prompts would better reflect real-world usage at the cost of being harder to compare across individuals and incorporate into a simulation framework.

\paragraph*{Main findings.} Our main results come from applying this method to \GPT, which we refer to as ``the LLM'' throughout. We chose this LLM because ChatGPT is the most commonly used model for financial advice in our survey (and others, e.g., \citealt{Lloyds2025_AI_Money,JDPower2025_AI_FinancialAdvice}) and \GPT was the latest model at the time we ran our analysis; we find quantitatively similar results using \Gemini and \GPTNew.

We document three facts about LLM advice over the life cycle. First, following LLM recommendations would move most individuals closer to the broad principles of life cycle theory relative to their observed behavior. On saving, many individuals across all age groups report having accumulated little financial wealth; by contrast, following the advice would lead virtually all simulated individuals to build savings buffers above \$10,000 by age 30. On portfolio choice, it would induce near-universal participation in the stock market, with equity shares that decline with age after 45, and risky holdings concentrated in diversified equity funds rather than individual stocks, crypto, gold, commodities, or collectibles. In addition to these broad patterns, the LLM's advice responds to the topics individuals raise in ways that are generally consistent with standard theory. For example, prompts that mention macroeconomic uncertainty lead the LLM to increase saving and reduce equity holdings, consistent with a stronger precautionary motive.

Second, we find that the LLM's advice departs from the more subtle prescriptions of life cycle theory in systematic ways. The advice implies an unusually high discount factor in a standard life cycle model. Instead of tailoring recommendations closely to individuals' evolving economic circumstances, the advice often relies on simple saving and withdrawal heuristics, including bunching at round numbers and recommending fixed withdrawal rules in retirement. Consistent with this reliance on heuristics, consumption is insufficiently smoothed: the LLM recommends too little decumulation in retirement and sharp consumption declines after job loss, even when simulated individuals have sufficient liquid wealth. Asset allocation advice exhibits a similar lack of active adjustment as portfolios drift passively with realized returns rather than being actively rebalanced. These departures appear to reflect, in part, the prompts that people write. A more structured ``academic'' prompt, which provides details on each agent's financial situation and explicitly references life cycle planning and acting in the user's best interest, improves consumption smoothing and reduces reliance on heuristics, though it does not increase active portfolio rebalancing.

Third, the LLM's advice varies systematically with individual characteristics, such as gender, financial literacy, and prior AI experience. For each of these characteristics, we split survey respondents into two groups and repeat our simulation, sampling only prompts written by individuals in each group (e.g., simulating life cycle paths using only prompts written by men versus women). The advice varies meaningfully along each dimension: average wealth at age 60 is around 5\% higher when sampling prompts written by men, individuals with high financial literacy scores, or those who report prior use of AI for financial guidance. These differences arise because the LLM consistently recommends lower equity shares for prompts written by women or individuals with lower financial literacy, and lower saving rates for prompts written by individuals without prior AI use.

To understand why LLM advice differs across individuals, we develop an approach to separate the roles of demand and supply. Focusing on gender, we use the set of survey prompts that do not explicitly identify the respondent's gender, and randomly add gender labels, such as ``I am a man.'' This separates demand-side differences (i.e., men and women write different prompts) from supply-side differences (i.e., the LLM gives different advice to the same prompt when the stated gender changes). We find that approximately two-thirds of the gender gap in recommended diversified equity shares is demand-driven: it reflects differences in the content of men's and women's prompts that remain even when both are assigned the same gender label. The remaining one-third is supply-driven: the same prompts receive higher equity recommendations when labeled as coming from a man rather than a woman. In contrast, gender differences in recommended holdings of non-diversified assets, such as individual stocks or crypto, are entirely demand-driven. These findings echo evidence from human financial advisors, where gender differences in advice also reflect both demand and supply \citep{bhattacharya2024women,bucherkoenen2025gender}. Applying the same approach to race, we find smaller differences and no supply-side effect: explicit race labels leave recommendations essentially unchanged.

\paragraph*{Related literature.} This paper contributes to two strands of literature. First, we contribute to the literature on financial advice and financial innovation by introducing a framework to quantitatively study LLM-generated financial advice. This literature is motivated by the evidence that individuals do not, on their own, always make optimal financial decisions.\footnote{This includes evidence that individuals do not take full advantage of available 401(k) employer matches \citep{choi2011100,choukhmane2025efficiency}, do not efficiently allocate assets across accounts \citep{bergstresser2004asset}, or fail to refinance fixed-rate mortgages when beneficial \citep{andersen2020sources}. For reviews, see \citet{Beshears2018a}, \citet{Gomes2021b}, and \citet{campbell2025household}.} While this creates a role for professional advice, substantial evidence shows that existing financial advice sources are limited by both demand- and supply-side constraints (see \citealt{reuter2024demand} for a recent review). For example, professional human advisors are often costly, and their advice can be low quality \citep{linnainmaa2021misguided,Andreis2025}, subject to conflicts of interest \citep{Mullainathan2012a}, and affected by misconduct \citep{Egan2019a}. Popular personal finance books are cheaper and more accessible, but their guidance often contains fallacies and departs from the prescriptions of normative economic models \citep{choi2022popular}. Automated advice, such as robo-advisors, can improve financial decision-making \citep{DAcunto2019,Rossi2020,Reher2024}, but, relative to LLMs, its uptake has been much slower. One possible reason is that robo-advisors may be more limited in their ability to provide personalized conversational advice, as well as the soft skills and emotional support that clients value in human advisors \citep{Greig2025a}. A natural question is therefore whether LLMs can offer advice that is at once affordable, widely accessible, personalized, and high-quality, a combination that no existing source fully provides.

Our contribution is to provide evidence on whether LLM-generated advice can fulfill this promise, taking into account that the advice depends on the questions people ask, the characteristics of the models, and the accumulated impact over the life cycle. In doing so, we extend a longstanding tradition of using quantitative life cycle models to evaluate financial products and policies, with recent applications to adjustable-rate mortgages \citep{Guren2021b,Campbell2021}, target-date funds \citep{duarte2024simple}, saving nudges \citep{choukhmane2025default,choukhmane2025efficiency}, and income-contingent student loans \citep{de2025insurance}. We apply this approach to what is perhaps the fastest-growing area of innovation in personal finance: generative AI.

Second, we contribute to the growing literature on how LLMs are changing economics and finance (see \citealt{Mo2025} for a review), and in particular to the emerging literature that studies LLMs as models of human behavior. This literature shows that LLMs often replicate human behavior and biases elicited in surveys and classic economic experiments \citep{Horton2023,Ross2024,Bini2025}.\footnote{Also related is the strand of this literature that studies agent-based economic models populated by LLM agents, including \cite{hao2025multi}, who study two-period consumption-saving problems, and \cite{douglas2024consumption}, who study a \cite{Krusell1998} economy.} A related strand studies LLM behavior in financial settings. Here too, LLMs both resemble and depart from human behavior: \cite{Ouyang2025} find that LLM risk-taking is stable across a variety of behavioral tasks; \cite{Fedyk2026} show that LLM preferences over asset allocations resemble those of younger, high-income individuals, but are less likely to violate transitivity; and \cite{rumpf2026humans} compare recommendations from individuals, professional advisors, and LLMs, and find that LLMs are far more sensitive to risk preferences but also exhibit substantial stochasticity. A related set of recent papers studies LLMs directly as providers of financial advice using researcher-designed personas or scenarios, examining whether LLMs can infer investor preferences and tailor recommendations while also documenting limitations such as stochasticity, biases, and sensitivity to demographic attributes \citep[e.g.,][]{Fieberg2025,takayanagi2025generative,foltyn2026worth}.

Relative to this literature, the methodology that we develop makes two contributions. First, instead of using LLMs to learn about, simulate, or replace human survey responses, we use surveys of humans to learn about LLMs: prompts written by a survey sample allow us to characterize both the supply \emph{and} demand sides of human--LLM interactions. In this sense, our approach uses humans to learn about AI. Second, and more specific to our financial advice application, we embed LLM advice within a dynamic, stochastic life cycle model. This allows us to characterize the cumulative effects of LLM advice over an entire lifetime and compare the resulting behavior with both observed behavior and a quantitative benchmark grounded in economic theory. This life cycle perspective distinguishes our analysis from studies that evaluate LLM advice primarily in static survey or experimental settings.

%% file: table_smm_results_data.tex
\begin{tabular}{llrr}
\toprule
\textbf{Prompt Type} & \textbf{LLM} & $\hat{\beta}$ & $\hat{\gamma}$ \\
\midrule
Survey & \GPT & 1.034 & 5.3 \\
Academic & \GPT & 0.990 & 4.7 \\
Survey & \Gemini & 1.054 & 5.1 \\
Survey & \GPTNew & 1.032 & 5.3 \\
\bottomrule
\end{tabular}

%% file: table_demographic_demand_supply.tex
\begin{table}[!t]
\caption{Separating Demand and Supply Drivers of Demographic Differences in Advice}
\label{tab:demographic_demand_supply}
\footnotesize
\makebox[\linewidth][c]{
\resizebox{\linewidth}{!}{
\begin{tabular}{l*{9}{c}}
\toprule
& \multicolumn{3}{c}{Diversified Equity Share} & \multicolumn{3}{c}{Nondiversified Asset Share} & \multicolumn{3}{c}{Net Saving Rate} \\
\cmidrule(lr){2-4}\cmidrule(lr){5-7}\cmidrule(lr){8-10}
& \shortstack{Author\\(Demand)} & \shortstack{Label\\(Supply)} & Total & \shortstack{Author\\(Demand)} & \shortstack{Label\\(Supply)} & Total & \shortstack{Author\\(Demand)} & \shortstack{Label\\(Supply)} & Total \\
\midrule
Female & -0.96 & -0.54 & -1.50 & -0.19 & 0.01 & -0.19 & -0.88 & -1.76 & -2.63 \\
(vs. Male) & (0.15) & (0.15) & (0.21) & (0.03) & (0.03) & (0.04) & (1.67) & (1.64) & (2.34) \\
\addlinespace
Black & 0.66 & 0.03 & 0.69 & -0.11 & -0.02 & -0.13 & 3.09 & 0.14 & 3.24 \\
(vs. White) & (0.21) & (0.18) & (0.28) & (0.05) & (0.04) & (0.06) & (3.39) & (2.89) & (4.45) \\
\addlinespace
Asian & -0.65 & -0.16 & -0.81 & -0.16 & -0.02 & -0.18 & 1.21 & -2.44 & -1.23 \\
(vs. White) & (0.26) & (0.18) & (0.31) & (0.05) & (0.04) & (0.07) & (4.13) & (2.89) & (5.04) \\
\bottomrule
\end{tabular}
}
}
\vskip 0.1in
\scriptsize{\emph{Notes:} This table separates demographic differences in LLM recommendations into differences associated with prompt content and effects of demographic labels, using the specification in (\ref{eq:gender_decomposition}). Each row reports one comparison: Female relative to Male, Black relative to White, or Asian relative to White. For each comparison, the sample excludes prompts that explicitly mention the relevant demographic characteristic, and labels are randomly assigned as described in \Cref{sec:gender}. The three sets of columns report changes in diversified equity share, non-diversified asset share, and the net saving rate, defined as post-tax earnings minus consumption divided by post-tax earnings. All regressions are restricted to working-life observations. ``Author'' reports $\beta_D$, the coefficient on the author's actual demographic characteristic, holding the assigned label fixed; it captures differences associated with prompt content (the demand channel). ``Label'' reports $\beta_S$, the coefficient on the demographic label inserted into the prompt, holding content fixed; it captures the causal effect of the label on the LLM's response (the supply channel). ``Total'' reports $\beta_D+\beta_S$, the combined effect of a prompt being both authored by and labeled as belonging to the demographic group. Standard errors are reported in parentheses.}
\end{table}

%% file: appendix.tex
\clearpage

\section{Prompt Construction and Design}\label{app:survey_description}

This appendix describes how the respondent-written prompts introduced in \Cref{sec:survey} are processed and integrated into the life cycle simulation framework. The survey questions are shown in \autoref{fig:survey_questions} and sample demographics are in \autoref{tab:survey_demographics}.

\subsection{Survey Prompt Variable Insertion}
Whenever an individual prompt mentions a state variable from the life cycle model (income, age, wealth, tenure, or asset allocation), we replace that mention with a placeholder that is updated to the simulated individual's state. \autoref{tab:prompt_variable_mapping} lists representative replacement patterns for each state variable. The rest of this section describes the rules governing these replacements.

\noindent \textbf{Household rules.}
Our model simulates outcomes at the individual level. When a prompt instead describes income or wealth at the household level, we scale up the simulated amount before inserting it: by default we double the income and wealth that are inserted unless the prompt mentions more than two working adults, in which case those numbers are scaled by the number of working adults. When a prompt provides information at the individual level, we use it directly. If the prompt mentions the income of another household member, we leave that income unchanged and insert household income as the sum of that income and the simulated individual income. If the prompt specifies each member's balance in an asset class, we assume holdings in that class are proportional to the state balances, so household holdings equal the simulated individual holdings divided by the respondent's share of total balance.

\noindent \textbf{Income values.}
We replace each income amount with its simulated counterpart, using gross or after-tax income depending on the tax treatment specified in the prompt. If the prompt does not specify the tax treatment, we insert the after-tax amount.\footnote{Phrases such as ``I bring home \$X'' or ``my monthly salary is \$X'' can be read as either gross or net income, by both the respondents and the LLM.} If the prompt reports an hourly wage and a number of hours worked per week, denoted by $N$, we compute hourly income as the simulated income divided by $52\times N$, and insert that amount. If hours or income are reported as a range, we use the midpoint.\footnote{When replacing a range with a single value would be ungrammatical, we replace both bounds with the same value.} If hours are not reported, we set $N=40$ if the job is full-time, or $N=20$ if the job is part-time. When income comes from multiple sources or is reported at multiple frequencies (e.g., monthly and hourly), we annualize each amount, preserve the implied ratios across sources, and rescale to the original frequency.

\noindent \textbf{Wealth values.} 
Many prompts describe financial wealth as dollar amounts held in several accounts. If the prompt mentions a single account with an ambiguous label such as ``savings,'' we interpret that amount as total wealth. If a single account clearly corresponds to one asset class, we interpret the amount as wealth in that class. For prompts with multiple accounts, let $\{D_n\}_{n\in N_D}$, $\{I_n\}_{n\in N_I}$, $\{S_n\}_{n\in N_S}$, and $\{O_n\}_{n\in N_O}$ denote dollar amounts in diversified stock, individual stock, non-stock, and other risky assets. We classify holdings as diversified when the prompt mentions exchange-traded funds, index funds, or tickers of assets in these categories, and as individual stocks when the prompt refers to specific stocks or a small bundle of stocks chosen by the writer. When the prompt refers to stock market investments without distinguishing between the two, we treat them all as diversified and assume no holdings of individual stocks. Instead of mentioning asset classes directly, individuals often mention holdings in accounts that can contain multiple asset classes, such as retirement or brokerage accounts. We refer to these accounts as ``mixed accounts'', denoting their holdings by $\{M_n\}_{n\in N_M}$. With these mixed accounts, total wealth is defined as $W \equiv \sum_{n\in N_D}D_n + \sum_{n\in N_I}I_n + \sum_{n\in N_S}S_n + \sum_{n\in N_O}O_n + \sum_{n\in N_M}M_n$. We also denote $\theta_x$ as the share of wealth in asset class $x$.

We now describe how we replace the dollar holdings in the different asset classes, distinguishing between two cases with and without mixed accounts. In the notation below, variables with a tilde (e.g., $\tilde{W}$) refer to the values in our simulation, while variables without a tilde (e.g., $W$) refer to values mentioned in prompts.

\begin{itemize}
    \item \textbf{Scenario 1: Well-defined asset classes.} If the prompt mentions only asset classes that fall within our four asset classes, then the wealth in each asset is distributed to all of the prompt-mentioned asset classes in ratios that match the original within-class ratios:
    \begin{equation}\label{eq:scenario1}
        \tilde{X}_n = \tilde{\theta}_x\frac{X_n}{\sum_{n\in N_X}X_n}\,\tilde{W}, \qquad X \in \{D, I, S, O\}.
    \end{equation}
    \textbf{Example:} Suppose that the prompt mentions that the original prompt writer has \$1,200 in a checking account and \$1,800 in a savings account. If the simulated agent that draws this prompt has \$10,000 in the non-stock asset class, then the prompt will state that there is \$4,000 in a checking account and \$6,000 in a savings account.
    \item \textbf{Scenario 2: Mixed accounts.} If the prompt mentions an asset that may contain multiple asset classes (e.g., a 401k or mutual fund), we first allocate all single-asset class holdings according to (\ref{eq:scenario1}), and then allocate whatever balance remains across the mixed accounts to match the original ratios:
    \begin{equation}\label{eq:scenario2_1}
        \tilde{X}_n = \min\!\left(\theta_x,\, \tilde{\theta}_x\right)\frac{X_n}{\sum_{n\in N_X}X_n}\tilde{W}, \qquad X \in \{D, I, S, O\},
    \end{equation}
    \begin{equation}\label{eq:scenario2_2}
        \tilde{M}_n = \left(\tilde{W} - \sum_{n\in N_D}\tilde{D}_n - \sum_{n\in N_I}\tilde{I}_n - \sum_{n\in N_S}\tilde{S}_n - \sum_{n\in N_O}\tilde{O}_n\right)\frac{M_n}{\sum_{n\in N_M}M_n}.
    \end{equation}
    \textbf{Example 1:} Suppose that the prompt mentions that the original prompt writer has \$1,200 in a checking account and \$1,800 in a 401(k). If the simulated agent has \$5,000 in the non-stock asset and \$5,000 in all other asset classes combined, then the prompt will state the following values:
    \begin{equation*}
    \begin{split}
        \text{Checking Account} &= \min\left(\frac{\$5,000}{\$10,000}, \frac{\$1,200}{\$1,200 + \$1,800}\right) \cdot \$10,000 = \$4,000, \\
        \text{401(k) Account} &= \$10,000 - \text{Checking Account} = \$6,000.
    \end{split}
    \end{equation*}
    \textbf{Example 2:} Suppose that the prompt mentions that the original prompt writer has \$1,800 in a checking account and \$1,200 in a 401(k). If the simulated agent has \$5,000 in the non-stock asset and \$5,000 in all other asset classes combined, then the prompt will state the following values:
    \begin{equation*}
    \begin{split}
        \text{Checking Account} &=  \min\left(\frac{\$5,000}{\$10,000}, \frac{\$1,800}{\$1,800 + \$1,200}\right) \cdot \$10,000 = \$5,000, \\
        \text{401(k) Account} &= \$10,000 - \text{Checking Account} = \$5,000.
    \end{split}
    \end{equation*}
    In this example, the minimum in (\ref{eq:scenario2_1}) binds: the simulated individual cannot have a 60\% allocation in checking accounts because that would exceed the \$5,000 held in the non-stock asset class.
\end{itemize}

\noindent \textbf{One-time lump sum payments.} 
When a prompt refers to a one-time lump-sum payment (e.g., an inheritance, settlement, or house sale), we replace the stated dollar value only if the payment has already occurred; expected future payments are left unchanged. We replace the value by first computing its share of the respondent's self-reported financial wealth, then applying that same share to simulated wealth.\footnote{Since self-reported financial wealth is categorical, we use the midpoint of each range, and \$50,000 when the respondent selected ``I don't know/prefer not to answer.''} For example, if a respondent reports having received a \$10,000 inheritance in their prompt, and separately reports total financial wealth of \$100,000, we replace the stated \$10,000 value with 10\% of total simulated wealth.

\subsection{Prompt Selection}\label{app:prompt_selection}
As described in \Cref{sec:survey}, out of 1,000 complete Qualtrics responses, we removed 33 that failed Prolific's platform-level authenticity checks and 15 that failed a minimum specificity requirement, leaving 952 responses. To pass the minimum specificity check, each of the respondent's three prompts had to address its assigned topic: the first the respondent's situation, the second general spending or saving, and the third investing.

\subsection{Prompt Bucketing}
\begin{figure}[!htbp]
    \caption{Prompt Bucketing Tiers}
    \vspace{-0.2in}
    \label{fig:prompt_bucketing}
    \makebox[\linewidth][c]{
	    \resizebox{0.7\linewidth}{!}{
		    \includegraphics[]{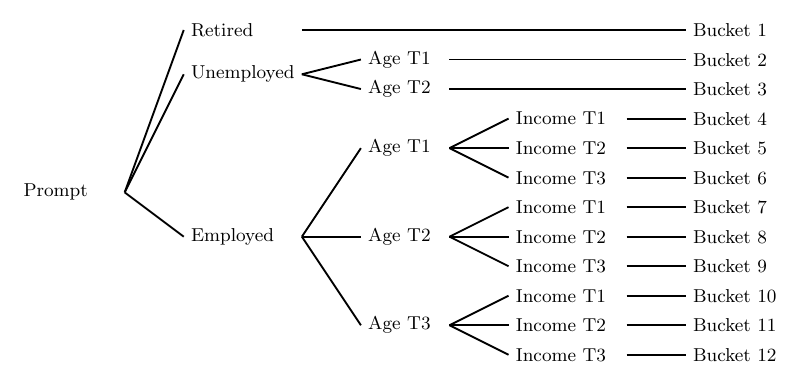}
	}}
    \scriptsize{\emph{Notes:} This figure outlines the tiered system used to assign survey prompts to buckets. Prompts are bucketed by respondent employment status, age tercile, and, for employed respondents, income tercile, yielding 12 buckets in total. The age and income cutoffs are heuristic thresholds based on the survey sample. In each period, the simulated individual's employment status, age, and income determine the relevant bucket, and a prompt is randomly drawn from that bucket.}
\end{figure}

We divide prompts into 12 buckets defined by employment status, age, and income. We first group prompts by the respondent's employment status (employed, unemployed, or retired). Within each employment group, we then split by age: into halves for the unemployed, into terciles for the employed, and into a single bucket for the retired. For the employed, we further split each age tercile into income terciles, yielding 2 (unemployed) + 3$\times$3 (employed) + 1 (retired) = 12 buckets. \autoref{fig:prompt_bucketing} illustrates the bucketing procedure, and \autoref{tab:prompt_bucketing_distribution} reports the distribution of prompt sets across buckets.\footnote{In our heterogeneity analysis (e.g., using only prompt sets written by women), we repeat the sorting and bucketing procedure separately for each subgroup.}
Because the survey elicits income in discrete brackets, ties are common and can make tercile divisions unbalanced. We assign tied respondents to the lower income tier, except when so many respondents select the highest bracket that this rule would leave the top tercile too small, in which case we assign ties to the higher tier instead.

\subsection{LLM Advice Pipeline}
We convert LLM responses into consumption and portfolio decisions in two steps. In the first step, we (i) assign each simulated agent-age pair to a prompt bucket based on employment status, age, and income, (ii) draw a prompt set at random from that bucket, (iii) construct the prepared message, and (iv) send it to the LLM, saving its response.\footnote{We instruct the LLM to provide responses in at most 200 words.} In the second step, we pass the response back to a second LLM along with deterministic instructions for converting the qualitative advice into quantitative choices (in JSON format). These quantitative outputs are then used to update the endogenous state variables in our simulation.

\clearpage

\begin{table}[!htbp]
    \caption{Mapping between State Variables and Respondent-Written Prompts}
    \label{tab:prompt_variable_mapping}
    \footnotesize
    \makebox[\linewidth][c]{
	    \begin{tabular}{ll}
	    \toprule
	    Original text & Replacement text \\
	    \midrule
	    \multicolumn{2}{c}{Age} \\
    \midrule
    ``X years old'' & ``\{age\} years old''\\
    ``in my Xs'' & ``in my \{math.floor(age/10)*10\}s''\\
    ``I retire in X years'' & ``I retire in \{65 - age\} years'' \\
    \midrule
    \multicolumn{2}{c}{Income} \\
    \midrule
    ``\$X a year before taxes'' & ``\$\{pretaxincome:,.0f\} a year before taxes'' \\
    ``\$X a month before taxes'' & ``\$\{pretaxincome/12:,.0f\} a month before taxes'' \\
    ``\$X biweekly before taxes'' & ``\$\{pretaxincome/26:,.0f\} biweekly before taxes'' \\
    ``\$X a week before taxes'' & ``\$\{pretaxincome/52:,.0f\} a week before taxes'' \\
    ``\$X a day before taxes'' & ``\$\{pretaxincome/(52*5):,.0f\} a day before taxes'' \\
    ``\$X an hour before taxes'' & ``\$\{pretaxincome/(52*40):,.2f\} an hour before taxes'' \\
    & \\
    ``\$X a year after taxes'' & ``\$\{income:,.0f\} a year after taxes'' \\
    ``\$X a month after taxes'' & ``\$\{income/12:,.0f\} a month after taxes'' \\
    ``\$X biweekly after taxes'' & ``\$\{income/26:,.0f\} biweekly after taxes'' \\
    ``\$X a week after taxes'' & ``\$\{income/52:,.0f\} a week after taxes'' \\
    ``\$X a day after taxes'' & ``\$\{income/(52*5):,.0f\} a day after taxes'' \\
    ``\$X an hour after taxes'' & ``\$\{income/(52*40):,.2f\} an hour after taxes'' \\
    \midrule
    \multicolumn{2}{c}{Stock Allocation} \\
    \midrule
    ``X\% of wealth in savings'' & ``\{100 - current\_stock\_allocation\}\% of wealth in savings''\\
    ``X\% of wealth in stock'' & ``\{current\_stock\_allocation\}\% of wealth in stock'' \\
    \midrule
    \multicolumn{2}{c}{Miscellaneous} \\
    \midrule
    ``X years into a job'' & ``\{tenure\} years into a job'' \\
    ``past annual income of \$X'' & ``past annual income of \$\{avgpastincome:,.0f\}'' \\
    \bottomrule
	    \end{tabular}
	    }
	    \vskip 0.1in
	    \scriptsize{\emph{Notes:} This table summarizes text replacement used to map respondent-written prompts to simulated state variables. The left column shows examples of original text in the survey prompts; the right column shows the corresponding replacement text used in the LLM prompt, where placeholders are filled with the simulated individual's age, income, portfolio allocation, or other state variables.}
\end{table}

\clearpage

\input{table_prompt_bucket_distribution.tex}

\clearpage

\subsection{LLM Prompts}\label{app:llm}

This subsection presents the full text of the two researcher-written prompts used in our analysis. The academic prompt (\Cref{sec:benchmarks}) is used to elicit financial advice from \GPT, while the translation prompt is used to convert the LLM's textual advice into quantitative choices via \GPTMini, as described in Step 4 of \Cref{sec:simulation}.

\input{prompt_academic}
\clearpage
\input{prompt_translation}

\clearpage

\section{Life Cycle Model Details}\label{app:life_cycle_model}

This appendix provides additional details on and the calibration of the life cycle model described in \Cref{sec:life_cycle_model}.

\subsection{Demographics and Preferences}

Each period corresponds to one year, and working life starts at $t=0$ and lasts for $T_w$ periods. Retirement starts at $t = T_w$, and agents can live at most $T$ periods. Before their certain death in period $t = T$, investors face age-dependent mortality risk with survival probability in period $t+1$ conditional on survival in period $t$ denoted by $m_t$. We denote an investor's age as $a_t = t + a_0$, where $a_0$ is the age investors enter working life. Investors have time-separable CRRA utility over consumption streams. We denote investors' annual discount factor as $\beta$ and relative risk aversion as $\gamma$.

\subsection{Labor Market}

At any point in time, investors can be in one of four employment states, denoted $emp_t$: $E$ = employed by the same employer as in the previous period, $JJ$ = employed by a different employer than in the previous period, $U$ = unemployed in the current period, and $Ret$ = retired. The fact that investors face uncertainty about their future employment status, in addition to earnings risk, is an important feature of our model because it introduces deviations in income shocks from normality, which \cite{Braxton2025} highlight as important empirically.

\noindent \textbf{Employment: $emp_t = E$.} While working, investors earn an exogenous income $w_t$. The log income process consists of a deterministic component that is cubic in age, a stochastic component that follows an AR(1) process with normally distributed innovations, and a transitory shock that is normally distributed:
\begin{align}
\ln w_t =& \delta_0 + \delta_1a_t + \delta_2a_t^2 + \delta_3a_t^3 + \eta_t + \varepsilon_t, \quad \eta_t = \rho \eta_{t-1} + \xi_t^E, \label{eq:wage_E} \\
\xi_0^E &\sim N(0, \sigma_{\xi_0}^2), \quad \xi_t^E \sim N(0,\sigma_\xi^2), \quad \varepsilon_t\sim N(0,\sigma^2_\varepsilon )\qquad \forall t > 0. \nonumber
\end{align}
Investors' tenure status evolves according to $ten_t = ten_{t-1} + 1$ if they remain employed by the same employer. We assume that the initial distribution of $\xi_0^E$ is different from that of $\xi_t^E$ for $t>0$ to account for heterogeneity in initial-period incomes.

\noindent \textbf{Job transition: $emp_t = JJ$.} While in the employed state ($E$), an investor may make a job-to-job transition with probability $\pi^{JJ}(t, ten_t)$ that depends on both their age and tenure at the current job. After a job-to-job transition, income evolves according to:
\begin{align}
\ln w_t = \delta_0 + \delta_1a_t + \delta_2a_t^2 + \delta_3a_t^3 + \eta_t + \varepsilon_t, \quad \eta_t = \rho \eta_{t-1} + \xi_t^{JJ}, \label{eq:wage_JJ}\\
\xi_t^{JJ} \sim N(\mu^{JJ}, \sigma_\xi^2), \qquad\varepsilon_t \sim N(0, \sigma_\varepsilon^2). \qquad\qquad\qquad \nonumber
\end{align}
This earnings process captures a wage premium associated with switching jobs. Investors' tenure is reset to $ten_t=0$ following a job-to-job transition.

\noindent \textbf{Unemployment: $emp_t = U$.} While in the employed state ($E$), an investor may become unemployed with probability $\pi^{EU}(t, ten_t)$ that depends on both their age and tenure at their current job. When investors are unemployed, they receive unemployment benefits equal to $ui_t = ui(\eta_t)$, where $ui(\eta_t)$ is described below. If investors become employed at $t+1$ after being unemployed in period $t$, income at $t+1$ evolves according to
\begin{align}
\ln w_{t+1} = \delta_0 + \delta_1a_{t+1} + \delta_2a_{t+1}^2 + \delta_3a_{t+1}^3 + \eta_{t+1} + \varepsilon_{t+1},\label{eq:wage_U}\\
\eta_{t+1} = \rho \eta_{t} + \xi_{t+1}^{U}, \quad \xi_{t+1}^{U} \sim N(-\mu^{EU}, \sigma_\xi^2), \quad\varepsilon_{t+1} \sim N(0, \sigma_\varepsilon^2). \nonumber
\end{align}
This earnings process captures the persistent wage reduction associated with experiencing unemployment.

\noindent \textbf{Retirement: $emp_t = Ret$.} In period $t=T_w$, all investors retire deterministically. During retirement in periods $t \in [T_w, T-1]$, investors earn public pension benefits denoted by $ss_t$, which are described below.

\subsection{Savings Account}

Investors start with zero assets at $t=0$ and cannot borrow. They can accumulate assets inside a savings account that can be invested in one of two financial assets for the life cycle model and one of four financial assets for the prompts to the LLMs. First, there is a risk-free bond that has a constant gross return of $R_t^B = R_f$ per year. Second, there is a risky asset that corresponds to a diversified stock market index and pays a stochastic i.i.d. gross return of $R_t^D$ per year, where
\begin{equation} \label{eq:R_stock_full}
\ln R_t^D = \ln R_f + \mu_D + u_t^D, \quad u_t^D \sim \mathcal{N}(0, \sigma_D^2).
\end{equation}
Additionally, there are two non-diversified assets, each with its own stochastic gross return, $R^I_t$ and $R^O_t$, corresponding to the individual stock bundle and other risky assets (e.g., crypto, gold, commodities, and collectibles), respectively. Their joint return process is defined relative to the diversified market return:
\begin{equation}
\ln R_t^x = \alpha_x + b_x\bigl(\ln R_t^D - \ln R_f\bigr) + \sigma_{\varepsilon,x}u_{x,t}, \quad u_{x,t} \sim \mathcal{N}(0,1), \quad \forall x \in \{I,O\},
\end{equation}
where $u_{I,t}$ and $u_{O,t}$ are drawn independently of each other and of the market shock $u_t^D$ in (\ref{eq:R_stock_full}). The balance of the savings account, denoted by $L_t$, then evolves according to:
\begin{equation} \label{eq:liquid_balance}
L_{t+1} = \left(L_t + s_t^l\right)\left[1 + (R_{t+1}^\theta - 1)(1-\tau_c)\right], \quad L_0 = 0,
\end{equation}
where $s^l_t$ is the net saving that the investor places in this account, $\tau_c$ is the rate of capital taxation, and $R_t^\theta$ is the investors' portfolio return that depends on their portfolio share in the risky assets.
\begin{equation}
  R_{t+1}^\theta = R_f + \theta^D_t\left(R^D_{t+1} - R_f\right) + \theta^I_t\left(R^I_{t+1} - R_f\right) + \theta^O_t\left(R^O_{t+1} - R_f\right), \quad \theta^D_t + \theta^I_t+\theta^O_t \in [0,1].
\end{equation}
When solving the life cycle model, we impose $\theta^I_t = \theta^O_t = 0$ because, by construction, neither non-diversified asset can improve the Sharpe ratio of a portfolio that combines the bond and diversified stock index (see \Cref{app:calibration}). While this is an approximation because the CAPM would not arise in equilibrium in our model, we make this assumption for parsimony when solving the model. In the LLM simulation, however, all four assets are available, and the LLM is free to allocate wealth to the non-diversified assets.

\subsection{Government}

\noindent \textbf{Unemployment benefits.} Investors receive an unemployment benefit of $ui(\eta_t)$ when their employment ends. This benefit depends on the labor productivity, $\eta_t$, from the last period in which the agent was employed.

\noindent \textbf{Retirement benefits.} After retirement, investors receive Social Security benefits, denoted by $ss_t = ss(ae_{T_w-1})$. $ae_{T_w-1}$ is the investor's average lifetime earnings at the time of retirement. Let \(ae_t\) denote the average of wages up to and including period \(t\). Initialize \(ae_0 = w_0\). Then
\begin{equation}\label{eq:average_earnings}
ae_{t+1} =
\begin{cases}
\frac{w_{t+1} + (t+1) \cdot ae_t}{t + 2}, & \text{ if } t < T_w-1, \\
ae_{T_w-1} & \text{ else.}
\end{cases}
\end{equation}

\noindent \textbf{Taxation.} Investors face a nonlinear income tax schedule $tax_i(\cdot)$, which depends on their taxable income. For employed individuals or those in job transitions, their taxable income is their wages. For unemployed or retired individuals, their taxable income is the benefits they receive from the government.

\subsection{Summary of Investors' Problem}

Investors face a dynamic optimization problem with seven state variables: $a_t$ = age, $\eta_t$ = labor productivity, $\varepsilon_t$ = transitory income shock, $emp_t$ = employment status, $ten_t$ = tenure, $ae_t$ = average lifetime income, and $L_t$ = savings. Denote the vector of these state variables as $\mathbf{x}_t$. Investors have three controls: $c_t$ = consumption, $\theta^D_t$ = portfolio share, and $s^l_t$ = savings. In choosing these controls, we restrict investors from borrowing and engaging in any margin trading (i.e., short-selling or taking leveraged positions):
\begin{equation} \label{eq:constraints}
L_t \geq 0, \quad \theta^D_t \in [0,1].
\end{equation}
When simulating LLM advice, the only difference is that there are five controls, including the two non-diversified asset shares $\theta^I_t$ and $\theta^O_t$.

\subsection{Calibration}
\label{app:calibration}

We calibrate our model parameters following \cite{Choukhmane2026} where possible. We provide a brief overview here and refer the reader to that paper for a more detailed discussion.

\noindent \textbf{Demographics.} We set the length of one period in the model to one year and set $a_0=22$, $T_w=43$, and $T=68$, such that workers enter working life at age 22, retire at 65, and live their final year of life at 89. For each age, we calibrate mortality risk to match the 2015 U.S.\ Social Security Actuarial Life Tables. We use the equivalence scale estimated in \cite{Lusardi2017} to capture changes in household composition over the life cycle.

\noindent \textbf{Labor income process.} We use data from the Survey of Income and Program Participation (SIPP) to estimate parameters of the labor income process and transition probabilities at the annual frequency. This income process has several components. First, we estimate an earnings process for workers staying in the same job, corresponding to (\ref{eq:wage_E}), which contains a deterministic and stochastic component.\footnote{Relative to \cite{Choukhmane2026}, we also estimate the variance of the transitory shock as part of this estimation.} Second, we use data on employment transitions from SIPP to estimate the median salary increase following a job-to-job transition, $\mu^{JJ}$, and the median salary decrease when workers transition back to employment after an unemployment spell, $-\mu^{EU}$. Third, we use SIPP microdata to estimate the three transition probabilities between the three working-age labor market states. Finally, we set the initial unemployment rate equal to 22\%, which is the share of age-22 individuals in SIPP who are unemployed.

\noindent \textbf{Tax and benefit system.} Investors' tax liability, $tax_i(\cdot)$, is calculated according to the 2025 U.S.\ federal income tax schedule for single filers who claim only the standard deduction. We calculate Social Security benefits according to the 2025 formula with a Supplemental Security Income program floor. Unemployment benefits are computed with a replacement rate of 40\%. We set the capital return tax rate, $\tau_c$, to 21\%.

\noindent \textbf{Bond and diversified stock returns.} We set the net risk-free rate to be constant at 2\%, a standard value in life cycle models \citep{Gomes2020}. We set the equity premium to be 6.4\%, which is equal to the average CPI-adjusted return on the CRSP Value-Weighted Index between 1925 and 2006 minus our 2\% risk-free rate. We set the volatility of log stock returns to 20\%, which matches that of the CRSP Value-Weighted Index. We assume that asset returns are uncorrelated with shocks to labor income and employment transition probabilities.

\noindent \textbf{Non-diversified asset returns.} Both non-diversified assets are calibrated to have the same underlying return process, but independent idiosyncratic shock realizations, so they represent distinct risky positions. We calibrate this process such that neither non-diversified asset can improve the Sharpe ratio of a portfolio that combines a bond and a diversified stock market index, implying that a CAPM investor would optimally hold zero portfolio shares in both assets. Formally, we start by taking the average arithmetic return and standard deviation, $\mu_x$ and $\sigma_x$, for individual stocks from \cite{Bessembinder2018}, which gives $\mu_x = 0.1474$ and $\sigma_x = 0.819$. Assuming the CAPM holds and denoting $\mu_m$ as the arithmetic mean diversified market return, we can compute the non-diversified asset beta using:
\begin{equation}
\beta_{x}^{\ast}
= \frac{(1+\mu_{x})-R_f}{(1+\mu_{m})-R_f}.
\end{equation}
Next, note that log return parameters for the non-diversified asset are given by:
\begin{align}
\sigma_{x,\log}
&= \sqrt{\log \left(1+\left(\frac{\sigma_{x}}{1+\mu_{x}}\right)^2\right)},\\
\mu_{x,\log}
&= \log(1+\mu_{x}) - \tfrac12\,\sigma_{x,\log}^2.
\end{align}
We assume $\beta_x^{\ast} = b_x$, and then compute the idiosyncratic log volatility using
\begin{equation}
\sigma_{\varepsilon,x}
= \sqrt{\sigma_{x,\log}^2 - b_x^2\sigma_D^2}
\end{equation}
and the log intercept as
\begin{equation}
\alpha_x
= \mu_{x,\log} - b_x\mu_D,
\end{equation}
where $\mu_D$ is the mean log excess return of the diversified asset in (\ref{eq:R_stock_full}).\footnote{We assume $b_x=\beta_x^{\ast}$ for simplicity. In reality, the arithmetic beta implied by a log-loading $b_x$ is $ \frac{\text{cov}(R_t^x, R_t^D)}{\text{var}(R_t^D)} = \frac{1+\mu_x}{1+\mu_m}\frac{\exp(b_x\sigma_D^2)-1}{\exp(\sigma_D^2)-1}$, which implies $b_x
= \frac{1}{\sigma_D^2}\log\left(1+\beta_x^{\ast}\frac{1+\mu_m}{1+\mu_x}\left(\exp(\sigma_D^2)-1\right)\right)$. However, given our calibration, these two are quite similar.} This then implies that, given a simulated market gross return $R_t^D$ and an independent draw $u_{x,t}\sim\mathcal{N}(0,1)$, we can construct each non-diversified asset's return as
\begin{equation}
R_t^x
= \exp\Bigl(
\alpha_x
+ b_x\bigl(\log R_t^D-\log R_f\bigr)
+ \sigma_{\varepsilon,x}\,u_{x,t}
\Bigr).
\end{equation}

\clearpage

\section{Textual Analysis Details}\label{app:textual_analysis}

This appendix provides methodological details for the dictionary-based textual analysis described in \Cref{sec:textual_analysis}.

\noindent \textbf{Text preprocessing.}
Each free-text response is processed through a three-step pipeline. First, finance-specific terms are normalized using pattern-based rewriting (e.g., ``t-bills'' becomes ``tbill''). Second, the text is tokenized and lemmatized using spaCy's \texttt{en\_core\_\hspace{0pt}web\_sm} model, which splits text into individual words and reduces them to root forms (e.g., ``worried'' becomes ``worry''). We use the small spaCy model because testing across all three model sizes (\texttt{sm}, \texttt{md}, \texttt{lg}) showed similar results. Third, we add a set of manual root reductions specific to financial words that the spaCy model misses (e.g., ``retirement'' to ``retire''). Importantly, finance-relevant strings---including account types (e.g., 401k), fund tickers (e.g., VTI), and acronyms (e.g., ETF)---are preserved as-is without lemmatization, but plural forms are normalized (e.g., ``401ks'' to ``401k,'' ``ETFs'' to ``ETF''). The output is one sequence of root-form words per respondent per prompt.

\noindent \textbf{Dictionary construction.}
Based on reading the prompts, we manually define 27 topic dictionaries covering the following categories: Debt, Savings, Retirement, Housing, Investment Strategy, Income, Employment Security, Budgeting, Education, Family, Gender, Financial Hardship, Financial Anxiety, Risk Preference, Medical, Long-Term Planning, Tax, Lifestyle Goals, Macroeconomic, Investment Assets, Providers, Products, Liquidity, Discipline, Diversification, Insurance, and Inheritance. Each category contains a list of keywords, which are listed in \autoref{tab:dictionaries}. Importantly, categories are not mutually exclusive: keywords can appear in multiple dictionaries (e.g., ``mortgage'' appears in both Debt and Housing). Additionally, keywords are processed through the same pipeline as respondent text, ensuring consistent matching between the prompts and categories. Multi-word keywords (e.g., ``credit card,'' ``high yield savings account'') are matched by checking for consecutive words in the respondent's processed text.

\begin{figure}[!htbp]
	\caption{Word Clouds by Individual Prompt}
	\vspace{-0.2in}
	\label{fig:wordclouds_by_prompt}
	\begin{center}
	\makebox[\linewidth][c]{%
	\begin{tabular}{@{}c@{\hskip 0.02\linewidth}c@{\hskip 0.02\linewidth}c@{}}
		\textbf{Financial Situation} & \textbf{Spending Advice} & \textbf{Investment Advice} \\[2pt]
		\includegraphics[width=0.32\linewidth]{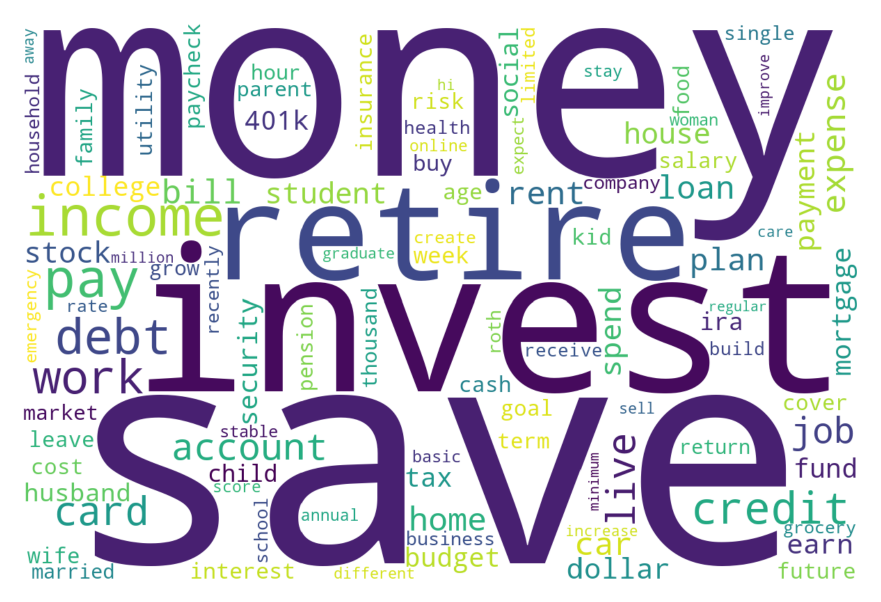} &
		\includegraphics[width=0.32\linewidth]{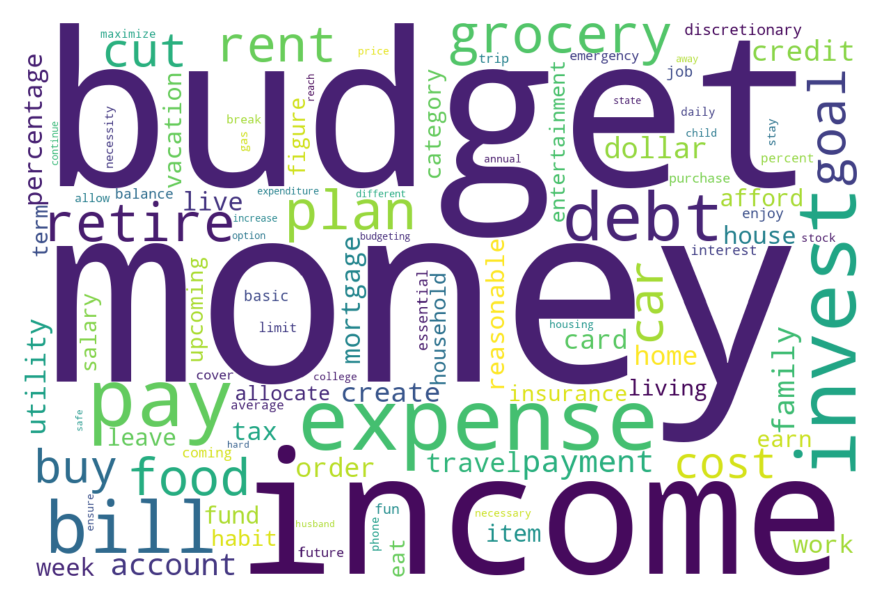} &
		\includegraphics[width=0.32\linewidth]{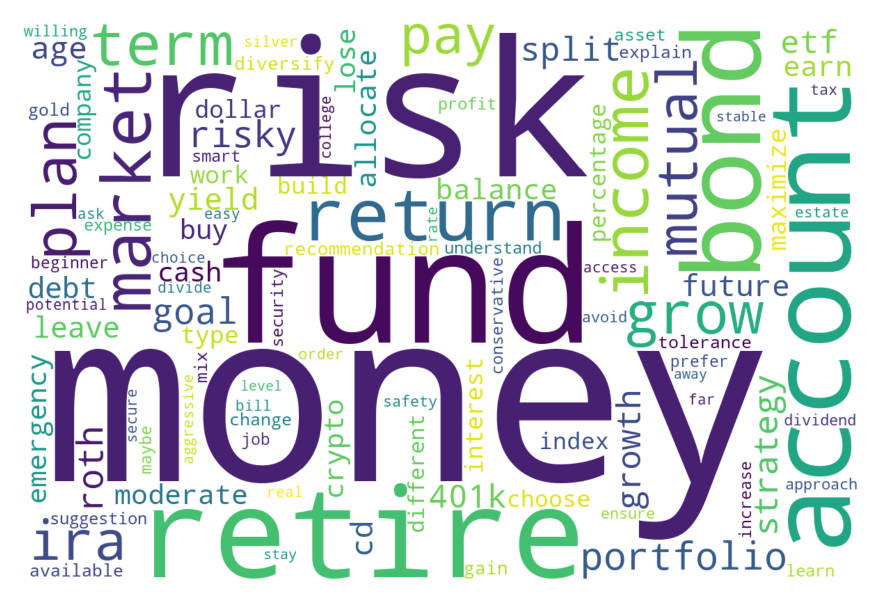}
	\end{tabular}%
	}
	\end{center}
	\scriptsize{\emph{Notes:} This figure reproduces the analysis in \autoref{fig:wordclouds}, separating out the three prompts that respondents write. \hyperlink{back:textual_topics}{Go back.}}
\end{figure}

\noindent \textbf{Stopwords.}
When making word clouds, a list of stopwords---a union of scikit-learn's default English stopwords, custom generic terms, and survey-instrument words (e.g., ``financial,'' ``advice,'' ``situation'')---is removed. Prompt-specific tautological terms are additionally removed from the prompt word clouds: ``spend'' is excluded from the spending prompt and ``invest'' from the investment prompt. The combined prompt word cloud (\autoref{fig:wordclouds}) excludes the union of all prompt-specific stopwords, while the individual prompt word clouds (\autoref{fig:wordclouds_by_prompt}) exclude only the stopwords specific to each prompt.



\clearpage
{\begin{landscape}
    \input{table_dictionary_keywords}
    \scriptsize{\hyperlink{back:dictionaries}{Go back.}}
\end{landscape}}

\clearpage

\section{Additional Details on Simulated Method of Moments Estimation}\label{app:smm}
To estimate the preference parameters $\beta$ and $\gamma$ for each version of the LLM output, we use simulated method of moments on a discrete grid. Our estimation uses a set of 324 moments. We use the 25th, 50th, and 75th percentiles of the wealth-to-income ratio from ages 23 to 64, and the 25th, 50th, and 75th percentiles of equity share from ages 23 to 88. Let $\mathcal{W}_{i,t}$ and $\theta_{i,t}$ be the wealth-to-income ratio and equity share, respectively, for agent $i$ at time $t$. The former is defined as $\mathcal{W}_{i,t}\equiv \frac{L_{i,t}}{w_{i,t} - tax_i(w_{i,t})}$. 

For the wealth-to-income ratio moments, we omit the first year, when wealth is zero by construction. We also restrict these moments to working ages, which isolates saving accumulation before retirement: LLM withdrawal advice in retirement bunches at fixed rules, such as 4\% of wealth, that the two-parameter life cycle model is not designed to reproduce. For the equity share moments, we omit the first and last periods of life.

We estimate the preferences by solving the life cycle model on a 201-by-121 grid of $\beta$ and $\gamma$ values ranging from 0.8 to 1.2 and 2 to 14, respectively. In the $\beta$ dimension, we use a step size of 0.002 while in the $\gamma$ dimension, the step size is 0.1. After solving for the life cycle moments for every point on this grid, we then calculate an error term between the targeted moments and every point on the grid as follows:
\begin{equation}
    Z(\beta,\gamma) = X_{\text{LC}}(\beta,\gamma) - X_{\text{LLM}}, \qquad \text{SMM}_{\text{error}}(\beta,\gamma) = Z(\beta,\gamma)' \cdot W \cdot Z(\beta, \gamma)
\end{equation}
In the above equation, $X_{\text{LLM}}$ is the vector of LLM moments and $X_{\text{LC}}(\beta,\gamma)$ is the corresponding vector of life cycle moments when solving the life cycle model using $\beta$ and $\gamma$ as the inputted preference parameters. $W$ is the weighting matrix. 
For our estimation, we use the identity matrix as the weighting matrix as it has better small sample properties than the optimal weighting matrix. 

\clearpage

\section{Additional Tables and Figures}\label{app:results}


\begin{figure}[!htbp]
\caption{Survey Questions for Eliciting LLM Prompts}
\label{fig:survey_questions}
\begin{center}
\makebox[\linewidth][c]{%
\scalebox{0.8}{%
\begin{minipage}{1.2\linewidth}

\begin{surveycard}{Prompt 1: Describe your situation}
\small
\textbf{Write a description of your situation}, including whatever details about yourself you think would help the AI tool give you useful financial advice. \textbf{Assume the AI tool does not have any information about you unless you provide it here.}
\surveytextbox[4.2cm]
\end{surveycard}

\vspace{8pt}

\begin{surveycard}{Prompt 2: Ask for spending advice}
\small
In your own words, write a prompt \textbf{asking an AI tool for advice on how much to spend} over the coming year.
\par\medskip
Assume the AI tool \underline{already has the information you provided} in your last response.
\surveytextbox[4.2cm]
\end{surveycard}

\vspace{8pt}

\begin{surveycard}{Prompt 3: Ask for investment advice}
\small
In your own words, write a prompt \textbf{asking an AI tool for advice on how to invest} your savings between stocks and safer options.
\par\medskip
Assume the AI tool \underline{already has the information you provided} in your last response.
\surveytextbox[4.2cm]
\end{surveycard}

\end{minipage}%
}}
\end{center}
\scriptsize{\emph{Notes:} This figure displays the three survey questions used to collect respondent-written LLM prompts. \hyperlink{back:survey_prompts}{Go back.}}
\end{figure}

\clearpage

\begin{figure}[!htbp]
\caption{Personal Finance Topics Discussed with AI Tools}
\vspace{-0.2in}
\label{fig:ai_topics}
\begin{center}
\makebox[\linewidth][c]{
    \resizebox{0.7\linewidth}{!}{
        \includegraphics[]{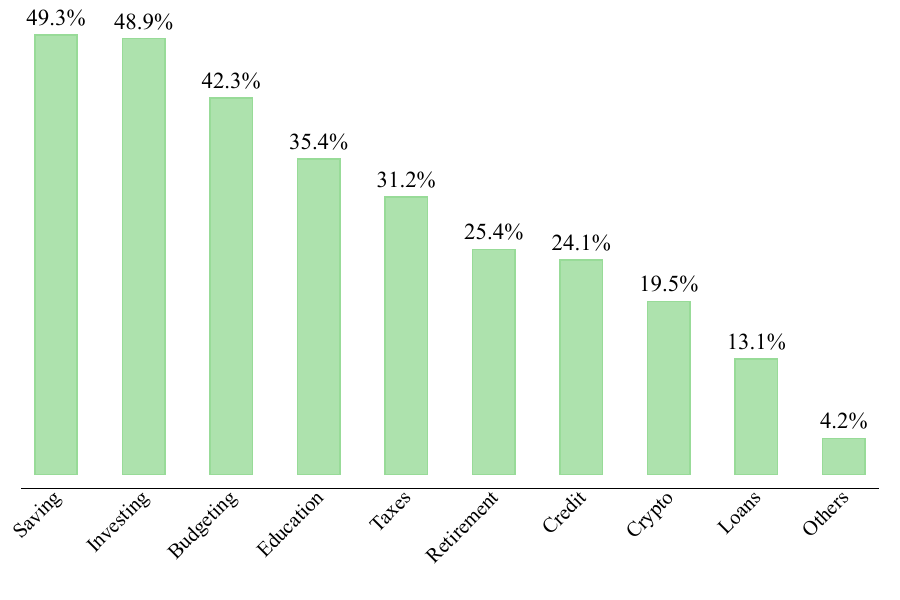}
}}
\end{center}
\vspace{-0.2in}
\scriptsize{\emph{Notes:} This figure shows the personal finance topics that respondents report discussing with AI tools, among those who have previously used AI for financial advice or information. Respondents could select multiple topics. \hyperlink{back:survey_prompts}{Go back.}}
\end{figure}


\clearpage
\input{table_survey_demographics}

\clearpage
\begin{table}[!htbp]
	\caption{Prompt Summary Statistics}
	\label{tab:summary_prompt}
	\footnotesize
	\makebox[\linewidth][c]{\resizebox{\linewidth}{!}{\input{table_prompt_summary_statistics_data}}}
	\vskip 0.1in
		\scriptsize{\emph{Notes:} This table reports descriptive statistics for the respondent-written individual prompts. The first row combines the three prompt questions for each respondent; the next three rows report statistics separately for the financial situation, spending advice, and investment advice prompts. The remaining rows pool the three prompts and report statistics by demographic subgroup. ``Share Numbers'' is the share of responses containing any digit. ``Share Dollars'' is the share containing ``\$'' or the word ``dollar.'' Income groups are low ($<$\$50k), middle (\$50k--\$100k), and high ($>$\$100k); income is not reported for 36 respondents. Financial literacy groups are low (0--3 correct), middle (4 correct), and high (5 correct out of 5 quiz questions). AI Use indicates whether the respondent has previously used AI for financial advice. \hyperlink{back:summary_prompt}{Go back.}}
\end{table}

\clearpage

\begin{figure}[!htbp]
\caption{AI Model Usage Among Survey Respondents}
\vspace{-0.2in}
\label{fig:ai_model_usage}
\begin{center}
\makebox[\linewidth][c]{
    \resizebox{\linewidth}{!}{
        \includegraphics[]{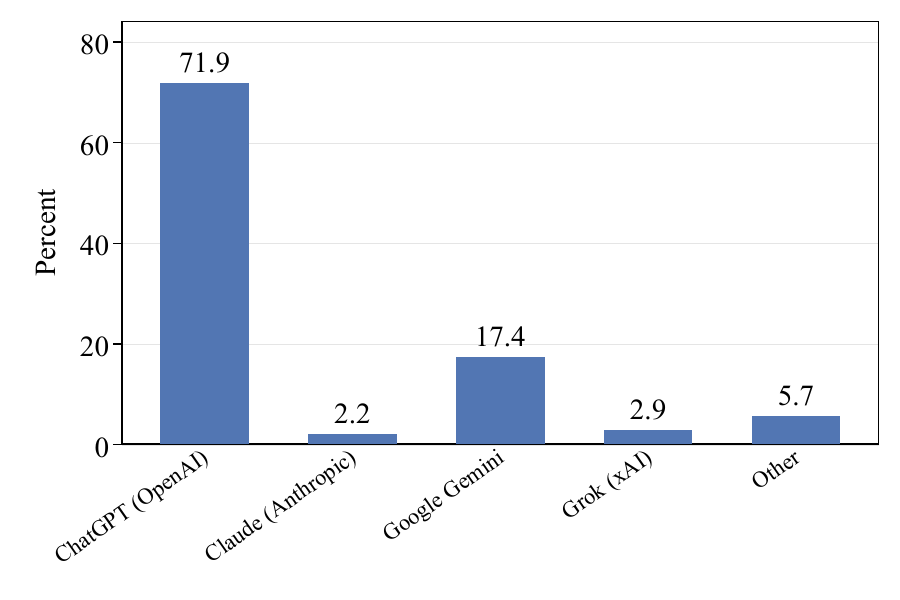}
}}
\end{center}
\scriptsize{\emph{Notes:} This figure reports the AI model used by respondents who have previously used AI for financial advice. Bars show the share of these respondents reporting each model. \hyperlink{back:model_selection}{Go back.}}
\end{figure}


\clearpage

\begin{figure}[!htbp]
	\caption{All Dictionary Categories by Prompt}
	\vspace{-0.2in}
	\label{fig:dict_all}
	\begin{center}
		\makebox[\linewidth][c]{
			\resizebox{\linewidth}{!}{
				\includegraphics[]{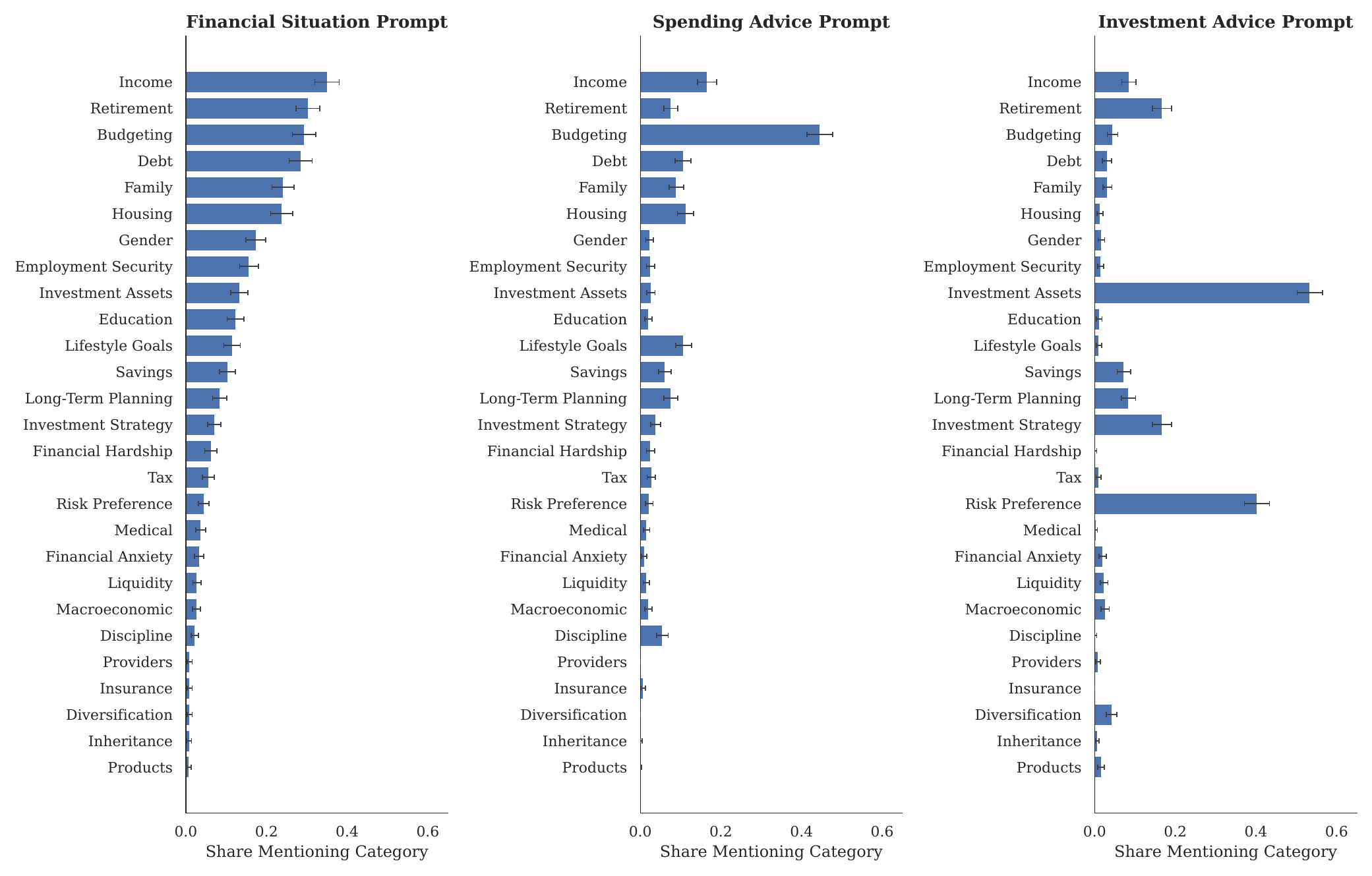}
		}}
	\end{center}
	\scriptsize{\emph{Notes:} This figure reports dictionary category mention rates separately for each of the three prompt questions. Bars show the share of respondents whose prompt contains at least one keyword from the category; capped lines denote 95\% confidence intervals. Categories are ordered by their mention rate in the financial situation prompt. \hyperlink{back:textual_topics}{Go back.}}
\end{figure}

\clearpage

\clearpage
\begin{figure}[!htbp]
	\caption{All Dictionary Categories: Prompts vs.\ Advice}
	\vspace{-0.2in}
	\label{fig:dict_combined_vs_advice}
	\begin{center}
		\makebox[\linewidth][c]{
			\resizebox{0.9\linewidth}{!}{
				\includegraphics[]{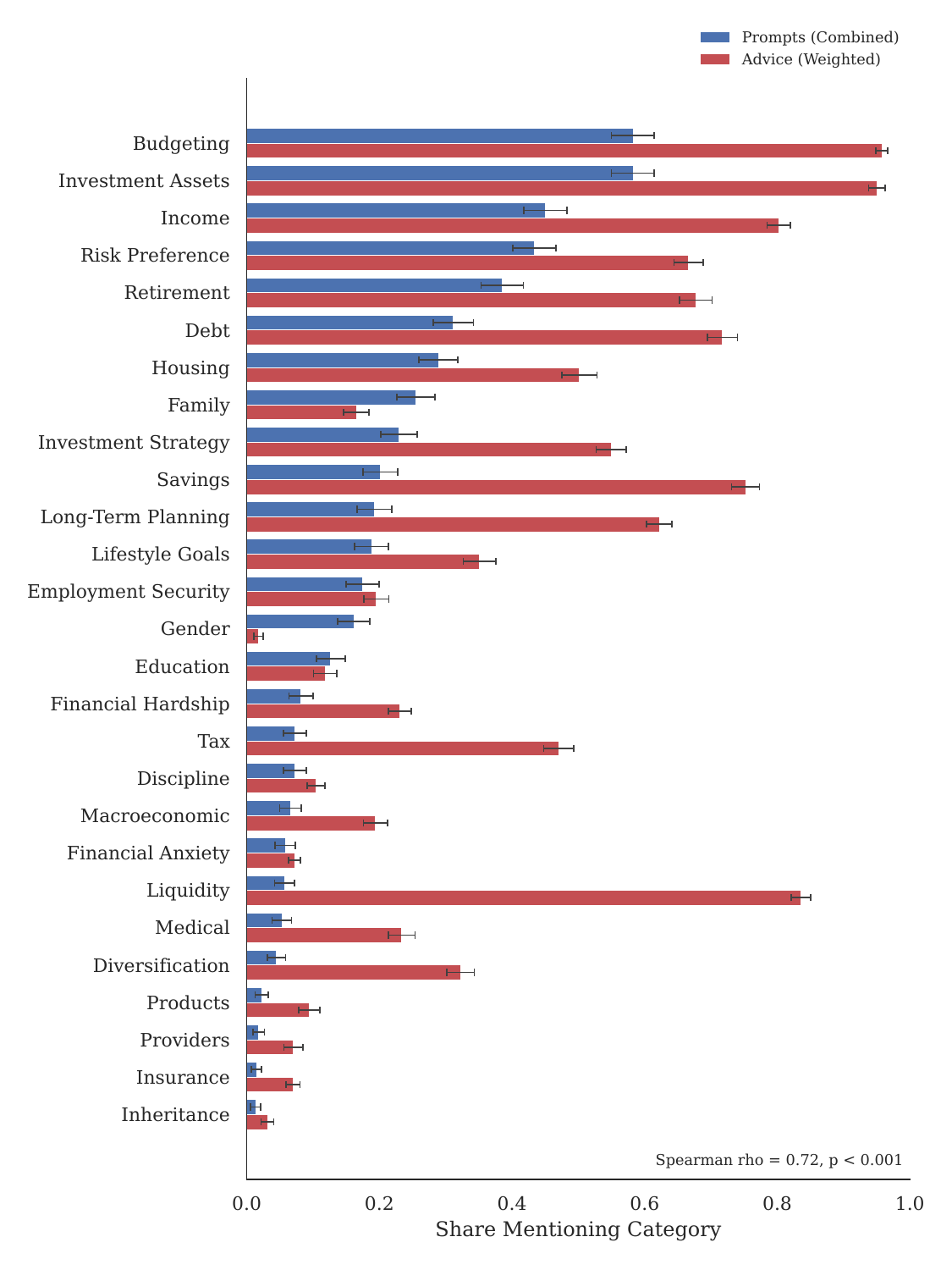}
		}}
	\end{center}
	\scriptsize{\emph{Notes:} This figure compares dictionary category mention rates in respondents' combined prompts with the corresponding LLM advice. Blue bars show the share of respondents whose combined prompt text contains at least one keyword from the category. Red bars show the share of advice responses mentioning the category, with each respondent's prompt weighted equally. Categories are ordered by their mention rate in the combined prompts. Capped lines denote 95\% confidence intervals. The Spearman rank correlation between prompt and advice mention rates is reported in the bottom-right corner. \hyperlink{back:prompt_advice_topics}{Go back.}}
\end{figure}


\clearpage

\begin{figure}[!htbp]
    \caption{Life Cycle Profiles of LLM-Recommended Consumption and Equity Shares: Alternative Models}
    \vspace{-0.2in}
    \label{fig:lifecycle_alternative_models}
    \begin{center}
        \makebox[\linewidth][c]{\resizebox{\linewidth}{!}{\includegraphics[]{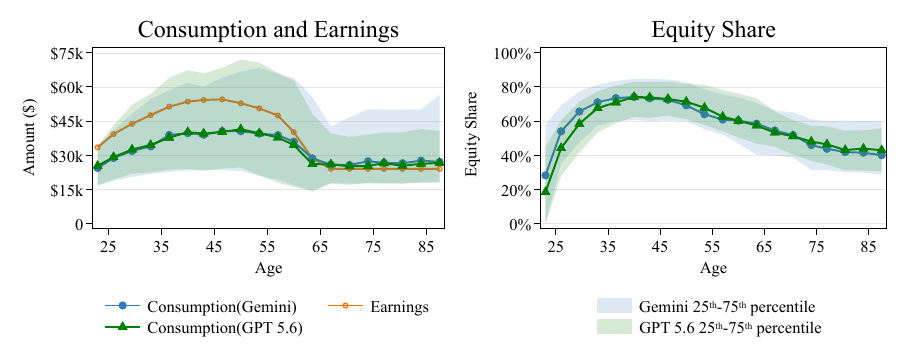}}}
    \end{center}
    \vspace{-0.2in}
    \singlespacing
    \scriptsize{\emph{Notes:} This figure plots the life cycle profiles for individuals following the LLM's recommended choices using \Gemini and \GPTNew with the survey prompts, as described in \Cref{sec:simulation} and \Cref{sec:robustness}. Ages (22--89) are grouped into 20 equally sized bins. The left panel shows consumption and post-tax earnings, and the right panel shows the equity share. Markers denote median values within each bin, and shaded areas represent the interquartile range (25$^{\text{th}}$--75$^{\text{th}}$ percentile). All values are in 2025 dollars. \hyperlink{back:lifecycle_profiles}{Go back.}}
\end{figure}

\clearpage

\begin{table}[!htbp]
    \caption{Summary Statistics from LLM and Life Cycle Model Simulations}
    \label{tab:summary_stats}
    \footnotesize
    \makebox[\linewidth][c]{\resizebox{0.95\linewidth}{!}{
        \input{table_life_cycle_summary_statistics_data}
    }}
    \vskip 0.1in
    \scriptsize{\emph{Notes:} This table reports mean values for key simulated inputs and outcomes by age group. The Total column reports averages over ages 22--64. Panel A presents exogenous inputs that are identical across both LLM and life cycle model simulations. Post-tax earnings correspond to wage and unemployment insurance earnings; employment rate is the share of individuals employed; and tenure is the average number of years at the current employer. Panels B and C report endogenous outcomes from the LLM and life cycle model, respectively, separately by employment status. Net saving rate is defined as post-tax earnings minus consumption, divided by post-tax earnings. All monetary values are in 2025 dollars. \hyperlink{back:summary_stats}{Go back.}}
\end{table}


\clearpage

\begin{figure}[!htbp]
    \caption{Non-Diversified Asset Participation and Portfolio Shares}
    \vspace{-0.2in}
    \label{fig:nondiv_lifecycle}
    \begin{center}
        \makebox[\linewidth][c]{\resizebox{\linewidth}{!}{\includegraphics[]{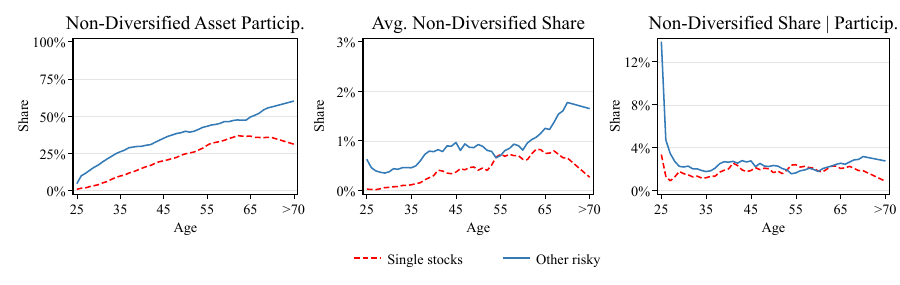}}}
    \end{center}
    \vspace{-0.2in}
    \singlespacing
    \scriptsize{\emph{Notes:} This figure plots the life cycle profiles of non-diversified asset holdings---individual stocks and other risky assets (e.g., crypto, gold, commodities, collectibles)---for individuals following the LLM's advice using the survey prompts, as described in \Cref{sec:simulation}. The left panel shows unconditional participation rates by age; the middle panel shows the average portfolio share of financial wealth allocated to each asset class; and the right panel shows the average portfolio share conditional on participation. Ages (22--89) are grouped into 20 equally sized bins. \hyperlink{back:nondiversified_assets}{Go back.}}
\end{figure}

\clearpage

\begin{figure}[!htbp]
    \caption{Observed Behavior vs.\ LLM-Recommended Behavior: Alternative Models}
    \vspace{-0.2in}
    \label{fig:status_quo_alternative_models}
    \begin{center}
        \makebox[\linewidth][c]{\resizebox{\linewidth}{!}{\includegraphics[]{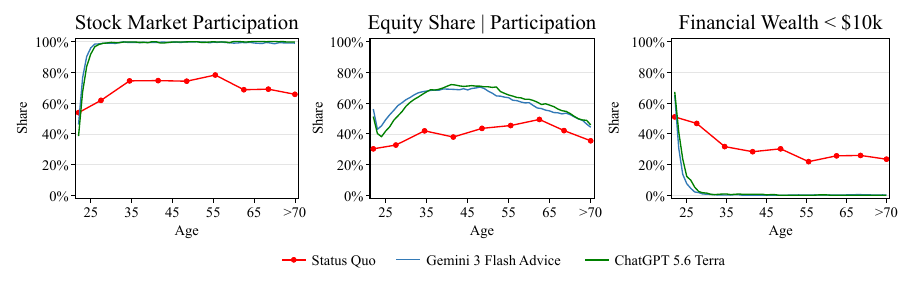}}}
    \end{center}
    \vspace{-0.2in}
    \scriptsize{\emph{Notes:} This figure compares survey respondents' observed financial behavior with the behavior implied by following advice from \Gemini and \GPTNew in the full life cycle simulation described in \Cref{sec:simulation}. The three panels show the average stock market participation rate (left), average equity share conditional on participation (middle), and the share of individuals with financial wealth below \$10,000 (right). Ages are grouped into eight 6-year bins and one last bin that captures every observation age 70 and above. All values are in 2025 dollars. \hyperlink{back:observed_behavior_alternative_models}{Go back.}}
\end{figure}

\clearpage

\begin{figure}[!htbp]
    \caption{Estimated Life Cycle Model Fit on Targeted Moments}
    \label{fig:model_fit}
    \parbox{\textwidth}{
        \small{\textbf{Panel A}: Survey Prompts, \GPT} \vskip 0.01in
        \makebox[\linewidth][c]{
            \resizebox{\linewidth}{!}{
                \includegraphics[]{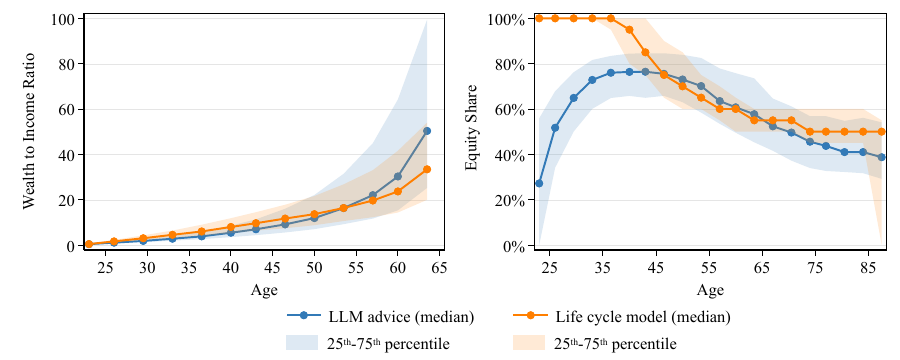}
        }}
    }
    \vskip 0.05in
    \parbox{\textwidth}{
        \small{\textbf{Panel B}: Academic Prompts, \GPT} \vskip 0.01in
        \makebox[\linewidth][c]{
            \resizebox{\linewidth}{!}{
                \includegraphics[]{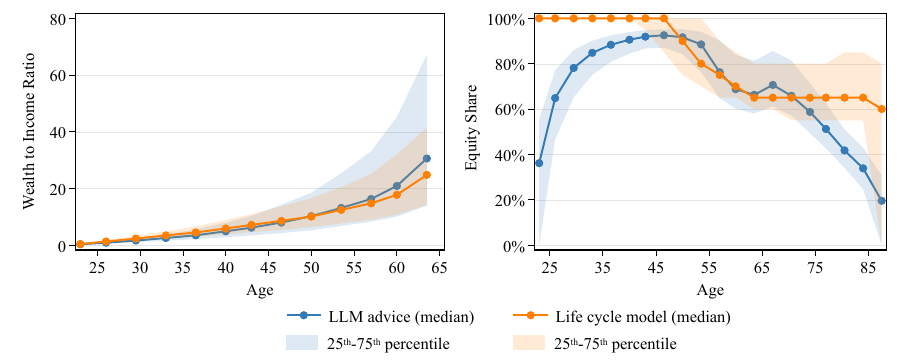}
        }}
    }
    \vskip 0.05in
    \parbox{\textwidth}{
        \small{\textbf{Panel C}: Survey Prompts, \Gemini} \vskip 0.01in
        \makebox[\linewidth][c]{
            \resizebox{\linewidth}{!}{
                \includegraphics[]{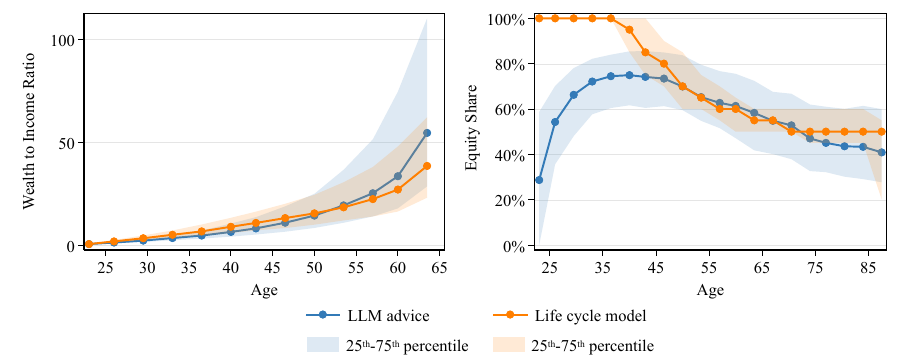}
        }}
    }
\end{figure}

\clearpage

\begin{figure}[!htbp]
    \ContinuedFloat
    \caption[]{Estimated Life Cycle Model Fit on Targeted Moments (Continued)}
    \label{fig:model_fit_alternative_models}
    \parbox{\textwidth}{
        \small{\textbf{Panel D}: Survey Prompts, \GPTNew} \vskip 0.01in
        \makebox[\linewidth][c]{
            \resizebox{\linewidth}{!}{
                \includegraphics[]{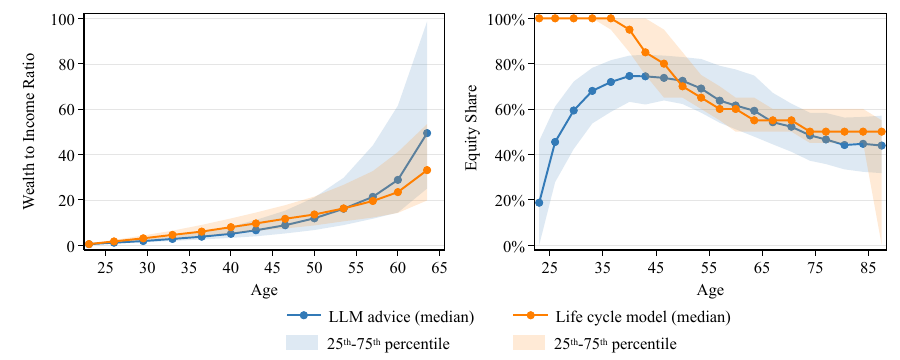}
        }}
    }
    \scriptsize{\emph{Notes:} This figure shows the fit of the estimated life cycle model, using the preference parameters reported in \autoref{tab:smms}, to the targeted SMM moments. Panels A and B use \GPT with the survey and academic prompts, respectively. Panels C and D use the survey prompts with \Gemini and \GPTNew, respectively. \hyperlink{back:model_fit}{Go back.}}
\end{figure}

\clearpage

\begin{figure}[!htbp]
    \caption{Responsiveness of LLM Advice to Shocks: Academic Prompt}
    \vspace{-0.2in}
    \label{fig:shocks_professor}
    \begin{center}
        \makebox[\linewidth][c]{\resizebox{\linewidth}{!}{\includegraphics[]{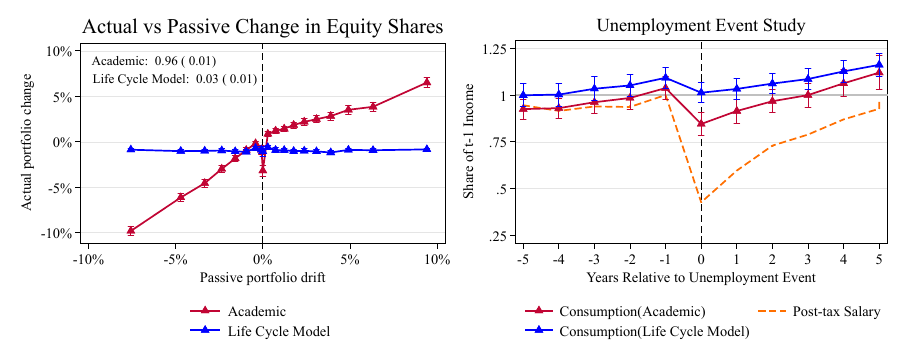}}}
    \end{center}
    \vspace{-0.2in}
\scriptsize{\emph{Notes:} This figure compares the responsiveness to shocks between the LLM advice and the life cycle model using the academic prompt described in \Cref{sec:benchmarks}. Red triangles denote the LLM advice, blue triangles denote the life cycle model results, and the orange dashed line indicates the post-tax earnings path implied by the life cycle model described in \Cref{sec:life_cycle_model}. The left panel presents a binscatter plot of average actual portfolio change against passive portfolio drift. Passive portfolio drifts are grouped into 20 equally sized bins. The right panel plots the unemployment event study, constructed in the same way as the right panel of \autoref{fig:smoothing}. \hyperlink{back:consumption_smoothing}{Go back.}}
\end{figure}

\clearpage

\begin{figure}[!htbp]
    \caption{Baseline Two-Step Pipeline vs.\ Direct Quantitative Pipeline}
    \label{fig:pipeline_comp}
    \parbox{\textwidth}{
        \small{\textbf{Panel A}: Consumption and Equity Share Life Cycle Profiles} \vskip 0.01in
        \makebox[\linewidth][c]{
            \resizebox{\linewidth}{!}{
                \includegraphics[]{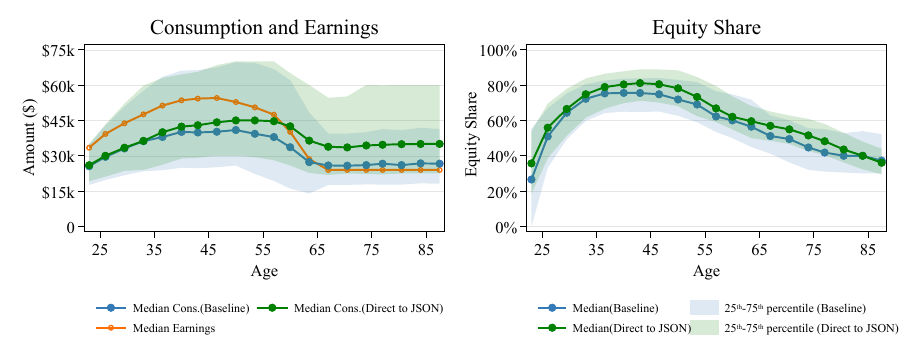}
        }}
    }
    \vskip 0.05in
    \parbox{\textwidth}{
        \small{\textbf{Panel B}: Saving and Withdrawal Heuristics} \vskip 0.01in
        \makebox[\linewidth][c]{
            \resizebox{\linewidth}{!}{
                \includegraphics[]{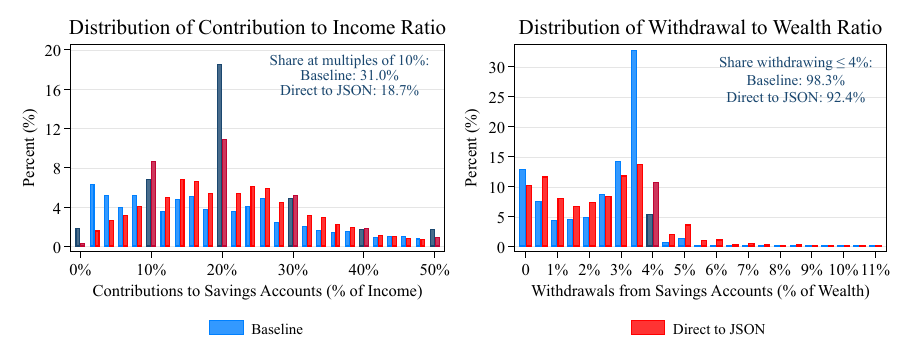}
        }}
    }
    \vskip 0.05in
    \parbox{\textwidth}{
        \small{\textbf{Panel C}: Shock Response} \vskip 0.01in
        \makebox[\linewidth][c]{
            \resizebox{\linewidth}{!}{
                \includegraphics[]{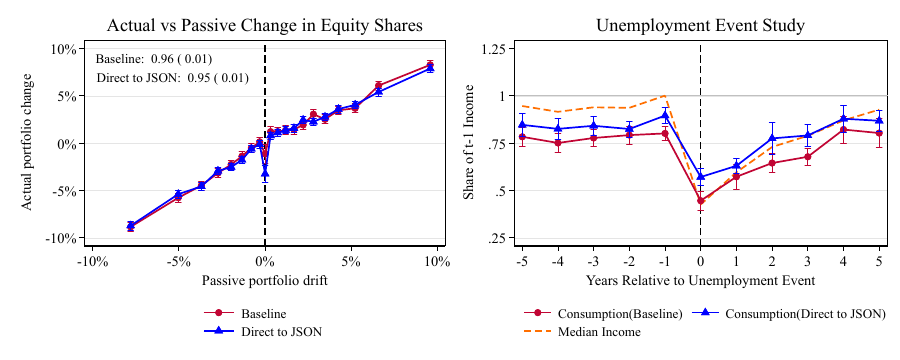}
        }}
    }
    \scriptsize{\emph{Notes:} This figure compares our main results under the baseline two-step pipeline with those from the alternative direct quantitative pipeline described in \Cref{sec:robustness}. \hyperlink{back:consumption_smoothing}{Go back.}}
\end{figure}

\clearpage

\begin{figure}[!htbp]
    \caption{Life Cycle Profiles of LLM-Recommended Consumption and Equity Shares: Academic Prompt}
    \vspace{-0.2in}
    \label{fig:lifecycle_professor}
    \begin{center}
        \makebox[\linewidth][c]{\resizebox{\linewidth}{!}{\includegraphics[]{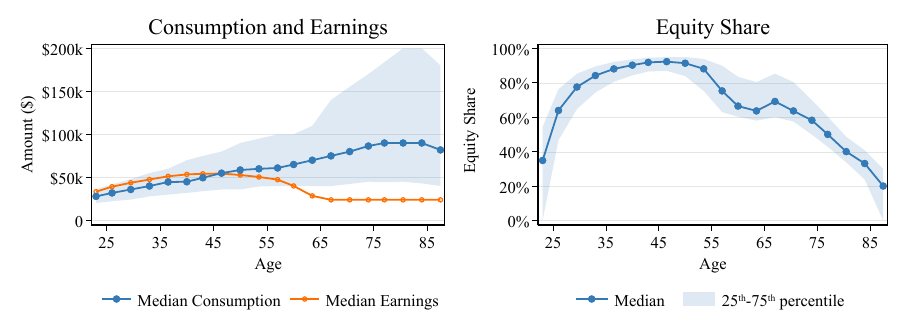}}}
    \end{center}
    \vspace{-0.2in}
    \singlespacing
    \scriptsize{\emph{Notes:} This figure plots the life cycle profiles for individuals following the LLM's recommended choices using the academic prompt, as described in \Cref{sec:benchmarks}. Ages (22--89) are grouped into 20 equally sized bins. The left panel shows consumption and post-tax earnings, and the right panel shows the equity share. Dots denote median values within each bin, and shaded areas represent the interquartile range (25$^{\text{th}}$--75$^{\text{th}}$ percentile). All values are in 2025 dollars. \hyperlink{back:structured_prompt_outputs}{Go back.}}
\end{figure}

\clearpage

\begin{figure}[!htbp]
    \caption{Observed Behavior vs.\ LLM-Recommended Behavior: Academic Prompt}
    \vspace{-0.2in}
    \label{fig:status_quo_professor}
    \begin{center}
        \makebox[\linewidth][c]{\resizebox{\linewidth}{!}{\includegraphics[]{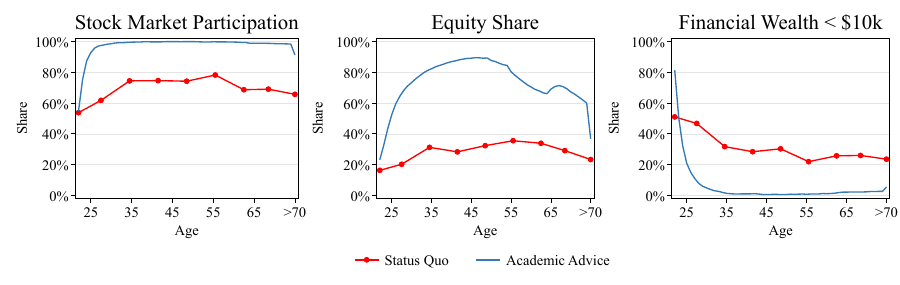}}}
    \end{center}
    \vspace{-0.2in}
    \scriptsize{\emph{Notes:} This figure compares survey respondents' observed financial behavior with the LLM's recommendations using the academic prompt described in \Cref{sec:benchmarks}. Observed behavior, as reported by individuals in our Prolific survey, is shown in red; recommendations using the academic prompt are shown in blue. Ages are grouped into eight 6-year bins and one last bin that captures every observation age 70 and above. All values are in 2025 dollars. \hyperlink{back:structured_prompt_outputs}{Go back.}}
\end{figure}

\clearpage

\begin{figure}[!htbp]
    \caption{Saving and Withdrawal Heuristics in LLM Advice: Alternative Models}
    \label{fig:heuristics_alternative_models}
    \parbox{\textwidth}{
        \small{\textbf{Panel A}: \Gemini} \vskip 0.01in
        \makebox[\linewidth][c]{
            \resizebox{\linewidth}{!}{
                \includegraphics[]{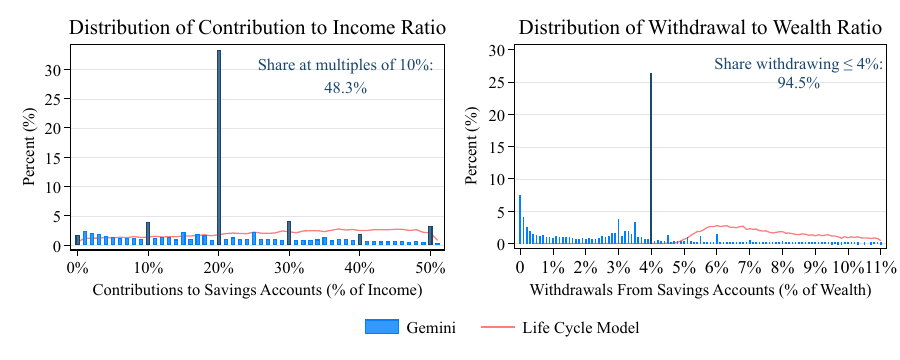}
        }}
    }
    \vskip 0.05in
    \parbox{\textwidth}{
        \small{\textbf{Panel B}: \GPTNew} \vskip 0.01in
        \makebox[\linewidth][c]{
            \resizebox{\linewidth}{!}{
                \includegraphics[]{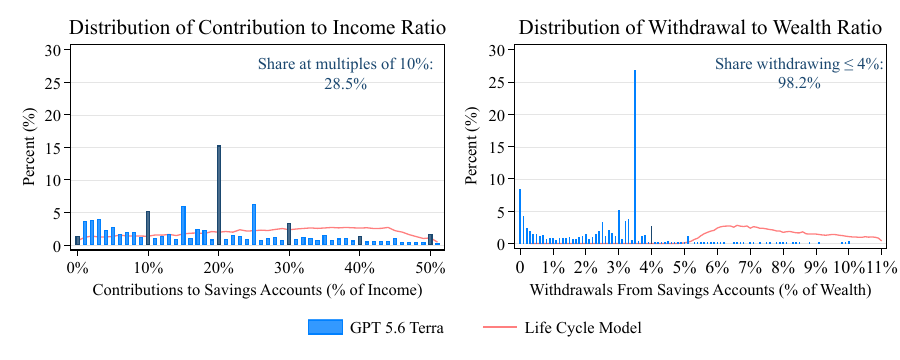}
        }}
    }
    \singlespacing
    \scriptsize{\emph{Notes:} This figure plots the distributions of saving contributions as a share of income (left) and retirement withdrawals as a share of wealth (right) implied by advice from \Gemini (Panel A) and \GPTNew (Panel B). The blue histograms show the choices implied by the LLM advice, dark blue bars highlight mass at common heuristic values, and red lines show the corresponding distributions generated by the estimated life cycle model described in \Cref{sec:life_cycle_model}. The LLM advice is generated using the survey prompts as described in \Cref{sec:simulation} and \Cref{sec:robustness}. \hyperlink{back:alternative_models}{Go back.}}
\end{figure}

\clearpage

\begin{figure}[!htbp]
    \caption{Responsiveness of LLM Advice to Shocks: Alternative Models}
    \label{fig:shocks_alternative_models}
    \parbox{\textwidth}{
        \small{\textbf{Panel A}: \Gemini} \vskip 0.01in
        \makebox[\linewidth][c]{
            \resizebox{\linewidth}{!}{
                \includegraphics[]{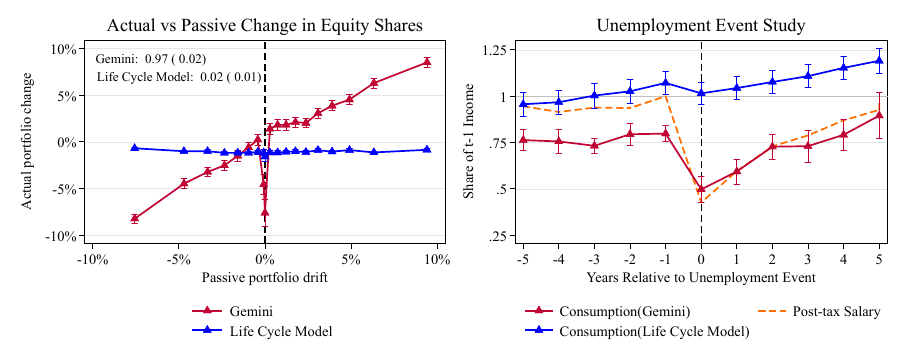}
        }}
    }
    \vskip 0.05in
    \parbox{\textwidth}{
        \small{\textbf{Panel B}: \GPTNew} \vskip 0.01in
        \makebox[\linewidth][c]{
            \resizebox{\linewidth}{!}{
                \includegraphics[]{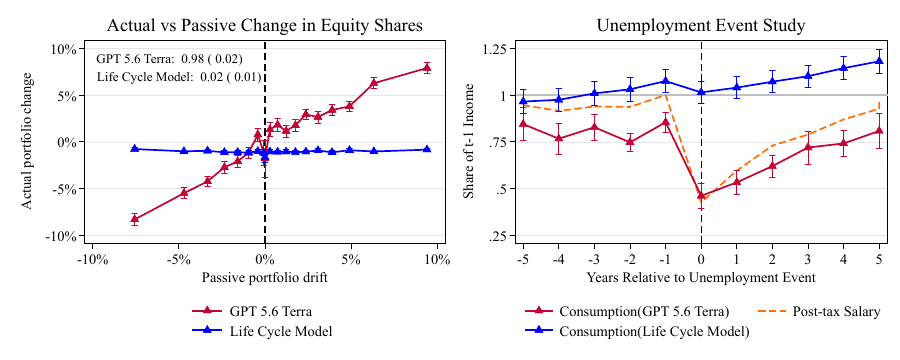}
        }}
    }
    \scriptsize{\emph{Notes:} This figure compares the responsiveness to shocks of advice from \Gemini (Panel A) and \GPTNew (Panel B) with the estimated life cycle model. Within each panel, the left plot shows average actual portfolio changes against passive portfolio drift, grouping observations into 20 equal-sized bins of passive drift. The right plot shows the unemployment event study constructed in the same way as the right panel of \autoref{fig:smoothing}. Red lines denote the LLM advice, blue lines denote the life cycle model, and the orange dashed line denotes the post-tax earnings path implied by the life cycle model described in \Cref{sec:life_cycle_model}. \hyperlink{back:alternative_models}{Go back.}}
\end{figure}

\clearpage

\begin{figure}[!htbp]
    \caption{Variation in Consumption and Equity Share Across LLM Queries}
    \vspace{-0.2in}
    \label{fig:LLM_variation}
    \begin{center}
        \makebox[\linewidth][c]{\resizebox{\linewidth}{!}{\includegraphics[]{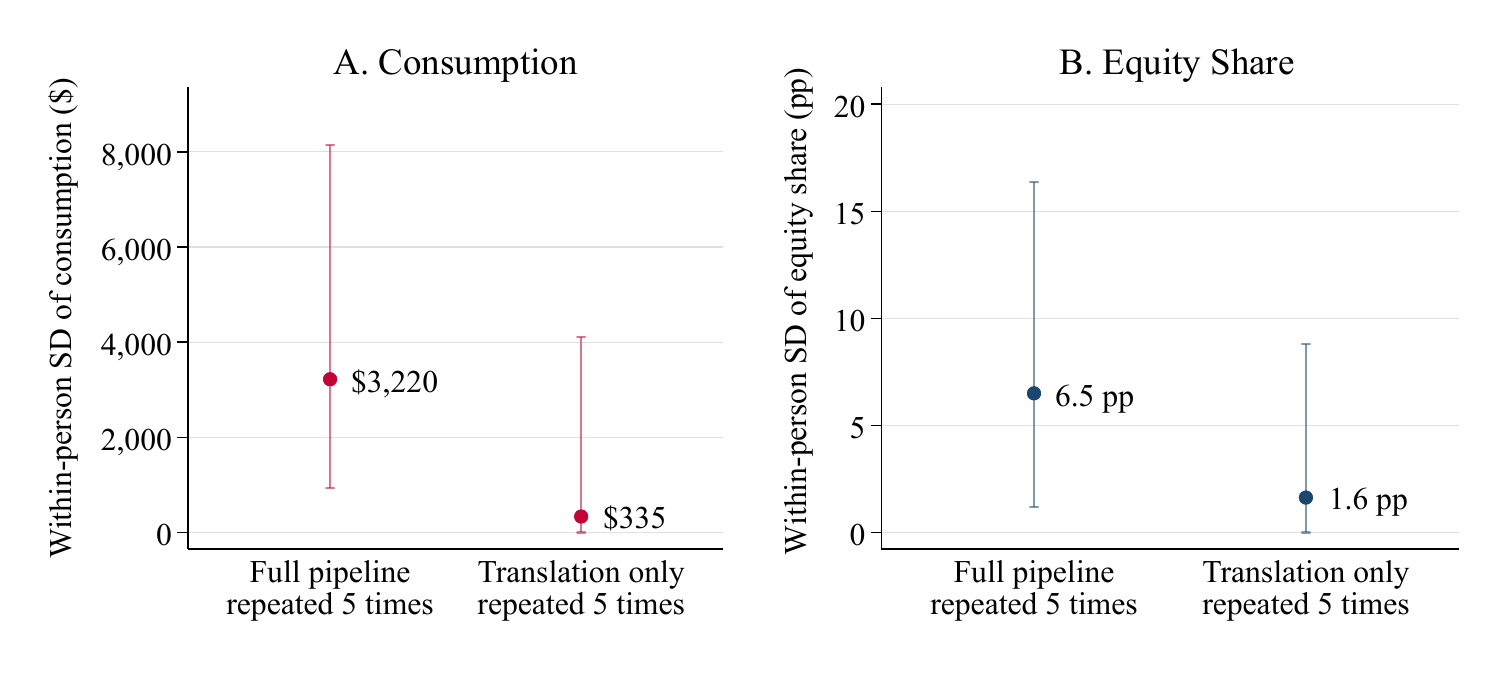}}}
    \end{center}
    \vspace{-0.2in}
    \singlespacing
    \scriptsize{\emph{Notes:} This figure measures the variability of the one-shot recommendations described in \Cref{sec:fact1}. We conduct two exercises for each of the 952 respondents. In the ``Full pipeline'' exercise, we hold the respondent's prompt fixed and repeat the entire procedure five times: each repetition generates new textual advice and then translates that advice into quantitative recommendations. In the ``Translation only'' exercise, we hold both the prompt and the textual advice fixed and repeat only the translation into quantitative recommendations five times. For each individual and exercise, we calculate the standard deviation across the five recommendations. Dots report the median of these individual-level standard deviations, and whiskers report the 25th and 75th percentiles. Panel A reports variation in recommended annual consumption in dollars. Panel B reports variation in the recommended equity share in percentage points. \hyperlink{back:llm_query_variation}{Go back.}}
\end{figure}

\clearpage

\begin{figure}[!htbp]
    \caption{Differences in Prompts and Advice by Financial Literacy}
    \label{fig:finlit_wordclouds}
    \parbox{\textwidth}{
        \small{\textbf{Panel A}: Words Overrepresented in Prompts by Financial Literacy} \vskip 0.05in
        \makebox[\linewidth][c]{%
        \begin{tabular}{@{}p{0.48\linewidth}@{\hskip 0.04\linewidth}p{0.48\linewidth}@{}}
            \centering\small{\textbf{High Financial Literacy}} & \centering\small{\textbf{Low Financial Literacy}} \tabularnewline
        \end{tabular}%
        }
        \vskip 0.05in
        \makebox[\linewidth][c]{
            \resizebox{\linewidth}{!}{
                \includegraphics[]{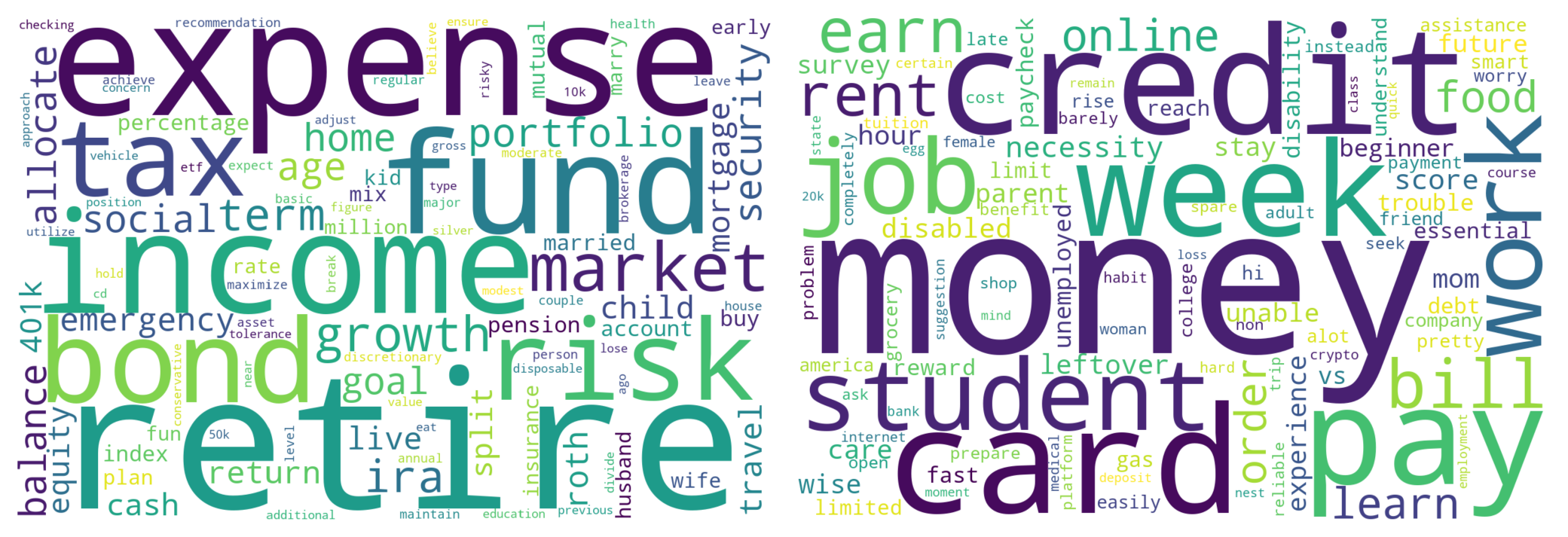}
        }}
    }
    \vskip 0.15in
    \parbox{\textwidth}{
        \small{\textbf{Panel B}: Words Overrepresented in Advice by Financial Literacy} \vskip 0.05in
        \makebox[\linewidth][c]{%
        \begin{tabular}{@{}p{0.48\linewidth}@{\hskip 0.04\linewidth}p{0.48\linewidth}@{}}
            \centering\small{\textbf{High Financial Literacy}} & \centering\small{\textbf{Low Financial Literacy}} \tabularnewline
        \end{tabular}%
        }
        \vskip 0.05in
        \makebox[\linewidth][c]{
            \resizebox{\linewidth}{!}{
                \includegraphics[]{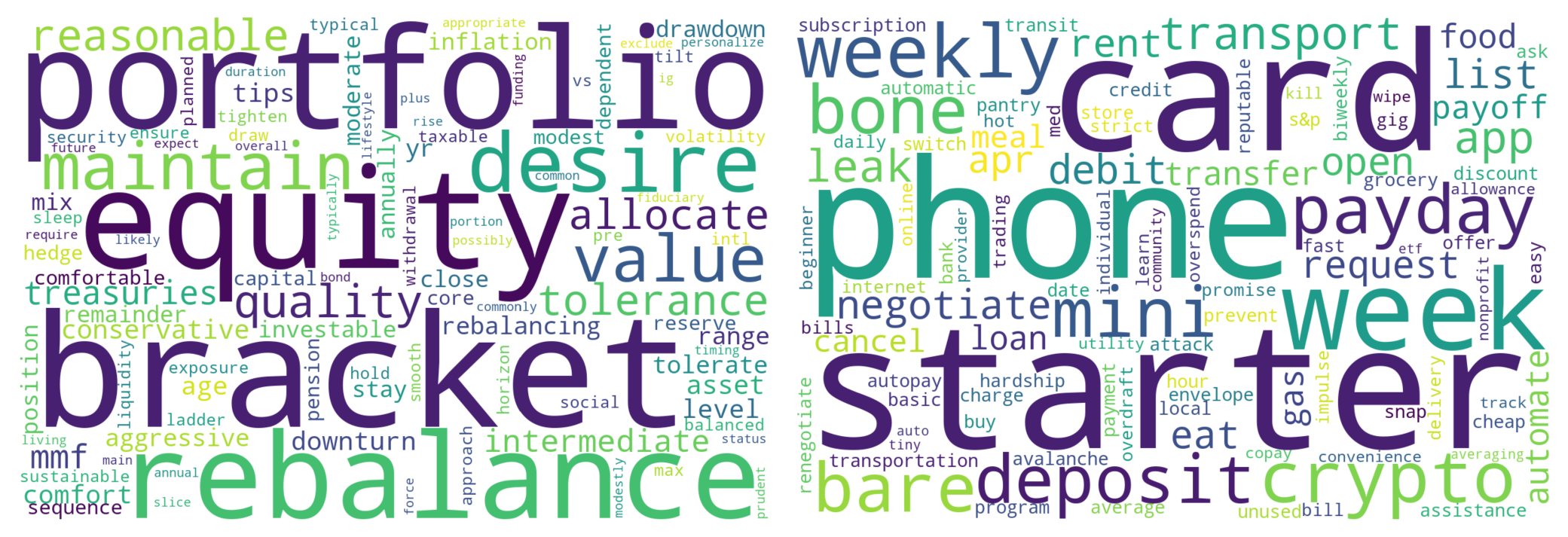}
        }}
    }
    \scriptsize{\emph{Notes:} This figure compares differences in respondent-written prompts and corresponding LLM advice by financial literacy. Panel A pools the three prompt questions and shows words overrepresented in prompts written by respondents with high financial literacy (left; 5 of 5 Big Five questions correct) and with low financial literacy (right; 0--4 correct). Panel B shows words overrepresented in the corresponding LLM advice for respondents with high financial literacy (left) and with low financial literacy (right). Word size is proportional to the difference in word frequency between groups. Common stopwords and words that mechanically reflect the survey questions are excluded. \hyperlink{back:financial_literacy_heterogeneity}{Go back.}}
\end{figure}

\clearpage

\begin{figure}[!htbp]
    \caption{Example Prompts by Financial Literacy}
    \vspace{-0.2in}
    \label{fig:finlit_examples}
    \begin{center}
        \makebox[\linewidth][c]{\resizebox{\linewidth}{!}{\includegraphics[]{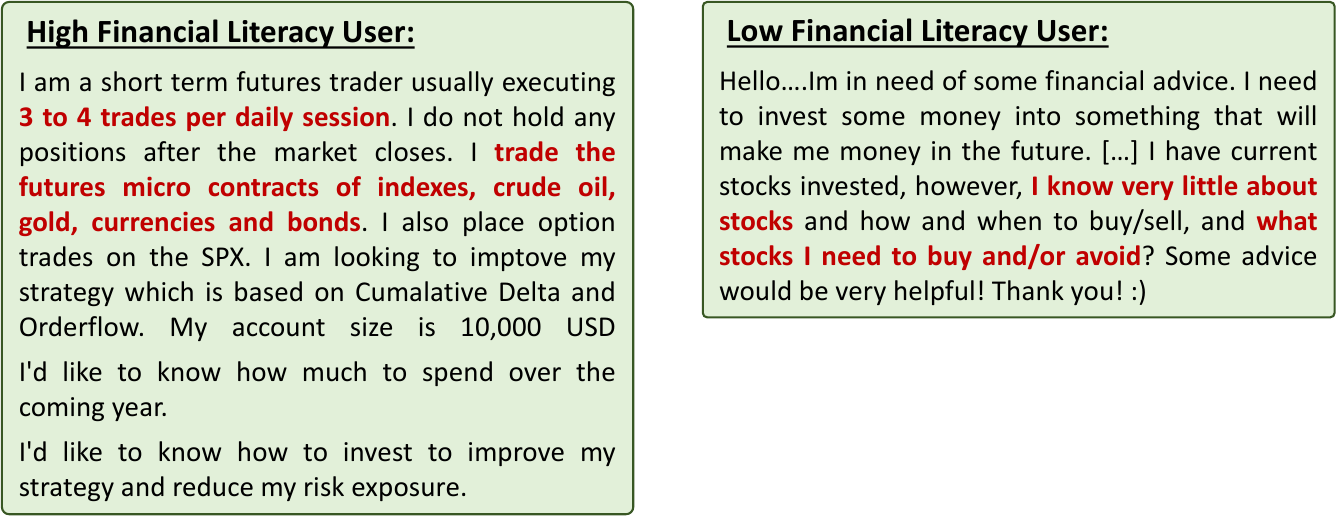}}}
    \end{center}
    \vspace{-0.2in}
    \scriptsize{\emph{Notes:} This figure presents examples of respondent-written prompts from individuals with high and low financial literacy, as measured by the Big Five financial literacy questions in the survey. Respondent text is reproduced verbatim; spelling and grammar errors are preserved. Red text highlights phrases related to financial knowledge and investment sophistication. \hyperlink{back:financial_literacy_heterogeneity}{Go back.}}
\end{figure}

\clearpage

\begin{figure}[!htbp]
\caption{Prior AI Use for Financial Advice Among Survey Respondents}
\vspace{-0.2in}
\label{fig:ai_incidence}
\begin{center}
\makebox[\linewidth][c]{
    \resizebox{0.7\linewidth}{!}{
        \includegraphics[]{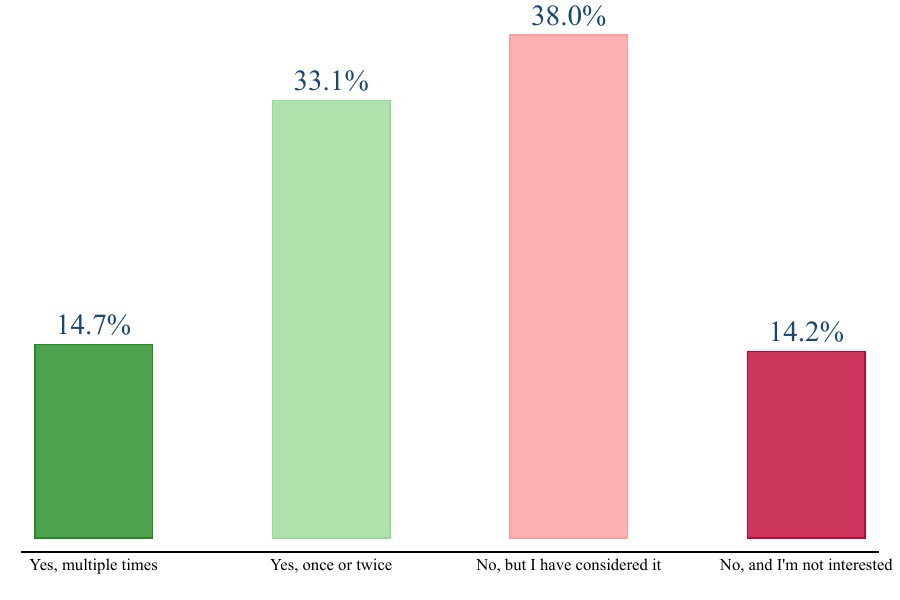}
}}
\end{center}
\vspace{-0.2in}
\scriptsize{\emph{Notes:} This figure reports responses to the survey question, ``In the past 3 months, have you used an AI tool to get financial advice or information?'' in the Prolific survey. Bars show the share of respondents selecting each response option. \hyperlink{back:prior_ai_heterogeneity}{Go back.}}
\end{figure}

\clearpage

\begin{figure}[!htbp]
    \caption{Example Prompts by Prior AI Experience}
    \vspace{-0.2in}
    \label{fig:aiuse_examples}
    \begin{center}
        \makebox[\linewidth][c]{\resizebox{\linewidth}{!}{\includegraphics[]{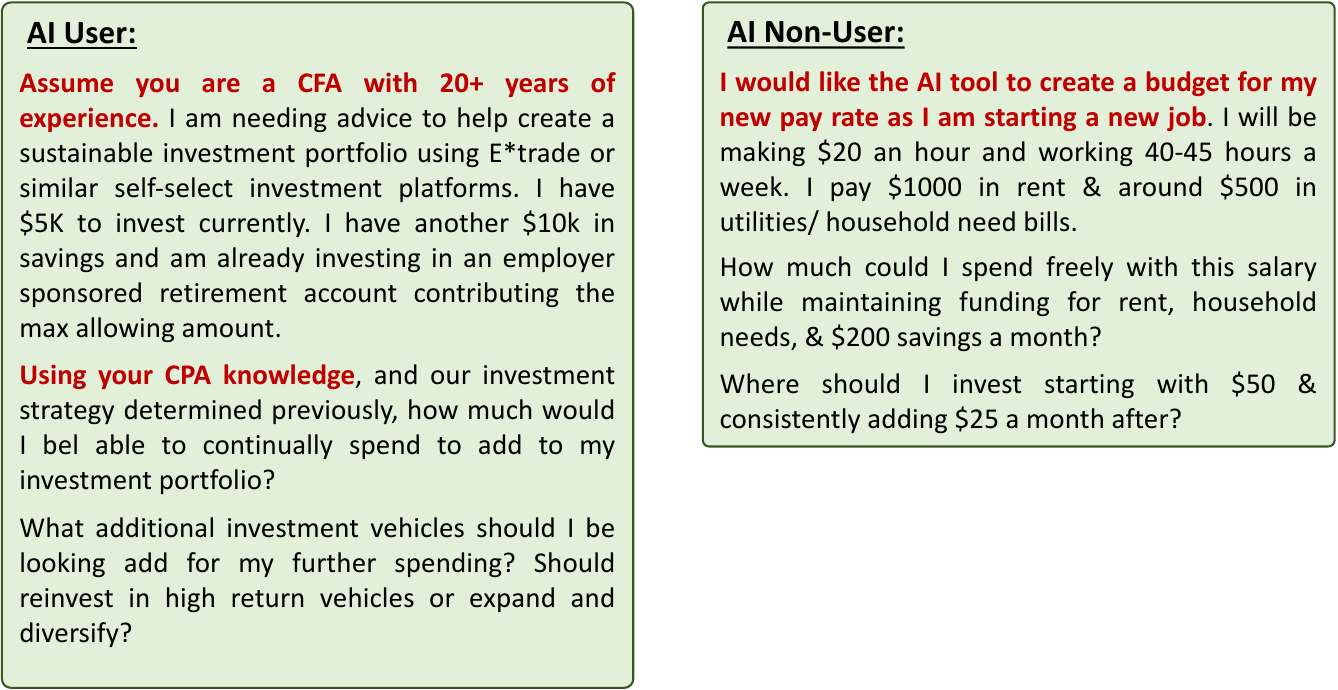}}}
    \end{center}
    \vspace{-0.2in}
    \scriptsize{\emph{Notes:} This figure presents examples of respondent-written prompts from individuals with and without prior AI use for financial advice. Respondent text is reproduced verbatim; spelling and grammar errors are preserved. Red text highlights phrases that directly reference AI or expert-advisor framing. \hyperlink{back:prior_ai_heterogeneity}{Go back.}}
\end{figure}

\clearpage

\begin{figure}[!htbp]
    \caption{Differences in Prompts and Advice by Prior AI Experience}
    \label{fig:aiuse_wordclouds}
    \parbox{\textwidth}{
        \small{\textbf{Panel A}: Words Overrepresented in Prompts by Prior AI Experience} \vskip 0.05in
        \makebox[\linewidth][c]{%
        \begin{tabular}{@{}p{0.48\linewidth}@{\hskip 0.04\linewidth}p{0.48\linewidth}@{}}
            \centering\small{\textbf{Prior AI Use}} & \centering\small{\textbf{No Prior AI Use}} \tabularnewline
        \end{tabular}%
        }
        \vskip 0.05in
        \makebox[\linewidth][c]{
            \resizebox{\linewidth}{!}{
                \includegraphics[]{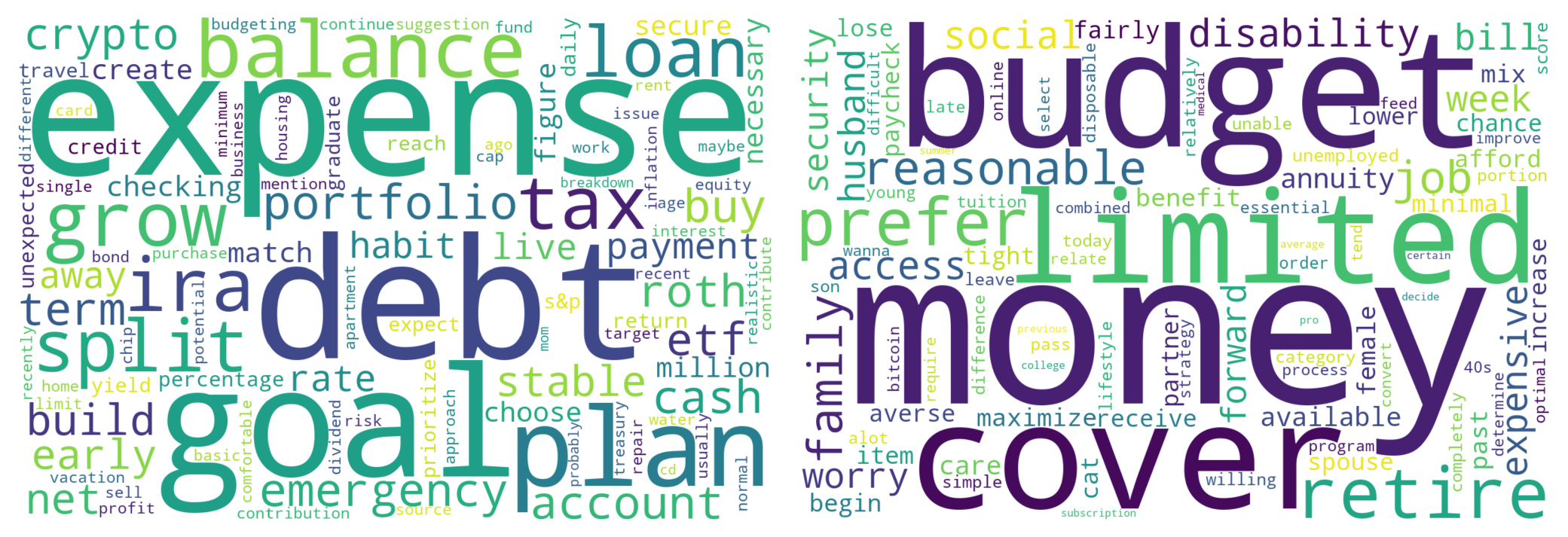}
        }}
    }
    \vskip 0.15in
    \parbox{\textwidth}{
        \small{\textbf{Panel B}: Words Overrepresented in Advice by Prior AI Experience} \vskip 0.05in
        \makebox[\linewidth][c]{%
        \begin{tabular}{@{}p{0.48\linewidth}@{\hskip 0.04\linewidth}p{0.48\linewidth}@{}}
            \centering\small{\textbf{Prior AI Use}} & \centering\small{\textbf{No Prior AI Use}} \tabularnewline
        \end{tabular}%
        }
        \vskip 0.05in
        \makebox[\linewidth][c]{
            \resizebox{\linewidth}{!}{
                \includegraphics[]{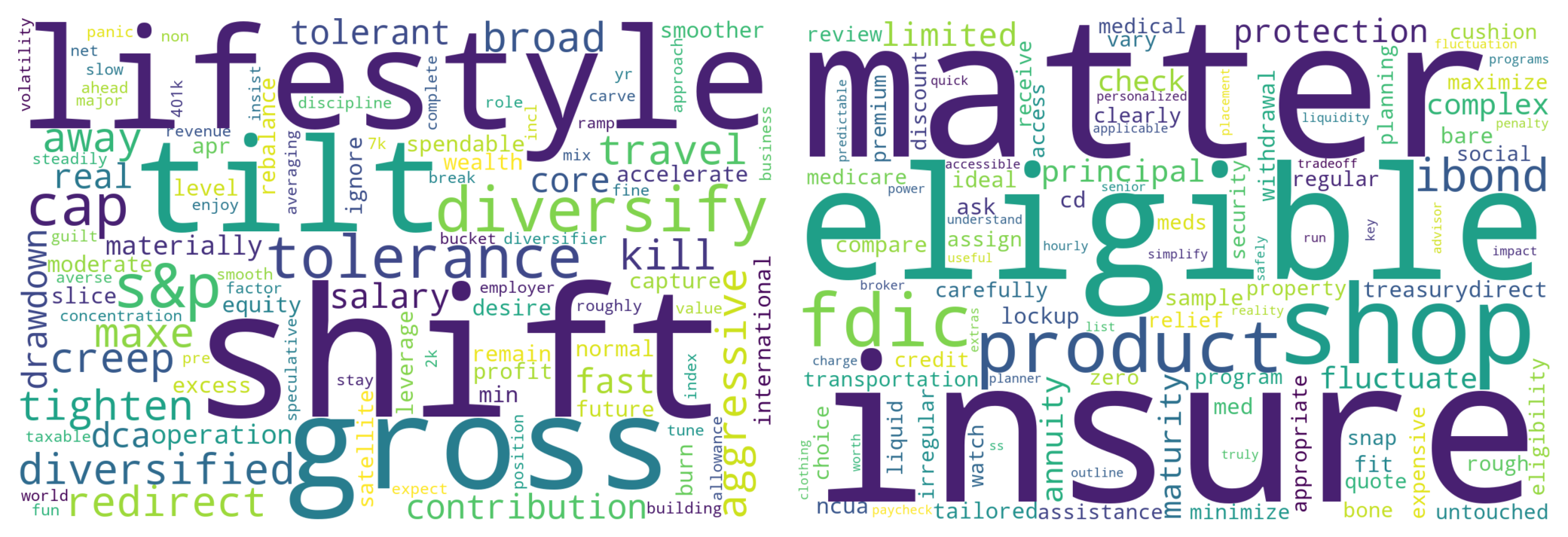}
        }}
    }
    \scriptsize{\emph{Notes:} This figure compares differences in respondent-written prompts and corresponding LLM advice by prior AI use for financial advice. Panel A pools the three prompt questions and shows words overrepresented in prompts written by respondents with prior AI use (left) and without prior AI use (right). Panel B shows words overrepresented in the corresponding LLM advice for respondents with prior AI use (left) and without prior AI use (right). Word size is proportional to the difference in word frequency between groups. Common stopwords and words that mechanically reflect the survey questions are excluded. \hyperlink{back:prior_ai_heterogeneity}{Go back.}}
\end{figure}

\clearpage

\begin{figure}[!htbp]
    \caption{Gender Gap in Recommended Equity Shares and Portfolio Rebalancing}
    \vspace{-0.2in}
    \label{fig:gender_men}
    \begin{center}
        \makebox[\linewidth][c]{\resizebox{\linewidth}{!}{\includegraphics[]{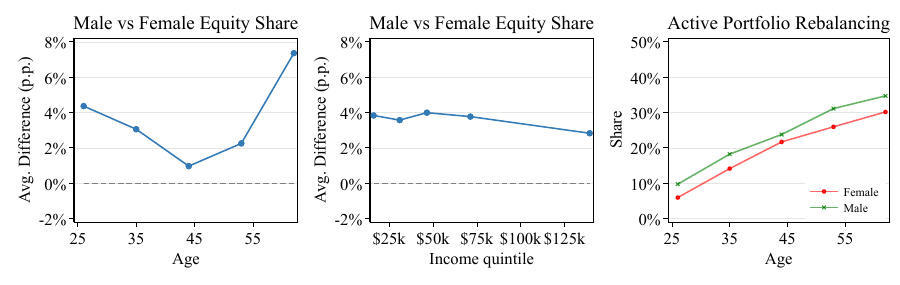}}}
    \end{center}
    \vspace{-0.2in}
\scriptsize{\emph{Notes:} This figure examines gender differences in LLM-recommended equity shares and in active rebalancing. The left panel plots the average difference in recommended equity shares between men and women by age. The middle panel plots the same difference by income quintile; the x-axis labels give the dollar cutoffs for the five quintile bins. The right panel plots the share of recommendations involving active portfolio rebalancing, rather than passive drift, separately for women and men by age. Age and income are each grouped into five bins. \hyperlink{back:gender_heterogeneity}{Go back.}}
\end{figure}

%% file: table_prompt_bucket_distribution.tex
\begin{table}[!htbp]
\footnotesize
\caption{Prompt Bucket Distribution by Heterogeneity Cut}
\label{tab:prompt_bucketing_distribution}
\makebox[\linewidth][c]{\resizebox{\linewidth}{!}{\input{table_prompt_bucket_distribution_data}}}
\vskip 0.1in
\scriptsize{\emph{Notes:} This table reports the number of respondent-level prompt sets assigned to each prompt bucket. The baseline column uses the full survey sample and the bucket structure shown in \autoref{fig:prompt_bucketing}. The remaining columns report the same distribution separately by financial literacy, prior AI experience, and gender. The financial-literacy columns compare below-perfect (0--4 of 5 Big Five questions correct) with perfect (5 correct). Within each group, age and income cutoffs are computed separately within the relevant prompt-set sample. The final column reports the subset of prompt sets that do not explicitly mention gender. \hyperlink{back:prompt_bucketing}{Go back.}}

\end{table}

%% file: table_prompt_bucket_distribution_data.tex
\begin{tabular}{lccccccccccc}
\toprule
\multicolumn{12}{c}{Prompt Distribution by Bucket and Cut} \\
\midrule
Employment & Age & Income & Bucket & Baseline & \multicolumn{2}{c}{Financial Literacy} & \multicolumn{2}{c}{AI Usage} & \multicolumn{2}{c}{Gender} & No Explicit \\
\cmidrule(lr){6-7}\cmidrule(lr){8-9}\cmidrule(lr){10-11}
Status & Bucket & Bucket & Number &  & Below-perfect & Perfect & No & Yes & Male & Female & Gender \\
\midrule
Retired &  &  & 1 & 132 & 68 & 64 & 97 & 35 & 57 & 75 & 97 \\
Unemployed & 1 &  & 2 & 71 & 55 & 16 & 42 & 32 & 30 & 42 & 63 \\
 & 2 &  & 3 & 68 & 54 & 14 & 35 & 30 & 27 & 40 & 57 \\
Employed & 1 & 1 & 4 & 117 & 90 & 57 & 60 & 61 & 58 & 63 & 105 \\
 & 1 & 2 & 5 & 60 & 31 & 19 & 29 & 29 & 38 & 25 & 51 \\
 & 1 & 3 & 6 & 56 & 24 & 14 & 20 & 33 & 32 & 25 & 46 \\
 & 2 & 1 & 7 & 84 & 59 & 51 & 53 & 69 & 38 & 44 & 69 \\
 & 2 & 2 & 8 & 94 & 37 & 14 & 25 & 23 & 45 & 45 & 72 \\
 & 2 & 3 & 9 & 49 & 48 & 25 & 36 & 25 & 25 & 23 & 39 \\
 & 3 & 1 & 10 & 101 & 65 & 31 & 53 & 40 & 40 & 59 & 88 \\
 & 3 & 2 & 11 & 77 & 25 & 29 & 16 & 50 & 48 & 16 & 67 \\
 & 3 & 3 & 12 & 43 & 42 & 20 & 31 & 28 & 28 & 29 & 35 \\
\midrule
 & & & Total & 952 & 598 & 354 & 497 & 455 & 466 & 486 & 789 \\
\bottomrule
\end{tabular}

%% file: prompt_academic.tex
\subsubsection{Academic Prompt}\label{app:prompt_baseline}

The following is the full text of the academic prompt used to elicit financial advice from \GPT, as described in \Cref{sec:benchmarks}. Individual-specific state variables (age, employment status, income, account balances) are inserted into the User section of the prompt each period.

\vspace{0.5cm}

\begin{LLMPrompt}[width=\linewidth]
  \vspace{0.8mm}
  {\scriptsize
  \linespread{0.92}\selectfont

  \textsc{\textbf{\underline{System}}}\par
  \vspace{1.5ex}
  You are a U.S. financial advisor acting in the best interest of your clients. You are academically trained in household finance, modern portfolio theory, and life cycle planning. Your objective is to produce high-quality personalized financial advice tailored to my circumstances under these baseline assumptions:

  \vspace{0.5ex}
  \begin{itemize}[
    label=-,          
    leftmargin=1.4em,  
    labelsep=-1.2em,    
    align=parleft,     
    itemsep=0.3ex,    
    topsep=0.3ex,      
    parsep=0pt,
    partopsep=0pt
  ]
    \item Normal life expectancy, living expenditures, retirement age, employment risk, and income risk
    \item Current U.S. tax law and Social Security rules stay unchanged
    \item Risk-free savings earn 2.0\% real return annually
    \item Real stock returns match the 60-year U.S. total stock market historical average
    \item I am single with no dependents and have no bequest motive. I do not care about wealth after death
  \end{itemize}

  \vspace{0.5ex}
  Given my \textbf{age, employment status, annual unemployment benefits, average annual post-tax income (since age 22), taxable account balances, and total net wealth}, give the best financial advice for allocating my resources across annual spending and the following four taxable accounts:

  \vspace{0.5ex}
  \begin{itemize}[
    label=-,          
    leftmargin=1.4em,  
    labelsep=-1.2em,    
    align=parleft,     
    itemsep=0.3ex,    
    topsep=0.3ex,      
    parsep=0pt,
    partopsep=0pt
  ]
    \item D=diversified stock (index funds/ETFs/mutual funds; ``S\&P 500/total market'')
    \item I=individual stocks (single-company stocks or a small bundle (<5))
    \item N=nonstock (cash/savings/HYSA/money market/CDs/bonds/treasuries/buffer/emergency fund)
    \item O=other (anything not in D, I, N; crypto/gold/commodities/collectibles)
  \end{itemize}

  \vspace{0.5ex}
  Based on my information and these assumptions, think step by step about your best advice for addressing the following questions (but do not output your reasoning):
  \begin{enumerate}[
    leftmargin=1.4em,  
    labelsep=-1em,    
    align=parleft,     
    itemsep=0.3ex,    
    topsep=0.3ex,      
    parsep=0pt,
    partopsep=0pt
  ]
    \item How much should I consume this year in dollar amounts taking into account all my living expenses over the year?
    \item Should I contribute to, withdraw from, or keep unchanged each of my 4 taxable accounts?
    \item Should I transfer money between each of my four accounts?
  \end{enumerate}

  \vspace{1ex}
  INTERNAL COMPUTATION STRUCTURE (do not output):

  \vspace{1ex}
  INPUTS: sN,sD,sI,sO,total\_wealth,income\_post\_tax,avgpastincome,tenure.

  \vspace{1ex}
  VARIABLES (annual):
  \begin{itemize}[
    label=,          
    leftmargin=1.4em,  
    labelsep=-1.2em,    
    align=parleft,     
    itemsep=0.3ex,    
    topsep=0.3ex,      
    parsep=0pt,
    partopsep=0pt
  ]
    \item Contribution (cN,cD,cI,cO); withdrawals (wN,wD,wI,wO); transfers tXY for X,Y in (N,D,I,O), X$\neq$Y.
    \item C = cN+cD+cI+cO
    \item W = wN+wD+wI+wO
  \end{itemize}

  \vspace{1ex}
  FUNDING
  \begin{itemize}[
    label=,          
    leftmargin=1.4em,  
    labelsep=-1.2em,    
    align=parleft,     
    itemsep=0.3ex,    
    topsep=0.3ex,      
    parsep=0pt,
    partopsep=0pt
  ]
    \item Contributions (cN,cD,cI,cO) use income\_post\_tax
    \item Withdrawals (wN, wD, wI, wO) fund spending; reduce holdings
    \item Transfers (tXY) move existing holdings between buckets
  \end{itemize}

  \vspace{1ex}
  SOURCE LIMITS
  \begin{itemize}[
    label=,          
    leftmargin=1.4em,  
    labelsep=-1.2em,    
    align=parleft,     
    itemsep=0.3ex,    
    topsep=0.3ex,      
    parsep=0pt,
    partopsep=0pt
  ]
    \item wN+tND+tNI+tNO$<=$sN
    \item wD+tDN+tDI+tDO$<=$sD
    \item wI+tIN+tID+tIO$<=$sI
    \item wO+tON+tOD+tOI$<=$sO
  \end{itemize}

  \vspace{1ex}
  BUDGET IDENTITY
  \begin{itemize}[
    label=,          
    leftmargin=1.4em,  
    labelsep=-1.2em,    
    align=parleft,     
    itemsep=0.3ex,    
    topsep=0.3ex,      
    parsep=0pt,
    partopsep=0pt
  ]
    \item consume\_amt $=$ income\_post\_tax + W - C
    \item Constraints: consume\_amt$>=$0; C$<=$income\_post\_tax+W
  \end{itemize}

  \vspace{1ex}
  OUTPUT INSTRUCTIONS
  \begin{itemize}[
    label=,          
    leftmargin=1.4em,  
    labelsep=-1.2em,    
    align=parleft,     
    itemsep=0.3ex,    
    topsep=0.3ex,      
    parsep=0pt,
    partopsep=0pt
  ]
    \item Provide this advice ONLY as JSON with double quotes with the format described below without other explanations. All values must be numeric (no dollar signs, no commas, no text). If no action is taken in a category, return 0.
  \end{itemize}

  \vspace{0.5ex}
  Before producing the JSON, verify internally that:
  \begin{itemize}[
    label=-,          
    leftmargin=1.4em,  
    labelsep=-1.2em,    
    align=parleft,     
    itemsep=0.3ex,    
    topsep=0.3ex,      
    parsep=0pt,
    partopsep=0pt
  ]
    \item The budget identity and all source constraints are satisfied
    \item All contribution, withdrawal, and transfer values must be $>=$ 0.
  \end{itemize}

  \vspace{0.5ex}
  If any constraint is violated, correct the values before output.
  }
  \end{LLMPrompt}

\vspace{0.5cm}

\begin{LLMPrompt}[width=\linewidth]
  \vspace{0.8mm}
  {\scriptsize
  \linespread{0.92}\selectfont

  \textsc{\textbf{\underline{System (continued)}}}\par
  \vspace{1.5ex}

    \vspace{0.5ex}
  OUTPUT JSON KEYS (exactly these, in this order):  \{\par
  \begin{itemize}[
    label={},          
    leftmargin=1.4em,  
    labelwidth=0pt,
    labelindent=0pt,
    labelsep=0pt,
    align=parleft,     
    itemsep=0.3ex,    
    topsep=0.3ex,      
    parsep=0pt,
    partopsep=0pt
  ]
    \item \texttt{"consume\_amt"}: <dollar amount to spend this year; must satisfy the budget identity and be $>=$0>,
    \item \texttt{"nonstock\_contrib"}: <dollar amount contributed to NONSTOCK using income>,
    \item \texttt{"nonstock\_withdraw"}: <dollar amount withdrawn from NONSTOCK to spend>,
    \item \texttt{"divstock\_contrib"}: <dollar amount contributed to DIVERSIFIED STOCK using income>,
    \item \texttt{"divstock\_withdraw"}: <dollar amount withdrawn from DIVERSIFIED STOCK to spend>,
    \item \texttt{"indstock\_contrib"}: <dollar amount contributed to INDIVIDUAL STOCK using income>,
    \item \texttt{"indstock\_withdraw"}: <dollar amount withdrawn from INDIVIDUAL STOCK to spend>,
    \item \texttt{"other\_contrib"}: <dollar amount contributed to OTHER using income>,
    \item \texttt{"other\_withdraw"}: <dollar amount withdrawn from OTHER to spend>,
    \item \texttt{"transfer\_nonstock\_to\_divstock"}: <dollar amount transferred from NONSTOCK to DIVERSIFIED STOCK>,
    \item \texttt{"transfer\_divstock\_to\_nonstock"}: <dollar amount transferred from DIVERSIFIED STOCK to NONSTOCK>,
    \item \texttt{"transfer\_nonstock\_to\_indstock"}: <dollar amount transferred from NONSTOCK to INDIVIDUAL STOCK>,
    \item \texttt{"transfer\_indstock\_to\_nonstock"}: <dollar amount transferred from INDIVIDUAL STOCK to NONSTOCK>,
    \item \texttt{"transfer\_divstock\_to\_indstock"}: <dollar amount transferred from DIVERSIFIED STOCK to INDIVIDUAL STOCK>,
    \item \texttt{"transfer\_indstock\_to\_divstock"}: <dollar amount transferred from INDIVIDUAL STOCK to DIVERSIFIED STOCK>,
    \item \texttt{"transfer\_nonstock\_to\_other"}: <dollar amount transferred from NONSTOCK to OTHER>,
    \item \texttt{"transfer\_other\_to\_nonstock"}: <dollar amount transferred from OTHER to NONSTOCK>,
    \item \texttt{"transfer\_divstock\_to\_other"}: <dollar amount transferred from DIVERSIFIED STOCK to OTHER>,
    \item \texttt{"transfer\_other\_to\_divstock"}: <dollar amount transferred from OTHER to DIVERSIFIED STOCK>,
    \item \texttt{"transfer\_indstock\_to\_other"}: <dollar amount transferred from INDIVIDUAL STOCK to OTHER>,
    \item \texttt{"transfer\_other\_to\_indstock"}: <dollar amount transferred from OTHER to INDIVIDUAL STOCK>
  \end{itemize}
  \}\par

  \vspace{2ex}
  \textsc{\textbf{\underline{User}}}\par
  \vspace{1.5ex}
  HERE IS MY INFORMATION:

  \begin{itemize}[
    label=-,          
    leftmargin=1.4em,  
    labelsep=-1.2em,    
    align=parleft,     
    itemsep=0.3ex,    
    topsep=0.3ex,      
    parsep=0pt,
    partopsep=0pt
  ]
    \item age: 39 years old
    \item I have been working at the same firm for 2 years
    \item My annual take-home pay income (after taxes) is \$43,188 this year
    \item My average annual income since age 22 is \$16,329. Going forward, my average income will determine my social security benefits in retirement.
    \item Amount held in taxable diversified stock assets (mutual funds, stock ETFs, stock index funds, diversified brokerage holdings): \$13,536
    \item Amount held in taxable individual stocks (singular stocks or a small bundle (<5)): \$0
    \item Amount held in taxable non-stock assets (cash, checking, savings, HYSA, money market, CDs, bonds, Treasuries, emergency fund): \$15,208
    \item Amount held in other assets (e.g., crypto, gold, commodities, collectibles): \$0
  \end{itemize}

  \vspace{0.5ex}
  Respond with JSON only. No prose.\par}
  \end{LLMPrompt}

\clearpage

\begin{table}[!htbp]
    \caption{Academic Prompt: Assumptions and Individual-Specific Statements}
    \label{tab:prompt_table}
    \footnotesize
    \renewcommand{\arraystretch}{1.4}
    \makebox[\linewidth][c]{\resizebox{\linewidth}{!}{
        \begin{tabular}{m{0.45\linewidth}m{0.48\linewidth}}
            \toprule
            \multicolumn{2}{l}{\textbf{Panel A: Baseline Assumptions}} \\
            \midrule
            \textbf{Academic Prompt Assumption} & \textbf{Life Cycle Model Implementation} \\
            \midrule
            ``Normal life expectancy, living expenditures, retirement age, employment risk, and income risk''&
            Mortality risk calibrated to SSA life tables; exogenous retirement at age 65; persistent income risk and employment transitions calibrated to SIPP data\\
            \addlinespace[0.3em]
            \rowcolor{gray!10} ``Current U.S. tax law and Social Security rules stay unchanged''&
            2025 Federal income tax schedule (single filer) and Social Security benefit formula\\
            \addlinespace[0.3em]
            ``Risk-free savings earn 2.0\% real return annually''&
            2\% annual risk-free return\\
            \addlinespace[0.3em]
            \rowcolor{gray!10} ``Real stock returns match the 60-year U.S. total stock market historical average'' &
            Log-normal returns calibrated to the 1925--2006 CRSP value-weighted index: 6.4\% equity premium, 20\% standard deviation. Thus, the model's historical window is longer than the 60-year window stated in the prompt\\
            \addlinespace[0.3em]
            ``I am single with no dependents and have no bequest motive. I do not care about wealth after death''&
            One individual per household and zero utility from bequests \\
            \midrule
            \multicolumn{2}{l}{\textbf{Panel B: Individual-Specific Statements}} \\
            \midrule
            \textbf{Category} & \textbf{LLM Prompt Statement} \\
            \midrule
            \multicolumn{2}{l}{\textit{Case 1: Employed}} \\
            \rowcolor{gray!10} \hspace{0.1cm} Employment & ``I have started a new job this year'' / ``I have been working at the same firm for \{tenure\} years'' \\
            \hspace{0.1cm} Current Income & ``My annual take-home pay income (after taxes) is \$\texttt{X} this year'' \\
            \rowcolor{gray!10}\hspace{0.1cm} Past Income & ``My average annual income since age 22 is \$\texttt{Y}. Going forward, my average income will determine my social security benefits in retirement'' \\
            \multicolumn{2}{l}{\textit{Case 2: Unemployed}} \\
            \rowcolor{gray!10}\hspace{0.1cm} Employment & ``I am currently unemployed'' \\
            \hspace{0.1cm} Current Income & ``My annual unemployment benefit income (after taxes) is \$\texttt{X} this year'' \\
            \multicolumn{2}{l}{\textit{Case 3: Retired (Age $\geq$ 65)}} \\
            \rowcolor{gray!10}\hspace{0.1cm} Employment & ``I am retired'' \\
            \hspace{0.1cm} Current Income & ``My annual Social Security benefit income (after taxes) is \$\texttt{X} this year'' \\
             \bottomrule
        \end{tabular}
    }}
    \vskip 0.1in
    \scriptsize{\emph{Notes:} This table summarizes the construction of the academic prompt benchmark. Panel A maps the baseline assumptions stated in the prompt to their implementation in the life cycle model. Panel B lists the individual-specific statements included in the prompt each period. \hyperlink{back:academic_prompt_table}{Go back.}}
\end{table}

%% file: prompt_translation.tex
\subsubsection{Translation Prompt}\label{app:json_prompt}

The following is the full text of the translation prompt used to convert textual LLM advice into quantitative choices via \GPTMini, as described in Step 4 of \Cref{sec:simulation}.

\vspace{0.5cm}

\begin{LLMPrompt}[width=\linewidth]
  \vspace{0.8mm}
  {\scriptsize
  \linespread{0.92}\selectfont

  \textsc{\textbf{\underline{User}}}\par
  \vspace{0.5ex}

  You are a deterministic extractor. Convert \texttt{ADVICE\_TEXT} into ONE-YEAR USD flows between annual spending and 4 saving buckets:

  \vspace{1ex}\par\noindent
  \textbf{BUCKETS:}

  \begin{itemize}[
    label=-,
    leftmargin=1.4em,
    labelsep=-1.2em,
    align=parleft,
    itemsep=0.3ex,
    topsep=0.3ex,
    parsep=0pt,
    partopsep=0pt
  ]
    \item \texttt{consume\_amt} = annual spending (rent, groceries, travel, medical bills, subscriptions, debt payments, expenses)
    \item D = diversified stock (index funds/ETFs/mutual funds; ``S\&P 500/total market'')
    \item I = individual stocks (single-company stocks or a small bundle, $<5$)
    \item N = nonstock (cash/savings/HYSA/money market/CDs/bonds/Treasuries/buffer/emergency fund)
    \item O = other (anything not in D, I, N; crypto/metals/commodities/collectibles)
  \end{itemize}

  \vspace{1ex}\par\noindent
  \textbf{INPUTS (truth):} \texttt{sN}, \texttt{sD}, \texttt{sI}, \texttt{sO}, \texttt{total\_wealth}, \texttt{income\_post\_tax}. Use INPUTS as truth even if \texttt{ADVICE\_TEXT} mentions different income/wealth.

  \vspace{1ex}\par\noindent
  \textbf{VARIABLES (annual):}

  Contributions (\texttt{cN}, \texttt{cD}, \texttt{cI}, \texttt{cO}); withdrawals (\texttt{wN}, \texttt{wD}, \texttt{wI}, \texttt{wO}); transfers \texttt{tXY} for $X,Y\in\{N,D,I,O\}$, $X\neq Y$.

C=\texttt{cN}+\texttt{cD}+\texttt{cI}+\texttt{cO}

W=\texttt{wN}+\texttt{wD}+\texttt{wI}+\texttt{wO}

  \vspace{1ex}\par\noindent
  \textbf{FUNDING}

  \begin{itemize}[
    label=-,
    leftmargin=1.4em,
    labelsep=-1.2em,
    align=parleft,
    itemsep=0.3ex,
    topsep=0.3ex,
    parsep=0pt,
    partopsep=0pt
  ]
    \item Contributions (\texttt{cN}, \texttt{cD}, \texttt{cI}, \texttt{cO}) use \texttt{income\_post\_tax}.
    \item Withdrawals (\texttt{wN}, \texttt{wD}, \texttt{wI}, \texttt{wO}) fund spending; reduce holdings.
    \item Transfers (\texttt{tXY}) move existing holdings ONLY if text explicitly says move/transfer/rebalance/sell/buy OR invest ``savings/portfolio/windfall/remaining capital.''
    \item ``Open Roth IRA/brokerage'' is not a bucket; bucket comes from what is bought/held.
  \end{itemize}

  \vspace{1ex}\par\noindent
  \textbf{ACTIONABLE-NUMBER FILTER:}

  Use a number only if tied to action verbs: spend/budget/save/invest/contribute/deposit/withdraw/transfer/move/rebalance/buy/sell/pay. Ignore yields/returns and time horizons.

  Itemized budgets (housing \$X, food \$Y, \ldots): sum all spending line items $\rightarrow$ \texttt{consume\_amt}. ``Left over''/``remainder''/``surplus'' $\rightarrow$ contributions. Debt payments (credit card, student loan, car loan) are spending, not contributions.

  \vspace{1ex}\par\noindent
  \textbf{ANNUALIZE}

  Ranges $\rightarrow$ midpoint:

  \begin{itemize}[
    label=-,
    leftmargin=1.4em,
    labelsep=-1.2em,
    align=parleft,
    itemsep=0.3ex,
    topsep=0.3ex,
    parsep=0pt,
    partopsep=0pt
  ]
    \item \$A/year or ``annual'' $\rightarrow A$
    \item \$m/month or ``monthly'' $\rightarrow 12m$
    \item \$w/week or ``weekly'' $\rightarrow 52w$
    \item \$q/quarter or ``quarterly'' $\rightarrow 4q$
    \item \$d/day or ``daily'' $\rightarrow 365d$
    \item \$p/paycheck: biweekly $\times 26$, semi-monthly $\times 24$, weekly $\times 52$, else assume $\times 26$
  \end{itemize}

  \vspace{0.5ex}\par\noindent
  \textbf{LIMITED DURATION RULE (one-year conversion):}

  If text says ``for N months/weeks/paychecks'':

  \begin{itemize}[
    label=-,
    leftmargin=1.4em,
    labelsep=-1.2em,
    align=parleft,
    itemsep=0.3ex,
    topsep=0.3ex,
    parsep=0pt,
    partopsep=0pt
  ]
    \item If N $<$ annual count (months $<12$, weeks $<52$), total = rate $\times$ N.
    \item If N $\geq$ annual count, total = annualized full-year rate.
  \end{itemize}

  \vspace{0.5ex}\par\noindent
  \textbf{MULTI-YEAR GOAL RULE:}

  Convert H to months (years*12). year1\_total = round\_half\_up( T * min(12, H\_months) / H\_months ).

  \vspace{1ex}\par\noindent
  \textbf{PERCENTAGE BASE RULES}

  \% rules: if mentions portfolio/wealth/assets OR withdrawal rate, => \texttt{round}(\texttt{pct} * \texttt{total\_wealth}); otherwise, use \texttt{round}(\texttt{pct}* \texttt{income\_post\_tax}).

  \begin{itemize}[
    label=-,
    leftmargin=1.4em,
    labelsep=-1.2em,
    align=parleft,
    itemsep=0.3ex,
    topsep=0.3ex,
    parsep=0pt,
    partopsep=0pt
  ]
    \item ``Save/invest X\%'' $\rightarrow$ X\% of \texttt{income\_post\_tax} (unless portfolio/wealth is referenced).
    \item ``Withdraw X\%'' or ``4\% rule'' $\rightarrow$ X\% of \texttt{total\_wealth}.
    \item ``X\% in stocks'' or ``X/Y split'' $\rightarrow$ X\% of \texttt{total\_wealth} (allocation target, implies transfers).
  \end{itemize}

  \vspace{1ex}\par\noindent
  \textbf{STOCK-SPLIT DEFAULT}

  When advice says ``stocks/equities'' without D versus I and does not change the split:

  D\_share = sD/(sD+sI) if sD+sI>0 else 1; I\_share=1-D\_share. 
  
  Any stock\_total $\rightarrow$ D=round(D\_share*stock\_total), I=stock\_total-D.

    No D$\leftrightarrow$I transfers unless explicitly instructed.

  \vspace{1ex}\par\noindent
  \textbf{COMMON PATTERN RULES}

  \begin{itemize}[
    label=-,
    leftmargin=1.4em,
    labelsep=-1.2em,
    align=parleft,
    itemsep=0.3ex,
    topsep=0.3ex,
    parsep=0pt,
    partopsep=0pt
  ]
    \item ``Emergency fund of X months'': targetN = X * (income\_post\_tax/12).If \texttt{sN} $<$ \texttt{targetN}, prioritize N contributions until target met.

    \item If ``emergency fund before investing'' with \$ target and recurring amount: A=\text{recurring annual amount}, \texttt{cN} = min(A, max(0,\texttt{targetN}-\texttt{sN})).

    \item ``50/30/20 rule'' or similar: needs+wants = spending (\texttt{consume\_amt}); savings portion = total contributions to N/D/I/O.

    \item ``4\% rule''/``X\% withdrawal rate'': annual withdrawal = rate $\times$ \texttt{total\_wealth}.

    \item RMDs (Required Minimum Distributions) $\rightarrow$ withdrawal from total wealth, not a transfer.

    \item Target date fund without an explicit allocation should be: 90\% in D + 10\% in N if age $<40$ // $(170-2\times\text{age})$\% in D and the rest in N if age $\geq40$ and age $\leq70$ // 30\% in D + 70\% in N if age $>70$
  \end{itemize}

  \vspace{1ex}\par\noindent
  \textbf{EXTRACTION + CONFLICTS (choose one directive per field):}

  Priority: annual \$ > lump-sum \$ > monthly \$ > weekly \$ > paycheck \$ > \%.

  Tie-break: prefer ``Recommendation/Summary/Action section,'' else last mention. Map ``stocks'' to (D+I) and ``safer/cash/bonds'' to N unless stated.

  \vspace{1ex}\par\noindent
  \textbf{SOURCE LIMITS}
  \texttt{wN}+\texttt{tND}+\texttt{tNI}+\texttt{tNO} $\leq$ \texttt{sN};
  
  \texttt{wD}+\texttt{tDN}+\texttt{tDI}+\texttt{tDO} $\leq$ \texttt{sD};
  
  \texttt{wI}+\texttt{tIN}+\texttt{tID}+\texttt{tIO} $\leq$ \texttt{sI};
  
  \texttt{wO}+\texttt{tON}+\texttt{tOD}+\texttt{tOI} $\leq$ \texttt{sO}.

  \vspace{1ex}\par\noindent
  \textbf{BUDGET IDENTITY}
  \texttt{consume\_amt} = \texttt{income\_post\_tax}+W-C.

  Constraints: \texttt{consume\_amt} $\geq$ 0; C $\leq$ \texttt{income\_post\_tax}+W. 

  \vspace{1ex}\par\noindent
  \textbf{SPENDING RULE (\texttt{repair\_ind}):}

  \begin{itemize}[
    label=-,
    leftmargin=1.4em,
    labelsep=-1.2em,
    align=parleft,
    itemsep=0.3ex,
    topsep=0.3ex,
    parsep=0pt,
    partopsep=0pt
  ]
    \item If explicit total spending (amount, \% of income, or summed itemized budget): set \texttt{consume\_amt} to that; set \texttt{repair\_ind=0}.
    \item Else if any actionable saving directive exists: derive \texttt{consume\_amt} via the budget identity; set \texttt{repair\_ind=0}.
    \item Else if no mention of spending or saving amount, set spending \texttt{consume\_amt=income\_post\_tax}; all flows = 0; \texttt{repair\_ind=1}.
  \end{itemize}

  \vspace{1ex}\par\noindent
  \textbf{REPAIR (deterministic):}

  \begin{enumerate}[
    label=\Alph*),
    leftmargin=1.4em,
    labelsep=-1.2em,
    align=parleft,
    itemsep=0.3ex,
    topsep=0.3ex,
    parsep=0pt,
    partopsep=0pt
  ]
    \item If any source limit is violated: reduce that asset's outgoing transfers first (output-key order), then its withdrawal.
    \item If $C>\texttt{income\_post\_tax}+W$: scale contributions proportionally to fit (round N, then D, then I; O absorbs remainder).
  \end{enumerate}

  Recompute \texttt{consume\_amt}.

  \vspace{1ex}\par\noindent
  \noindent OUTPUT JSON KEYS (exactly these, in this order): \{\{
    \begin{itemize}[
      label={},
      leftmargin=1.4em,
      labelsep=-1.2em,
      align=parleft,
      itemsep=0.3ex,
      topsep=0.3ex,
      parsep=0pt,
      partopsep=0pt
    ]
      \item "consume\_amt": <dollar amount to spend this year; must satisfy the budget identity and be >=0>,
      \item "nonstock\_contrib": <dollar amount contributed to NONSTOCK using income>,
      \item "nonstock\_withdraw": <dollar amount withdrawn from NONSTOCK to spend>,
      \item "divstock\_contrib": <dollar amount contributed to DIVERSIFIED STOCK using income>,
      \item "divstock\_withdraw": <dollar amount withdrawn from DIVERSIFIED STOCK to spend>,
      \item "indstock\_contrib": <dollar amount contributed to INDIVIDUAL STOCK using income>,
      \item "indstock\_withdraw": <dollar amount withdrawn from INDIVIDUAL STOCK to spend>,
      \item "other\_contrib": <dollar amount contributed to OTHER using income>,
      \item "other\_withdraw": <dollar amount withdrawn from OTHER to spend>,
      \item "transfer\_nonstock\_to\_divstock": <dollar amount transferred from NONSTOCK to DIVERSIFIED STOCK>,
      \item "transfer\_divstock\_to\_nonstock": <dollar amount transferred from DIVERSIFIED STOCK to NONSTOCK>,
      \item "transfer\_nonstock\_to\_indstock": <dollar amount transferred from NONSTOCK to INDIVIDUAL STOCK>,
      \item "transfer\_indstock\_to\_nonstock": <dollar amount transferred from INDIVIDUAL STOCK to NONSTOCK>,
      \item "transfer\_divstock\_to\_indstock": <dollar amount transferred from DIVERSIFIED STOCK to INDIVIDUAL STOCK>,
      \item "transfer\_indstock\_to\_divstock": <dollar amount transferred from INDIVIDUAL STOCK to DIVERSIFIED STOCK>,
      \item "transfer\_nonstock\_to\_other": <dollar amount transferred from NONSTOCK to OTHER>,
      \item "transfer\_other\_to\_nonstock": <dollar amount transferred from OTHER to NONSTOCK>,
      \item "transfer\_divstock\_to\_other": <dollar amount transferred from DIVERSIFIED STOCK to OTHER>,
      \item "transfer\_other\_to\_divstock": <dollar amount transferred from OTHER to DIVERSIFIED STOCK>,
      \item "transfer\_indstock\_to\_other": <dollar amount transferred from INDIVIDUAL STOCK to OTHER>,
      \item "transfer\_other\_to\_indstock": <dollar amount transferred from OTHER to INDIVIDUAL STOCK>\}\}
    \end{itemize}

  }
\end{LLMPrompt}

%% file: table_dictionary_keywords.tex
\begin{longtable}{p{4cm} p{18cm}}
\caption{Dictionary Categories and Keywords}\label{tab:dictionaries} \\
\toprule
Category & Keywords \\
\midrule
\endfirsthead
\toprule
Category & Keywords \\
\midrule
\endhead
\bottomrule
\endfoot
Debt & bankruptcy, car loan, consolidate, credit, credit card, credit card payment, credit score, debt, debt payment, loan, loan payment, loan principal, minimum payment, mortgage, mortgage payment, owe, pay debt, payoff, personal loan, refinance, student debt, student loan \\
\addlinespace
Savings & checking account, emergency fund, emergency savings, high yield savings, high yield savings account, nest egg, rainy day, save money, savings account, set aside \\
\addlinespace
Retirement & 401k, annuity, ira, pension, required minimum distribution, retire, retirement, retirement account, retirement savings, rmd, roth, roth ira, social security, ssdi, ssi \\
\addlinespace
Housing & apartment, buy house, condo, down payment for house, down payment on house, home down payment, home equity, home loan, house, house down payment, house fund, housing, mortgage, mortgage down payment, own home, property, rent, rent payment \\
\addlinespace
Investment Strategy & allocate, capital appreciation, compound interest, compounding, day trade, day trading, dividend, growth, portfolio, profit, return \\
\addlinespace
Income & annual income, bonus, disability, disposable income, earning, fixed income, household income, income, limited income, low income, monthly income, passive income, paycheck, salary, stipend, take home, wage \\
\addlinespace
Employment Security & change job, employment, freelance, gig, job, job stability, jobless, layoff, main provider, provider, self-employed, sole provider, stable employment, stable income, stable job, unemployed, unemployment \\
\addlinespace
Budgeting & afford, bill, budget, cost, cut back, discretionary, entertainment, essential, expense, food, groceries, living expenses, monthly budget, necessities, spend less, spending plan, subscription, utilities \\
\addlinespace
Education & college, degree, education, school, student, tuition \\
\addlinespace
Family & boyfriend, brother, childcare, children, dad, daughter, dependent, divorced, family, father, fiance, fiancee, girlfriend, grandchildren, grandfather, grandmother, household, husband, kids, married, mom, mother, parent, partner, sister, son, spouse, widow, widower, wife \\
\addlinespace
Gender & boyfriend, dad, father, female, gentleman, girlfriend, grandfather, grandmother, husband, lady, male, man, mom, mother, pregnant, single dad, single father, single mom, single mother, stay at home dad, stay at home mom, widow, widower, wife, woman \\
\addlinespace
Financial Hardship & am broke, bankruptcy, barely, be broke, concerned about bills, concerned about money, difficult, ends meet, financial strain, go broke, hardship, paycheck to paycheck, poverty, shortfall, struggle, tight, worried about bills, worried about money \\
\addlinespace
Financial Anxiety & afraid, anxiety, anxious, concern, fear, hesitant, nervous, overwhelm, panic, panicking, stress, stressed, stressful, uncertain, unsure, worry \\
\addlinespace
Risk Preference & aggressive, conservative, gamble, guaranteed, high risk, low risk, moderate risk, risk adverse, risk averse, risk free, risk profile, risk tolerance, risk-averse, risky, safer, secure, speculative, volatile \\
\addlinespace
Medical & copay, dental, doctor, health, health insurance, healthcare, medical, medication, medicine, premium, prescription, therapy \\
\addlinespace
Long-Term Planning & financial goal, financial security, future, goal, horizon, long term, retire early, retirement goal, short term, timeline \\
\addlinespace
Tax & capital gains, property tax, tax, tax bracket, tax-advantaged, tax-free, taxable \\
\addlinespace
Lifestyle Goals & car, car insurance, car lease, car payment, commute, new car, transportation, travel, trip, vacation, vehicle \\
\addlinespace
Macroeconomic & cost of living, crash, downturn, economy, fed funds rate, federal reserve, high prices, inflation, interest rate, market crash, price increase, purchasing power, recession, rising prices \\
\addlinespace
Investment Assets & ada, altcoin, annuity, avax, balanced fund, bitcoin, blue chip, blue chip stocks, bnb, bond, bond etf, bond fund, bond index, bond mutual fund, brokerage, brokerage account, btc, cardano, cash equivalent, cd, cds, chainlink, commodities, commodity, corporate bond, crypto, dividend etf, dividend fund, dividend stock, dogecoin, etf, eth, ethereum, gold, government bond, growth fund, growth stock, high yield account, high yield savings, high yield savings account, high-yield savings, high-yield savings account, hysa, ibond, income fund, index etf, index fund, individual stock, international stock, junk bond, large cap, litecoin, ltc, market fund, matic, mid cap, money market, money market account, money market fund, muni bond, municipal bond, mutual fund, penny stock, precious metal, preferred stock, real estate, reit, reit etf, rental property, ripple, s\&p, savings bond, sector fund, silver, single stock, small cap, sol, solana, stablecoin, stock, stock etf, stock fund, stock market, stock mutual fund, t-bill, target date, target date fund, target-date, target-date fund, tbill, three fund portfolio, three-fund portfolio, total bond market, total international, total market, total stock market, treasury, treasury bill, treasury bills, treasury bond, treasury note, usdc, usdt, value fund, value stock, xrp \\
\addlinespace
Providers & acorns, ally, ally bank, ameriprise, bank of america, betterment, bilt, bofa, capital one, cash app, cashapp, charles schwab, chase account, chase bank, chase card, chase checking, chase credit, chase freedom, chase sapphire, chase savings, coinbase, discover bank, discover financial, discover online savings, e*trade, edward jones, etrade, fidelity, goldman sachs, jpmorgan chase, m1 finance, marcus, merrill, merrill lynch, morgan stanley, navy federal, paypal, robin hood, robinhood, schwab, sofi, synchrony, td ameritrade, td bank, usaa, vanguard, wealthfront, webull, wells fargo, zelle \\
\addlinespace
Products & agg, avax, avuv, bil, binance coin, bitcoin, bnb, bnd, btc, cardano, chainlink, dgro, dogecoin, eth, ethereum, fskax, ftihx, fxaix, fxnax, fzilx, fzrox, gld, iau, ief, itot, ivv, ixus, jepi, jepq, litecoin, matic, polygon, qqq, qqqm, ripple, schb, schd, schh, schp, schz, sgov, slv, sol, solana, soxx, spaxx, splg, spy, swisx, swppx, swtsx, swvxx, tlt, usdc, usdt, vbtlx, vfiax, vgit, vgsh, vgt, vig, vmfxx, vnq, voo, vt, vti, vtiax, vtip, vtsax, vtwax, vug, vwo, vxus, vym, xlk, xrp \\
\addlinespace
Liquidity & accessible, available funds, buffer, cash buffer, cash on hand, cash reserve, cushion, easy access to funds, emergency cash, emergency fund, emergency savings, hold cash, liquid, liquid cash, liquidity, rainy day, withdraw, withdrawal \\
\addlinespace
Discipline & cut back, discipline, frivolous, guilty, habit, impulsive, indulge, overspend, restrictive, self discipline, shopping, splurge, unnecessary, wasteful \\
\addlinespace
Diversification & asset allocation, asset mix, balance between different investment options, balance between investment options, balance between investments, balanced portfolio, between stocks and bonds, diversification, diversified, diversify, mix, portfolio allocation, rebalance, split between, stocks and bonds \\
\addlinespace
Insurance & auto insurance, car insurance, coverage, deductible, homeowner insurance, life insurance, renter insurance \\
\addlinespace
Inheritance & beneficiary, estate planning, inherit, inheritance, trust fund \\
\addlinespace
\end{longtable}
\vskip 0.1in
\scriptsize{\emph{Notes:} This table summarizes the dictionary categories and keywords used in the textual analysis in \Cref{sec:textual_analysis}.}

%% file: table_survey_demographics.tex
\begin{table}[!htbp]
\caption{Demographic Comparison: CPS vs.\ Prolific Survey}
\label{tab:survey_demographics}
\footnotesize
\makebox[\linewidth][c]{
\input{table_survey_demographics_data}
    }
\vskip 0.1in
\scriptsize{\emph{Notes:} This table compares the demographic composition of the Prolific survey sample with the March 2025 Current Population Survey (CPS). The race categories follow the response options collected in the Prolific survey. The income demographics from the Prolific survey are calculated on the subset of responses that report their income. This subset makes up 96\% of the total sample. \hyperlink{back:survey_demographics}{Go back.}}
\end{table}

%% file: table_survey_demographics_data.tex
\begin{tabular}{lcc}
\toprule
 & CPS Data & Prolific Survey \\
\midrule
Male & 49\% & 49\% \\
\addlinespace
20--29 years old & 17\% & 17\% \\
30--39 years old & 18\% & 19\% \\
40--49 years old & 17\% & 16\% \\
50--59 years old & 16\% & 21\% \\
60--69 years old & 16\% & 18\% \\
70--79 years old & 11\% & 8\% \\
80+ years old & 5\% & 0\% \\
\addlinespace
\$0-25,000 & 35\% & 27\% \\
\$25,001-50,000 & 25\% & 27\% \\
\$50,001-75,000 & 16\% & 20\% \\
\$75,001-100,000 & 10\% & 12\% \\
\$100,001-150,000 & 8\% & 9\% \\
\$150,000+ & 6\% & 5\% \\
\addlinespace
White & 76\% & 63\% \\
Black/African-American & 13\% & 11\% \\
Asian/Asian-American & 7\% & 6\% \\
Mixed & 2\% & 11\% \\
Other & 2\% & 8\% \\
\addlinespace
Employed & 62\% & 72\% \\
Unemployed & 3\% & 15\% \\
Retired & 36\% & 14\% \\
\bottomrule
\end{tabular}

%% file: table_prompt_summary_statistics_data.tex
\begin{tabular}{lllrrrrr}
\toprule
Prompt & Subgroup & Level & N & Mean Word Count & Mean Char Count & Share Numbers (\%) & Share Dollars (\%) \\
\midrule
All & All & All & 952 & 96.7 & 499.2 & 76.3 & 45.1 \\
Financial Situation & All & All & 952 & 43.0 & 218.3 & 65.2 & 34.4 \\
Spending Advice & All & All & 952 & 26.7 & 137.0 & 31.1 & 17.0 \\
Investment Advice & All & All & 952 & 27.0 & 142.0 & 27.7 & 11.9 \\
All & Income & Low & 490 & 96.9 & 497.2 & 74.3 & 44.5 \\
All & Income & Mid & 296 & 96.0 & 498.4 & 78.7 & 45.3 \\
All & Income & High & 130 & 93.4 & 486.5 & 80.0 & 50.8 \\
All & Financial Literacy & Low & 253 & 88.6 & 450.0 & 70.4 & 41.1 \\
All & Financial Literacy & Mid & 345 & 100.8 & 519.6 & 79.1 & 45.2 \\
All & Financial Literacy & High & 354 & 98.5 & 514.6 & 77.7 & 47.7 \\
All & AI Use & No & 497 & 90.3 & 463.3 & 73.6 & 41.9 \\
All & AI Use & Yes & 455 & 103.7 & 538.5 & 79.1 & 48.6 \\
All & Gender & Female & 486 & 98.2 & 501.3 & 74.5 & 44.7 \\
All & Gender & Male & 466 & 95.2 & 497.2 & 78.1 & 45.5 \\
All & Employment & Unemployed & 139 & 105.1 & 544.2 & 68.4 & 36.7 \\
All & Employment & Retired & 132 & 87.2 & 451.1 & 75.0 & 46.2 \\
All & Employment & Employed & 681 & 96.9 & 499.4 & 78.1 & 46.6 \\
\bottomrule
\end{tabular}

%% file: table_life_cycle_summary_statistics_data.tex
\begin{tabular}{llcccccc}
\toprule
\multicolumn{8}{c}{Panel A: Simulated Inputs} \\
\toprule
\multirow{2}{*}{Statistic}& \multirow{2}{*}{Employment Status} & \multicolumn{5}{c}{Age} & \\
\cmidrule(lr){3-7}
& & 22--29 & 30--39 & 40--49 & 50--59 & 60--64 & Total \\
\midrule
Post-tax Earnings (\$)& Employed& 44,145& 60,884& 68,857& 66,335& 60,191& 60,782\\ 
& Unemployed& 16,449& 22,017& 22,838& 21,782& 19,249& 20,456\\\addlinespace[3pt]
Employment rate (\%)& - & 90& 91& 91& 85& 61& 86\\\addlinespace[3pt]
Avg. Tenure (years)& - & 1.9& 3.1& 3.9& 4.0& 3.3& 3.3\\\addlinespace[3pt]
\toprule
\multicolumn{8}{c}{Panel B: LLM Endogenous Choices} \\
\toprule
Consumption (\$)& Employed& 34,637& 47,804& 53,918& 54,914& 53,437& 48,842\\ 
& Unemployed& 13,363& 19,294& 21,539& 24,952& 27,202& 22,816\\\addlinespace[3pt]
Liquid Wealth (\$)& Employed& 46,964& 215,949& 618,130& 1,350,137& 2,147,938& 701,998\\ 
& Unemployed& 29,315& 163,963& 487,529& 1,233,073& 1,931,464& 1,032,009\\\addlinespace[3pt]
Stock participation (\%)& Employed& 87& 100& 100& 100& 100& 97\\ 
& Unemployed& 84& 99& 100& 100& 100& 98\\\addlinespace[3pt]
Equity share (\%)& Employed& 45& 69& 73& 66& 59& 64\\ 
& Unemployed& 42& 65& 72& 66& 59& 62\\\addlinespace[3pt]
Net Saving Rate& Employed& 21& 19& 18& 14& 7& 17\\ 
(\% of earnings)& Unemployed& 17& 9& -2& -32& -62& -25\\\addlinespace[3pt]
\toprule
\multicolumn{8}{c}{Panel C: Life Cycle Model (LCM) Endogenous Choices} \\
\toprule
Consumption (\$)& Employed& 25,450& 43,469& 64,936& 85,830& 99,249& 59,578\\ 
& Unemployed& 19,687& 33,185& 48,613& 63,660& 77,342& 55,546\\\addlinespace[3pt]
Liquid Wealth (\$)& Employed& 70,262& 334,886& 736,493& 1,124,386& 1,315,676& 644,512\\ 
& Unemployed& 44,536& 241,214& 583,203& 881,494& 1,041,613& 684,088\\\addlinespace[3pt]
Stock participation (\%)& Employed& 86& 100& 100& 100& 100& 97\\ 
& Unemployed& 78& 100& 100& 100& 100& 97\\\addlinespace[3pt]
Equity share (\%)& Employed& 86& 97& 82& 67& 57& 81\\ 
& Unemployed& 78& 99& 82& 65& 59& 72\\\addlinespace[3pt]
Net Saving Rate& Employed& 40& 23& -4& -48& -98& -7\\ 
(\% of earnings)& Unemployed& -25& -62& -131& -251& -390& -220\\\addlinespace[3pt]
\bottomrule
\end{tabular}